\documentclass[a4paper,11pt]{article}
\pdfoutput=1 

\usepackage{jheppub}
\usepackage{epsfig}
\usepackage{amsmath}
\def\be{\begin{equation}}
\def\ee{\end{equation}}
\def\bea{\begin{array}}
\def\beqa{\begin{eqnarray}}
\def\eeqa{\end{eqnarray}}
\def\beqas{\begin{eqnarray*}}
\def\eeqas{\end{eqnarray*}}

\def\bp{\begin{picture}}
\def\ep{\end{picture}}
\def\bc{\begin{center}}
\def\ec{\end{center}}
\def\bfig{\begin{figure}}
\def\efig{\end{figure}}

\def\bit{\begin{itemize}}
\def\eit{\end{itemize}}
\def\nn{\nonumber}
\def\f{\frac}

\def\[{\left[}
\def\]{\right]}
\def\({\left(}
\def\){\right)}

\def\..{\left.}
\def\.{\right.}
\def\tl{\tilde}
\def\ra{\rightarrow}
\def\la{\leftarrow}

\def\tm{\times}
\def\gev{\rm GeV}
\def\tev{\rm TeV}

\def\la{\lambda}

\def\al{\alpha}

\def\ep{\epsilon}

\def\ga{\gamma}
\def\pa{\partial}
\def\pr{\prime}

\title{Gluino-SUGRA scenarios in light of FNAL muon g-2 anomaly}
\author[a]{Zhuang Li,}
\author[a]{Guo-Li Liu,}
\author[a,b]{Fei Wang*,\note{*Corresponding author.}}
\author[b,c]{Jin Min Yang,}
\author[a]{Yang Zhang,}
\affiliation[a]{School of Physics, Zhengzhou University,\\450000,ZhengZhou
P.R.China}
\affiliation[b]{Institute of Theoretical Physics,Chinese Academy of Sciences,\\Beijing 100080, P. R. China}
\affiliation[c]{School of Physics, University of Chinese Academy of Sciences,\\ Beijing 100049, P. R. China}
\emailAdd{lizhuang@gs.zzu.edu.cn}
\emailAdd{guoliliu@zzu.edu.cn}
\emailAdd{feiwang@zzu.edu.cn}
\emailAdd{jmyang@itp.ac.cn}
\emailAdd{zhangyangphy@zzu.edu.cn}
\abstract{Gluino-SUGRA ($\tl{g}$SUGRA), which is an economical extension of the predictive mSUGRA, adopts much heavier gluino mass parameter than other gauginos mass parameters and universal scalar mass parameter at the unification scale. It can elegantly reconcile the experimental results on the Higgs boson mass, the muon $g-2$,
the null results in search for supersymmetry at the LHC and the results from B-physics.
In this work, we propose several new ways to generate large gaugino hierarchy (i.e. $M_3\gg M_1,M_2$) for $\tl{g}$SUGRA model building and then discuss in detail the implications of the new muon $g-2$ results with the updated LHC constraints on such $\tl{g}$SUGRA scenarios. We obtain the following observations: (i) For the most interesting $M_1=M_2$ case at the GUT scale with a viable bino-like dark matter, the $\tl{g}$SUGRA can explain the muon $g-2$ anomaly at $1\sigma$ level and be consistent with the updated LHC constraints for $6\leq M_3/M_1 \leq 9$ at the GUT scale; (ii) For $M_1:M_2=5:1$ at the GUT scale with wino-like dark matter, the $\tl{g}$SUGRA model can explain the muon $g-2$ anomaly at $2\sigma$ level and be consistent with the updated LHC constraints for $3\leq M_3/M_1 \leq 3.2$ at the GUT scale; (iii) For $M_1:M_2=3:2$ at the GUT scale with mixed bino-wino dark matter, the $\tl{g}$SUGRA model can explain the muon $g-2$ anomaly at $1\sigma$ level and be consistent with the updated LHC constraints for $6.9\leq M_3/M_1 \leq 7.5$ at the GUT scale.  Although the choice of heavy gluino will always increase the FT involved, some of the $1\sigma/2\sigma$ survived points of $\Delta a_\mu^{combine}$ can still allow low EWFT of order several hundreds and be fairly natural. Constraints from (dimension-five operator induced) proton decay are also discussed.}

\begin{document}
\maketitle
\flushbottom
\section{Introduction}
The E989 muon $g-2$ experiment in Fermilab recently reported its first measurement. Combining with the BNL result, it gives a value for the muon anomalous magnetic moment $a_\mu\equiv (g-2)_\mu/2$~\cite{FNAL:gmuon}
\beqa
a^{\rm FNAL+BNL}_\mu = (11659206.2 \pm 4.1) \tm 10^{-10}~,
\eeqa
which is in full agreement with the previous BNL measurement~\cite{BNL:gmuon}. Using the new combined result of FNAL and BNL, the world average deviation from the Standard Model prediction~\cite{SMmuong-2}
 \beqa
\Delta a^{\rm{FNAL+BNL}}_\mu =(25.1 \pm  5.9)  \tm 10^{-10}
 \eeqa
is pushed to a significance of $4.2\sigma$. Such a deviation may indicate the existence of new physics beyond the standard model (SM)~\cite{Athron:muong-2}.

Low energy supersymmetry (SUSY) is a leading candidate for physics beyond the SM.
In such a framework, not only the quadratic divergence of the Higgs boson can be eliminated, but also the dark matter puzzle can be explained elegantly. However, the low energy SUSY is facing many challenges from the LHC experiments, especially the null search results of superpartners.  Recent analyses based on the Run-2 of 13 TeV LHC data of 139 fb$^{-1}$ integrated luminosity~\cite{Run2} constrain the gluino mass $m_{\tl{g}}$ above 2.2 TeV~\cite{CMSSM:gluino} and the top squark mass $m_{\tl{t}_1}$ above 1.1 TeV~\cite{CMSSM:stop} in the context of simplified models. As the low energy SUSY spectrum is determined by the SUSY breaking mechanism, we need to adopt a proper SUSY breaking mechanism in the UV theory so that such an intricated low energy SUSY spectrum can be generated naturally. Depending on the way the visible sector $'feels'$ the SUSY breaking effects from the hidden sector, the SUSY breaking mechanisms can be classified into gravity mediation~\cite{SUGRA}, gauge mediation~\cite{GMSB}, anomaly mediation~\cite{AMSB}, etc.

Since the new result by the Fermilab has come out, the SUSY interpretation for muon $g-2$ anomaly have been discussed in many works~\cite{muong-2inSUSY}. Generally speaking, it is not easy to explain the muon $g-2$ anomaly in SUGRA-type models~\cite{2104.03262,GUT-SUSY} (unlike the MSSM which can readily give sufficient contributions to the muon $g-2$ \cite{Abdughani:2019wai,MSSM:mg2}). Large SUSY contributions to $\Delta a_\mu$ in general require light sleptons and light electroweakinos \cite{Abdughani:2019wai}. With universal gaugino mass input at the GUT scale, the gaugino mass ratio is predicted to be $M_1:M_2:M_3\approx 1:2:6$ at the electroweak (EW) scale. For gluino mass heavier than 2.2 TeV, the bino and wino cannot be very light with such a gaugino mass ratio. Besides, the universal input for sfermion masses at the GUT scale also sets stringent constraints on the slepton masses at the EW scale. Given the stringent LHC constraints on colored squarks, the slepton masses cannot be sufficiently light at the EW scale for the universal sfermion mass input at the GUT scale. So, despite of the very predictive nature of mSUGRA, we have to seek for various extension of mSUGRA scenario to explain the new muon $g-2$ data.

To accommodate the muon $g-2$ anomaly in the framework of supergravity with grand unification, there is an interesting economical extension of mSUGRA called gluino-SUGRA ($\tl{g}$SUGRA) ~\cite{gSUGRA:nath}, which is a special case of non-universal gaugino mass realization~\cite{NUGM,Gogoladze:2014cha}. By relaxing the gaugino mass ratio and at the same time keeping the sleptons light, the gluino mass can be much heavier than other gauginos and sfermions at the unification scale. Therefore, the gaugino mass ratio at the EW scale does not constrain the electroweakino masses for a heavy gluino mass. Besides, the gluino mass enters in the renormalization group equations (RGE) for the squark masses and thus the squark masses will be driven to values proportional to the gluino mass as they run down from the GUT scale to the electroweak scale, ameliorating the stringent constraints from the LHC squark searches. The sleptons, which carry no color charge, will stay light. So, the RGE evolution will split the squark masses from slepton masses at the electroweak scale for a common $m_0$ at the GUT scale, which is just needed to account for the muon $g-2$ anomaly~\cite{gSUGRA:nath}. The work of~\cite{Gogoladze:2014cha}, which is in spirit a similar realization to $\tl{g}$SUGRA, also leads to a heavy gluino of order TeV with the slepton masses around a few hundred GeV, assuming that the gluino soft SUSY breaking mass term $M_3$ at $M_{GUT}$ is a few times larger than the bino ($M_1$) and wino ($M_2$) soft mass terms.

It is well known that the minimal renormalizable SUSY SU(5) GUT model is ruled out by proton decay constraints, assuming superpartner masses of the order of a few TeV. Especially, it was argued that the minimal SUSY SU(5) GUT can not survive the constraints from proton lifetime not even raising the SUSY scalar masses of the first two families~\cite{murayama:pierce}. There are a number of ways to overcome the conclusion of~\cite{murayama:pierce}, for example, the addition of Planck scale corrections due to higher dimensional operators~\cite{Babu:2020ncc,Bajc:2002pg}. Including the Planck-suppressed operators, possibly in the GUT-breaking sector of the superpotential and in the gauge kinetic terms, can not only correct the wrong mass predictions of minimal SUSY SU(5) but also modify the value of the color-triplet Higgsino mass~\cite{Babu:2020ncc} so as that the proton decay constraints from dimension-5 operators can be relaxed. Besides, the consequential gravitational smearing effects can modify the unification of gauge coupling constants as well as affect analysis of proton decay. At the same time, non-universal gaugino masses (such as $M_3\gg M_1,M_2$ for $\tl{g}$SUGRA at the GUT scale) can be generated naturally with such higher dimensional operators.

In this work we will propose new realizations to generate gaugino mass hierarchy $M_3\gg M_1,M_2$ for $\tl{g}$SUGRA and then discuss in detail the implications of the new FNAL muon $g-2$ results on $\tl{g}$SUGRA. We expect that the new muon $g-2$ results can shed new light on SUSY searches at the LHC and dark matter detections. There are other interesting solutions to muon $g-2$ anomaly in the SUSY framework, such as the NUHM2 scenario, in which the third family squarks and sleptons have different masses compared to the first two families~\cite{Ibe:2013oha,Babu:2014lwa}.

This paper is organized as follows. In Sec ~\ref{sec-2}, we propose new realizations to generate gaugino hierarchy $M_3\gg M_1,M_2$ for $\tl{g}$SUGRA. In Sec ~\ref{sec-3}, the implications of new FNAL muon $g-2$ experimental results on the $\tl{g}$SUGRA scenarios are discussed. Sec ~\ref{conclusion} contains our conclusions.

\section{Status of  mSUGRA/CMSSM and New ways to generate gaugino mass hierarchy in $\tl{g}$SUGRA}
\label{sec-2}

\begin{figure}[htb]
\begin{center}
\includegraphics[width=2.9in]{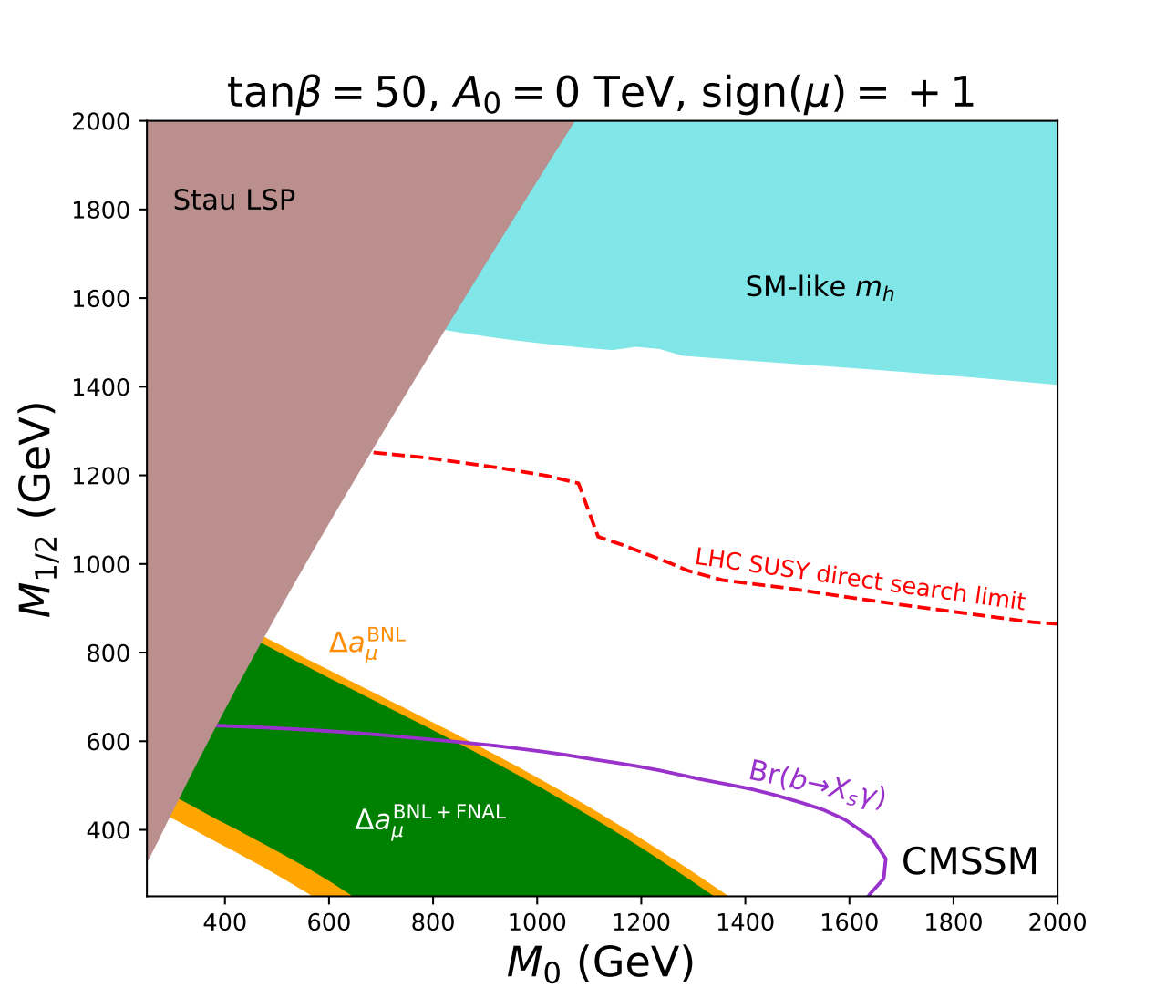}
\includegraphics[width=2.9in]{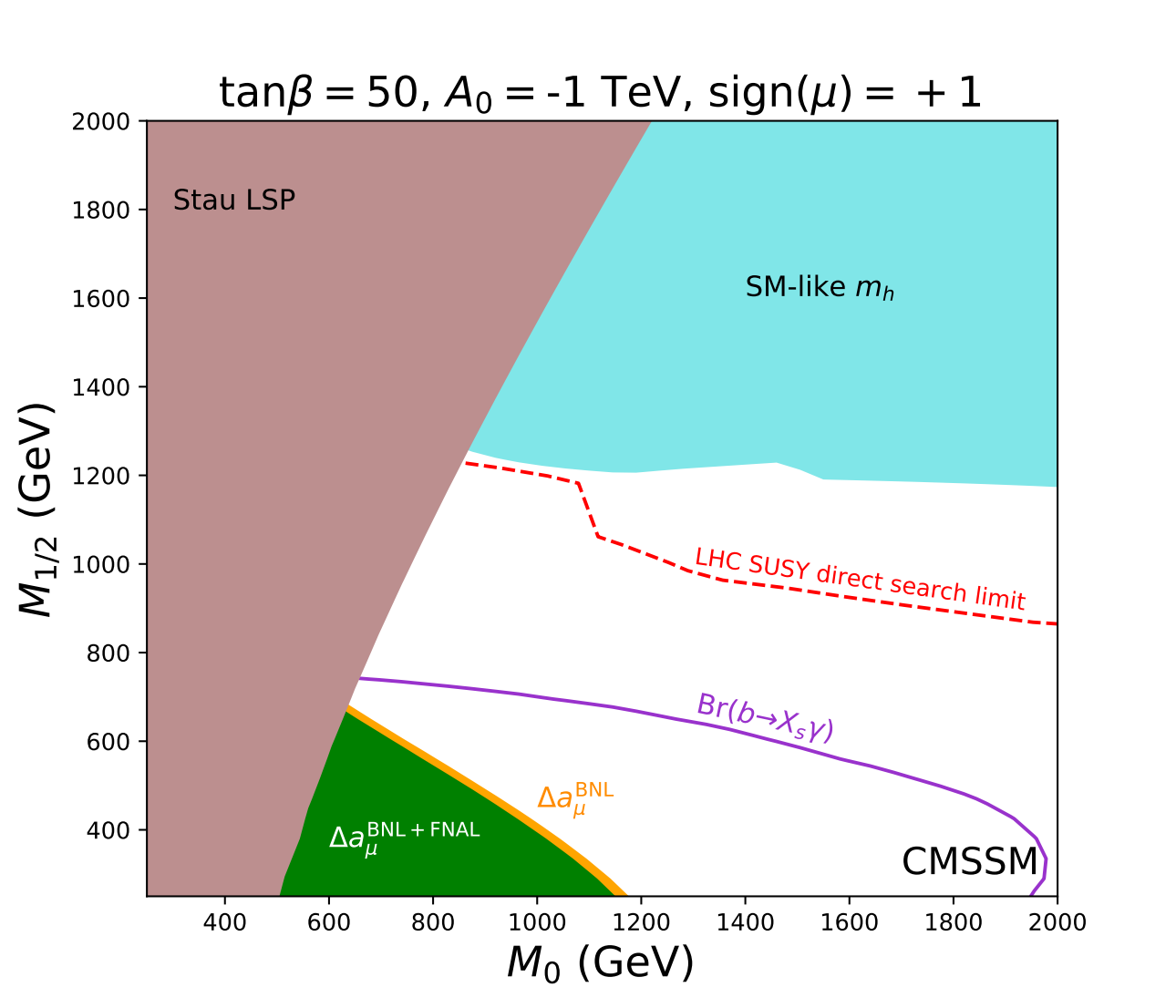}\\
\includegraphics[width=2.9in]{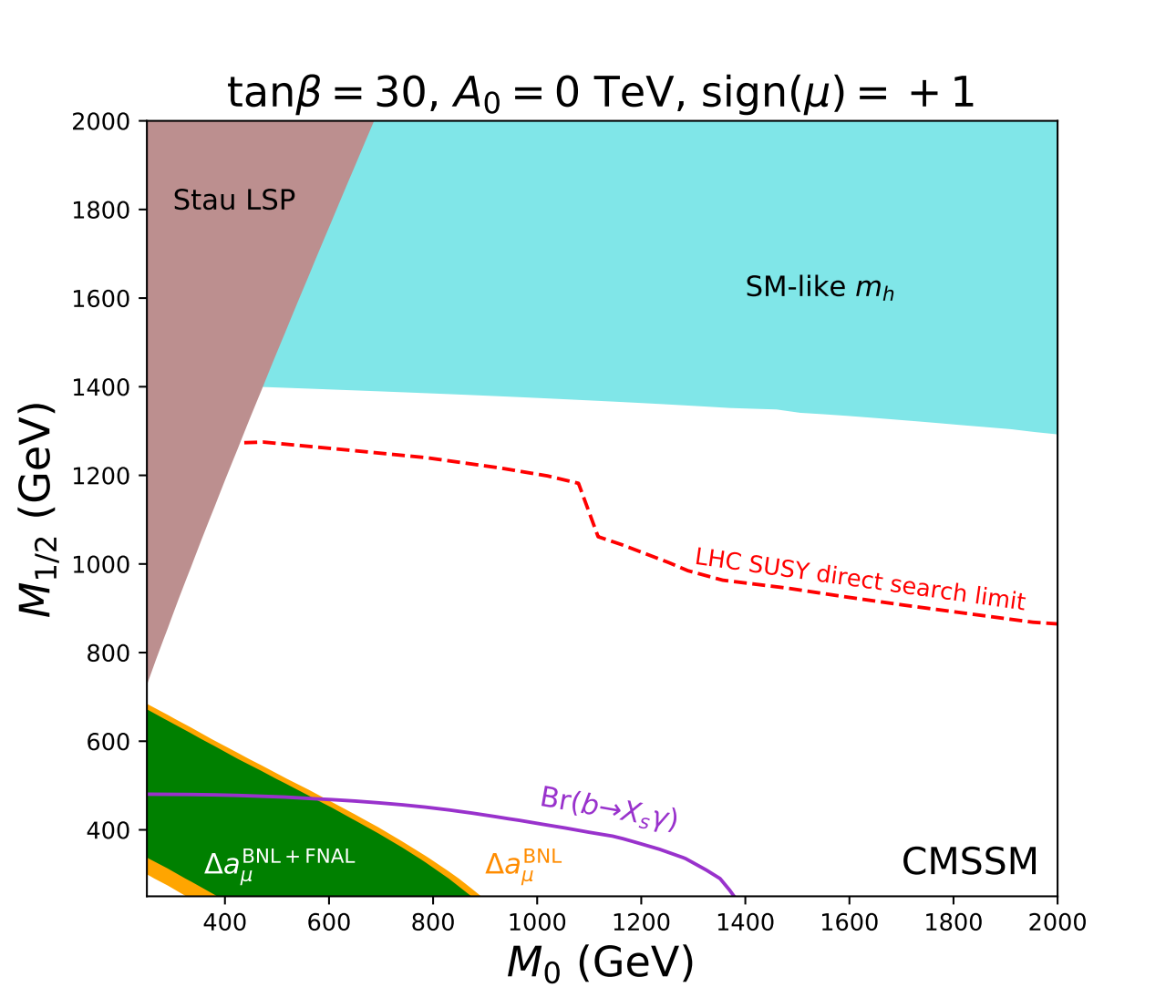}
\includegraphics[width=2.9in]{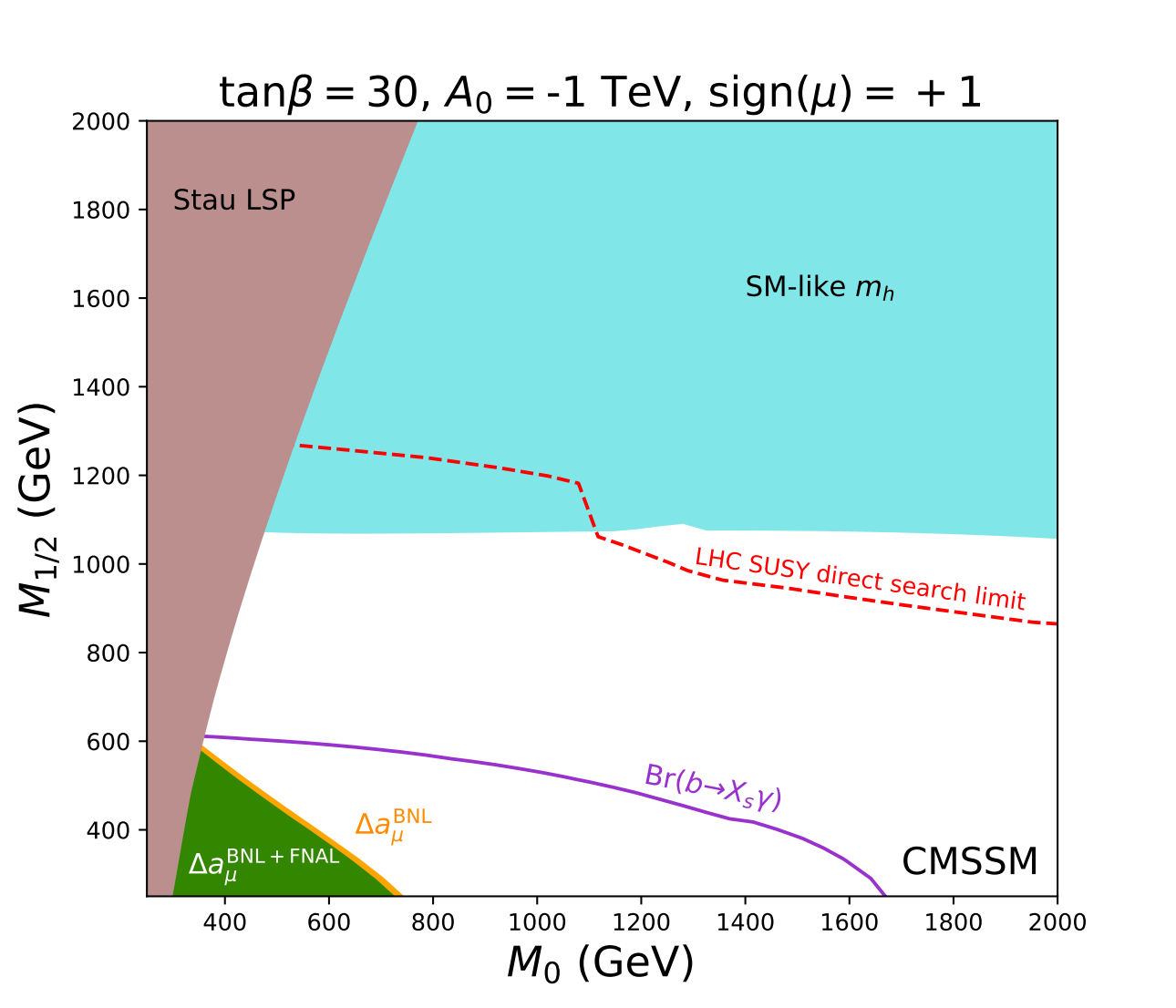}\\
\vspace{-.5cm}\end{center}
\caption{ Tensions between the $\Delta a_{\mu}^{\rm FNAL+BNL/BNL}$ explanation and the relevant experimental limits in CMSSM, showing on the plane of $(M_0,M_{1/2})$ for sign$(\mu)=+1$, $A_0=0$~\tev~(left), 1~\tev~(right) and $\tan\beta=50$~(upper), 30~(lower). The panels are taken from our previous work ~\cite{2104.03262}. In the panels, the green regions satisfy the combined FNAL+BNL results, while the orange regions plus the green regions are consistent with the previous BNL measurements. The areas below the red dash and purple curves are excluded by LHC direct search bounds for sparticles and Br($b\to s \gamma$) bound, respectively. The blue shaded regions are consistent with the present Higgs mass measurement, with 2~\gev theoretical uncertainty. In the brown shaded areas, LSP is the charged $\tilde{\tau}_1$, which is excluded. All the limits are shown at $2\sigma$ CL.
}
\label{fig0}
\end{figure}

  It is well known that mSUGRA/CMSSM is one of the most predictive SUSY models in which the soft supersymmetry breaking scalar masses $m_0$ , gaugino masses $m_{1/2}$ and trilinear terms $A_0$ are assumed to have universal values at some UV input mass scale. On the other hand, experimental data from low energy collider and dark matter direct detections give very stringent constraints on the very predicitve mSUGRA model. As noted in the introduction, CMSSM/mSUGRA with universal inputs at the GUT scale can not explain the muon $g-2$ anomaly with various LHC constraints. In our previous work~\cite{2104.03262}, we had carry out detailed discussions on the possible status of CMSSM/mSUGRA in light of recent muon $g-2$ anomaly. We show the results of~\cite{2104.03262} again in Fig.\ref{fig0}. It can be seen that the region which can account for the recent $g-2$ anomaly is excluded by LHC direct
searches for sparticles (and also be constrained stringently by Br($b \to s \ga$) bound).  Besides, the region of parameter space, which can explain the 125 GeV Higgs, has no overlap with the survived region which can account for the recent $g-2$ anomaly.

We should note that this conclusion is hold only when the universal inputs are given at the GUT scale. As noted in~\cite{super-GUT:mSUGRA}, it is not necessarily the case that $M_{in}=M_{GUT}$, since either the primordial SUSY breaking mechanism or its communication to the observable sector may involve a dynamical scale below $M_{GUT}$.  For sub-GUT~\cite{sub-GUT:mSUGRA} or super-GUT~\cite{super-GUT:mSUGRA} inputs for mSUGRA, in which the universal inputs can be lower than or higher than the GUT scale, parameter spaces that can explain the muon $g-2$ anomaly with various LHC constraints can possibly exist. In this work, we concentrate only on the scenarios with inputs of soft parameters at the GUT scale.

A salient feature of $\tl{g}$SUGRA is the mass hierarchy among gauginos $M_3\gg M_1,M_2$ at the GUT scale. In~\cite{gSUGRA:nath}, a much heavier gluino can be generated by certain combination of {\bf 75} and {\bf 24} representation Higgs. Two of us also proposed to generate a much heavier $M_3$ parameter via proper wave-function normalizations for vector superfields~\cite{gSUGRA:WWY,gSUGRA:WWYZ}. In this section, we propose alternative ways to generate the hierarchy between $M_3$ and $M_1,M_2$:
\bit
\item Hierarchy among the gauginos can be realized by introducing an additional $U(1)_H$ symmetry with non-trivial charge assignment to the higher-dimensional Higgs field. A possible $U(1)_H$ charge-assignment choice is given as
\beqa
Q_H: S({\bf 1})=n,~~Q_H: \Phi({\bf 24})=-1,~~Q_H: T({\bf 1})=-1.
\eeqa
To preserve the $U(1)_H$ symmetry, the superpotential should take the following form
\beqa
{\cal L}\supseteq \int d^2\theta \(W^a W^a+ c_S\f{S}{\Lambda^{n+1}} W^a \(c_0T\delta_{ab}+c_1\Phi_{ab}\)^n W^b\)~,\label{case1}
\eeqa
The non-renormalizable theorem of SUSY guarantees that this form will not be spoiled
by higher order perturbative corrections.

With $c_0=-c_1=1$, we have
\beqa
&&\langle T\rangle\delta_{ab}-\langle\Phi_{ab}\rangle\\
&&\sim ~{\rm diag}
\left(v_T+\f{v_U}{\sqrt{15}}, v_T+\f{v_U}{\sqrt{15}},v_T+\f{v_U}{\sqrt{15}},
v_T-\f{\sqrt{15}v_U}{10},v_T-\f{\sqrt{15}v_U}{10} \right)\nn\\
&&+
\theta^2{\rm diag}
\left(F_T+\f{F_U}{\sqrt{15}}, F_T+\f{F_U}{\sqrt{15}},F_T+\f{F_U}{\sqrt{15}},
F_T-\f{\sqrt{15}F_U}{10},F_T-\f{\sqrt{15}F_U}{10} \right)~.\nn
\eeqa
The gaugino masses at the GUT scale are given by
\beqa
M_3&=&\f{n\langle S \rangle}{\(\Lambda\)^{n+1}} (v_T+\f{v_U}{\sqrt{15}})^{n-1}(F_T+\f{F_U}{\sqrt{15}}),\\
M_2&=&\f{n\langle S \rangle}{\(\Lambda\)^{n+1}} (v_T-\f{\sqrt{15}v_U}{10})^{n-1}(F_T-\f{\sqrt{15}F_U}{10}),\\
M_1&=&\f{n\langle S \rangle}{\(\Lambda\)^{n+1}}\f{6}{5}\[\f{1}{3}(v_T+\f{v_U}{\sqrt{15}})^{n-1}(F_T+\f{F_U}{\sqrt{15}})\right. \nn \\
&& ~~~~~~~~~~~~\left. +\f{1}{2}(v_T-\f{\sqrt{15}v_U}{10})^{n-1}(F_T-\f{\sqrt{15}F_U}{10})\].
\eeqa
With $F_T\gtrsim F_U$, we can generate the hierarchy $M_3\simeq \f{5}{2} M_1 \gg M_2$ when
\beqa
1\gg \(\f{v_T+\f{v_U}{\sqrt{15}}}{\Lambda}\)^{n-1}\gg \(\f{v_T-\f{\sqrt{15}v_U}{10}}{\Lambda}\)^{n-1}~.
\eeqa
Similarly, we can adopt the most general combination of the singlet $T$, the {\bf 24} representation Higgs field and the {\bf 75} representation Higgs field. The VEV of the combination can be given by a $10\tm 10$ matrix in the form
\beqa
\langle c_0T+c_1\Phi_{\bf 24}+c_2\Phi_{\bf 75}\rangle =\(v_0+\theta^2 F_0\)
diag(\underbrace{~a,\cdots,~a}_3,\underbrace{~b,\cdots,~b}_{6},~c)~,
\eeqa
which can keep the $SU(3)_c\tm SU(2)_L\tm U(1)_Y$ gauge symmetry after SU(5) symmetry breaking.
The gaugino masses at the GUT scale can be given by
\beqa
M_3&=&\f{n\langle S \rangle}{\(\Lambda\)^{n+1}}\f{1}{18}\left\{6 a^{n-1}
+12 b^{n-1}\right\}F_0,\\
M_2&=&\f{n\langle S \rangle}{\(\Lambda\)^{n+1}} \f{1}{6}\left\{6b^{n-1}\right\}F_0,\\
M_1&=&\f{n\langle S \rangle}{\(\Lambda\)^{n+1}}\f{1}{90}\left\{48a^{n-1}
+6b^{n-1}+36c^{n-1}\right\}F_0 .
\eeqa
With suitable choices of $c_0,c_1$ and $c_2$ to give $4a^{n-1}=-3c^{n-1}$ and $a>b$, we can obtain the gaugino mass hierarchy $M_3\gg M_2\simeq 15 M_1$ for a not too small $n$. Other gaugino mass ratios can be obtained by introducing combinations involving other higher dimensional representation Higgs fields, for example, the {\bf 200} representation of SU(5).

\item  An alternative way to generate hierarchy among the gaugino masses are presented in \cite{gSUGRA:WWYZ} by two of us. We can introduce the following Lagrangian, which contains the kinetic term for gauge bosons as well as a non-renormalizable term involving a new SU(5) singlet $S$ suppressed by a scale $\Lambda> M_{GUT}$, for example, the Planck scale $M_P$
\beqa
{\cal L}\supseteq \int d^2\theta\[ W^a W^a+ c_S\f{S}{\Lambda}W^a W^a+ \f{c_\Phi}{\Lambda}W^a \Phi_{ab} W^b\]~.
\label{VEV:24}
\eeqa
We assume that the SU(5) gauge singlet $S$ develops a VEV of order $\Lambda$ and the higher dimensional Higgs field $\Phi$ acquires both the F-term VEV and the lowest component VEV.
Tuning $c_S$ so that the coefficient in front of the kinetic term for SU(3) almost vanishes, we can obtain the gaugino mass hierarchy by wavefunction normalization. For example, we assume that the {\bf 24} representation Higgs field acquires a VEV of the following forms
\beqa
\langle \Phi_{\bf 24} \rangle= \(v_U+\theta^2 F_U\)
 {\sqrt {\frac{3}{5}}} ~{\rm diag}
\left(-\frac{1}{3}, -\frac{1}{3}, -\frac{1}{3},
\frac{1}{2}, \frac{1}{2} \right)~.
\label{case2}
\eeqa
We can tune the coefficients to satisfy
\beqa\label{wavefunction:I}
1+c_S\f{\langle S \rangle}{\Lambda}-c_{\Phi_{\bf 24}}\f{v_U}{\Lambda}\f{2}{\sqrt{15}}= \f{\epsilon}{\sqrt{15}}\ll 1.
\eeqa
Then, the gaugino mass ratios will be given by
\beqa
M_1:M_2:M_3=\f{\f{1}{\sqrt{15}}}{Z_{U(1)_Y}}:
\f{\f{3}{\sqrt{15}}}{Z_{SU(2)_L}}:\f{-\f{2}{\sqrt{15}}}{Z_{SU(3)_c}}
\eeqa
after taking into account the wavefunction normalization factor
\beqa
Z_{U(1)_Y}:Z_{SU(2)_L}:Z_{SU(3)_c}=\f{3}{\sqrt{15}}:\f{5}{\sqrt{15}}:\f{\epsilon}{\sqrt{15}}~.
\eeqa
So, we have the gaugino mass ratios at the GUT scale
\beqa
M_1:M_2:M_3=\f{1}{3}:\f{3}{5}:-\f{1}{\epsilon}~,~~
\eeqa
for $\epsilon\ll1$.

Similarly, if we adopt {\bf 75} representation Higgs to have both the F-term VEV and the lowest component VEV, we will obtain the gaugino ratio
\beqa
M_1:M_2:M_3=\f{20}{24}:\f{12}{8}:-\f{1}{\epsilon}
\eeqa
at the GUT scale.
For the {\bf 200} representation Higgs with both the F-term VEV and the lowest component VEV, we will obtain
\beqa
M_1:M_2:M_3=\f{10}{9}:\f{2}{1}:\f{1}{\epsilon}
\eeqa
at the GUT scale.

\item Similar to the previous settings, we can introduce the following Lagrangian involving two different higher dimensional representation Higgs fields
\beqa
{\cal L}\supseteq \int d^2\theta\[ W^a W^a+ c_S\f{S}{\Lambda}W^a W^a+ \f{c_{\Phi_1}}{\Lambda}W^a \Phi_{1,ab} W^b+\f{c_{\Phi,2}}{\Lambda}W^a \Phi_{2,ab} W^b\]
\label{case3}
\eeqa
 If the two Higgs fields acquire either the lowest component VEV (with $\Phi_1$) or the F-component VEV (with $\Phi_2$), the wavefunction normalization factor can be independent of the unrescaled gaugino mass ratio. For example, we can assume that the {\bf 24} representation Higgs field and the {\bf 75} representation Higgs field acquire VEVs of the form
\beqa
\langle \Phi_{\bf 24} \rangle&=& v_{\bf 24}
 {\sqrt{\frac{3}{5}}} ~{\rm diag}
\left(-\frac{1}{3}, -\frac{1}{3}, -\frac{1}{3},
\frac{1}{2}, \frac{1}{2} \right)~,\\
\langle \Phi_{\bf 75} \rangle&=& \theta^2
\f{F_{\bf 75}}{2\sqrt{3}}
{\rm diag}
\left(\underbrace{~1,\cdots,~1}_3,\underbrace{-1,\cdots,-1}_{6},
3\right)~.~ \,
\eeqa
If we assume the condition
\beqa
\label{wavefunction:II}
1+c_S\f{\langle S \rangle}{\Lambda}-c_{\Phi_{\bf 24}}\f{v_U}{\Lambda}\f{2}{\sqrt{15}}= \f{\epsilon}{\sqrt{15}}\ll 1,
\eeqa
we can have the gaugino mass ratios at the GUT scale
\beqa
M_1:M_2:M_3=\f{\f{20}{5\sqrt{3}}}{Z_1}:-\f{\f{12}{5\sqrt{3}}}{Z_2}:-\f{\f{4}{5\sqrt{3}}}{Z_3}
=\f{5}{3}:-\f{3}{5}:-\f{1}{\epsilon},
\eeqa
with
\beqa
Z_1:Z_2:Z_3=\f{3}{\sqrt{15}}:\f{5}{\sqrt{15}}: \f{\epsilon}{\sqrt{15}}.
\eeqa
Similar techniques can be applied to other combinations of Higgs fields, including the combination in which the singlet $S$ acquires F-term VEV while the higher dimensional representation Higgs acquires the lowest component VEV.
\eit

 We need to add proper additional GUT symmetry breaking sector to trigger the VEVs of various higher dimensional representation Higgs fields. For example, in minimal SU(5) GUT, to break the $SU(5)$ gauge group to $SU(3)_c\tm SU(2)_L\tm U(1)_Y$, we can add
\beqa
W_{SB}&\supseteq& \f{f}{3} Tr(\Phi_{\bf 24}^3)+\f{m_{24}}{2} Tr(\Phi_{\bf 24}^2)
+\bar{H}_{\bf \overline{10}}(M^\pr+\la_1 \Phi_{\bf 24}) H_{\bf 10}+\cdots,
\label{SU5:minimal}
\eeqa
to trigger the GUT breaking by $\Phi_{\bf 24}$ Higgs~\cite{hagiwara}. In $\tl{g}$SUGRA scenario, to realize much heavier $M_3$ at the GUT scale, a complicate GUT symmetry breaking sector beyond eq.(\ref{SU5:minimal}) for SU(5) GUT is necessary.


The thresholds at the GUT scale depend on the spectrum of the GUT-scale particles. It is known that GUT threshold correction of order -4\% is needed for gauge coupling unification with the current low energy data. Such a requirement implies certain relations among the masses of the various thresholds.
The weak triplet $\Sigma_3$ and color octet $\Sigma_8$ supermultiplets in the adjoint Higgs {\bf 24} can play an important part in superhigh-scale physics. In minimal renormalizable supersymmetric SU(5) GUT model, the weak triplet and color octet will acquire GUT-scale masses. With threshold corrections from GUT-scale weak triplet and color octet, gauge coupling unification set a constraint for the triplet Higgs masses $M_{H_C}$ to be $M_{H_C}\lesssim 3.15\tm 10^{16} {\rm GeV}$~\cite{Babu:2020ncc}, which can not be consistent with the $p\to K^+ \bar{\nu}$ proton decay bounds from Super-Kamiokande assuming superpartner masses of order TeV~\cite{murayama:pierce}. However, it is possible that the masses of $\Sigma_8$ and $\Sigma_3$ can naturally be much lighter than the GUT scale, for example, originated from dimension-4 Planck scale induced terms in the superpotential with negligible renormalizable cubic term. Besides, large mass-splitting can also appear between $\Sigma_8$ and $\Sigma_3$. With the present low energy data, such as the value of $\al_S(M_Z )$ and weak mixing angle $\theta_W$, such a large mass splitting among $\Sigma_8$ and $\Sigma_3$ is found to be fairly ample for gauge coupling unification to be completely adapted with relatively larger $M_{H_C}$ so that current proton decay constraints are satisfied~\cite{Chkareuli:1998wi,Bajc:2002pg,Bajc:2002bv}.


In some extended SU(5) GUT models, additional {\bf 75} representation Higgs (and ${\bf 50+\overline{50}}$ Higgs) are introduced to realize doublet-triplet (D-T) splitting by missing partner mechanism. With the VEV by {\bf 75} representation Higgs to break the GUT group instead of {\bf 24} representation Higgs, the color-triplet masses within ${\bf 5+\overline{5}}$ can be heavy while the doublet within ${\bf 5+\overline{5}}$ can be light because no weak doublet is contained in ${\bf 50+\overline{50}}$ Higgs.
 Requiring gauge coupling unification with low energy data, we can define proper effective GUT scale with much larger effective triplet Higgs masses by proper forms of superpotential (for example, see ~\cite{Zheng:2012pt}), which can be consistent with proton decay constraints. Although the introduction of ${\bf 75}$ representation Higgs with the additional ${\bf 50 + \overline{50}}$ pairs would lead to the Landau pole between the GUT and Planck scales, this may signal the presence of a non-perturbative UV fixed point~\cite{fix-point}. In general, the GUT threshold correction can get significant contributions from the GUT breaking sector, which can meet the requirements of gauge coupling unification (in addition to possible additional uncertainties due to gravitational smearing). Therefore, GUT symmetry breaking with the presence of high dimensional representation Higgs can be consistent.
\section{Implications of new FNAL g-2 results on $\tl{g}$SUGRA}\label{sec-3}
\subsection{General Discussions}
The SUSY contributions to the muon $g_\mu-2$ are dominated by the chargino-sneutrino and the neutralino-smuon loops. At the leading order of $m_W/m_{SUSY}$ and $\tan\beta$, with $m_{SUSY}$  the SUSY-breaking masses and the Higgsino mass $\mu$, various loop contributions give~\cite{gmuon}
 \begin{align}
 \Delta a_{\mu }(\tilde{W}, \tilde{H}, \tilde{\nu}_\mu)
 &= \frac{\alpha_2}{4\pi} \frac{m_\mu^2}{M_2 \mu} \tan\beta\cdot
f_C
 \left( \frac{M_2 ^2}{m_{\tilde{\nu }}^2}, \frac{\mu ^2}{m_{\tilde{\nu }}^2}  \right) ,
 \label{eq:WHsnu} \\
 \Delta a_{\mu }(\tilde{W}, \tilde{H},  \tilde{\mu}_L)
 &= - \frac{\alpha_2}{8\pi} \frac{m_\mu^2}{M_2 \mu} \tan\beta\cdot
 f_N
 \left( \frac{M_2 ^2}{m_{\tilde{\mu }_L}^2}, \frac{\mu ^2}{m_{\tilde{\mu }_L}^2} \right),
 \label{eq:WHmuL}  \\
 \Delta a_{\mu }(\tilde{B},\tilde{H},  \tilde{\mu }_L)
  &= \frac{\alpha_Y}{8\pi} \frac{m_\mu^2}{M_1 \mu} \tan\beta\cdot
 f_N
 \left( \frac{M_1 ^2}{m_{\tilde{\mu }_L}^2}, \frac{\mu ^2}{m_{\tilde{\mu }_L}^2} \right),
 \label{eq:BHmuL}\\
 \Delta a_{\mu }(\tilde{B}, \tilde{H},  \tilde{\mu }_R)
  &= - \frac{\alpha_Y}{4\pi} \frac{m_{\mu }^2}{M_1 \mu} \tan \beta \cdot
  f_N \left( \frac{M_1 ^2}{m_{\tilde{\mu }_R}^2}, \frac{\mu ^2}{m_{\tilde{\mu }_R}^2} \right), \label{eq:BHmuR}\\
 \Delta a_{\mu }(\tilde{\mu }_L, \tilde{\mu }_R,\tilde{B})
 &= \frac{\alpha_Y}{4\pi} \frac{m_{\mu }^2 M_1 \mu}{m_{\tilde{\mu }_L}^2 m_{\tilde{\mu }_R}^2}  \tan \beta\cdot
 f_N \left( \frac{m_{\tilde{\mu }_L}^2}{M_1^2}, \frac{m_{\tilde{\mu }_R}^2}{M_1^2}\right), \label{eq:BmuLR}
\end{align}
with $m_\mu$ being the muon mass, $\alpha_Y=g_1^2/4\pi$ and $\alpha_2=g_2^2/4\pi$ for the $U(1)_Y$ and  $SU(2)_L$ gauge groups, respectively. The loop functions are defined as
\begin{align}
&f_C(x,y)= xy
\left[
\frac{5-3(x+y)+xy}{(x-1)^2(y-1)^2}
-\frac{2\log x}{(x-y)(x-1)^3}
+\frac{2\log y}{(x-y)(y-1)^3}
\right]\,,
\\
&f_N(x,y)= xy
\left[
\frac{-3+x+y+xy}{(x-1)^2(y-1)^2}
+\frac{2x\log x}{(x-y)(x-1)^3}
-\frac{2y\log y}{(x-y)(y-1)^3}
\right]\,.
\label{moroi3}
\end{align}
They satisfy $0\le f_{C,N}(x,y) \le 1$ and are monochromatically increasing for $x>0$ and $y>0$.
In the limit of degenerate masses, they satisfy $f_C(1,1)=1/2$ and $f_N(1,1)=1/6$. It can be seen that large SUSY contributions to $\Delta a_\mu$ require light sleptons, light electroweakinos and large $\tan\beta$. However, as noted previously, the current LHC experiments have already set stringent constraints on colored sparticles, such as the 2.2 TeV bound for gluino and the 1.4 TeV bound for squarks. The $\tl{g}$SUGRA can naturally generate a hierarchy between light sleptons and heavy squarks as well a hierarchy between colored gluino and electroweakinos, satisfying the stringent bounds from the LHC. Therefore, we anticipate that $\tl{g}$SUGRA can elegantly solve the muon $g-2$ discrepancy.

We have the following remarks on fine-tuning~(FT) and GUT constraints related to the $\tl{g}$SUGRA scenarios:
\bit
\item In $\tl{g}$SUGRA, the choice of heavy gluino can possibly increase the FT involved. The EWSB condition gives
 \be
\frac{m_Z^2}{2} = \frac{(m_{H_d}^2+\Sigma_d^d)-(m_{H_u}^2+\Sigma_u^u)\tan^2\beta}{(\tan^2\beta -1)}
-\mu^2~,
\ee
where the $\Sigma_u^u$ and $\Sigma_d^d$ terms come from derivatives
of $\Delta V$ evaluated at the potential minimum and
$\tan\beta\equiv\frac{v_u}{v_d}$. As the largest radiative corrections to $H_u$ mass
come from top squarks $\Sigma_u^u\sim 3y_t^2/(16\pi^2)\times F(m_{\tl{t}_{1,2}}^2)$ with $F(m^2)=m^2(\log \frac{m^2}{Q^2}-1)$,  naturalness prefers a spectrum of light top squarks. On the other hand, the top squark masses will receive contributions from the gluino loop, which goes like $\delta m_{\tl{t}_i}\sim 2g_3^2/(3\pi^2) m_{\tl{g}}^2 \tm \log$ for $\log\sim 1$. So, larger values of gluino mass will in general increase the FT unless the low energy value of trilinear coupling $A_t$ increases accordingly in a proper mass region due to radiative naturalness~\cite{radiative:natural}, which states that increasing $m_{\tl{t}_i}^2$ while at the same time increasing $A_t$ can
possibly make the involved FT unchanged.

Two types of FT measurements are always used in the literatures. The Barbieri-Giudice fine-tuning~(BGFT) measure with respect to certain input parameter $'a_i'$ is defined as
          \beqa
          \Delta_{BG}(a_i)\equiv \left|\f{\pa \ln M_Z^2}{\pa \ln a_i}\right|~,
          \eeqa
 while the total fine-tuning is defined to be $ \Delta_{BG}= \max\limits_{i}|\Delta_{BG}(a_i)|$, with $\{i\}$ the set of parameters defined at the input scale. As the BGFT in general will overestimate the FT involved~\cite{Baer:2013gva}, the electroweak fine-tuning~(EWFT) measure $\Delta_{EW}$ is also widely used in various studies. The EWFT measure $\Delta_{EW}$ is defined as~\cite{radiative:natural}
   \beqa
   \Delta_{EW}\equiv \max\limits_{i}(C_i)/\(\f{m_Z^2}{2}\),
   \eeqa
with the relevant terms
\beqa
C_{H_u}&=&\left|-\f{m^2_{H_u}\tan^2\beta}{\tan^2\beta-1}\right|~,~
C_{H_d}=\left|\f{m^2_{H_d}}{\tan^2\beta-1}\right|~,~
C_{\mu}=\left|-\mu_{eff}^2\right|~,~\nn\\
C_{\Sigma_{u}^u(\tl{t}_{1,2})}&=&
\f{\tan^2\beta}{\tan^2\beta-1}\left|\f{3}{16\pi^2}F(m_{\tl{t}_{1,2}}^2)\left[y_t^2-g_Z^2\pm \f{y_t^2 A_t^2-8g_Z^2(\f{1}{4}-\f{2}{3}x_w)\Delta_t}{m_{\tl{t}_{2}}^2-m_{\tl{t}_{1}}^2}\right]\right|~,\nn\\
C_{\Sigma_{d}^d(\tl{t}_{1,2})}&=&\f{1}{\tan^2\beta-1}\left|\f{3}{16\pi^2}F(m_{\tl{t}_{1,2}}^2)\left[g_Z^2\pm \f{y_t^2 \mu_{eff}^2+8g_Z^2(\f{1}{4}-\f{2}{3}x_w)\Delta_t}{m_{\tl{t}_{2}}^2-m_{\tl{t}_{1}}^2}\right]\right|~,
\eeqa
where $x_w=\sin^2\theta_W$ and
\beqa
\Delta_t&=&\f{(m_{\tl{t}_{L}}^2-m_{\tl{t}_{R}}^2)}{2}+M_Z^2\cos2\beta(\f{1}{4}-\f{2}{3}x_w)~,\nn\\
F(m^2)&=& m^2\(\log\f{m^2}{m_{\tl{t}_{1}}m_{\tl{t}_{2}}}-1\)~.
\eeqa
  In our numerical study, we show both the EWFT and BGFT involved in the region that can explain the recent muon $g-2$ anomaly.

\item $\tl{g}SUGRA$, which adopts the GUT-scale inputs, can be constrained by proton decay experiments. The instability of nucleons is one of the most striking consequence of the GUT. Various experiments can test such a hypothesis of grand unification. For the $p\to K^+ \bar{\nu}$ decay mode, the JUNO~\cite{JUNO}, DUNE~\cite{DUNE} and Hyper-Kamiokande~\cite{hyperK} experiments can reach the sensitivity at the level of a lifetime $\tau_p\sim 10^{34}$ year after operating for 10 to 20 years~\cite{ellis:proton decay}.  The nucleon decay via the exchange of GUT-breaking heavy gauge bosons is strongly suppressed by $M_{GUT}^{-4}$ with the typical SUSY GUT scale $M_{GUT}\sim 10^{16} {\rm GeV}$. On the other hand, the nucleon decay via the exchange of color-triplet Higgs is suppressed by $M_H^{-2}$, which is the dominant proton decay mode for many SUSY SU(5) GUT models even with its additional suppressions by light fermion Yukawa couplings and a loop factor. Bounds on proton decay lifetime by super-Kamiokande~\cite{superK} for the dimension-five proton decay effective operators had already set very stringent constraints on the minimal SUSY GUT model and the scale of low energy SUSY spectrum. The proton decay constraints for sub-GUT and super-GUT SUGRA models can be seen in~\cite{ellis:proton decay}.

We need to integrate out the colored triplets to obtain the dimension-five effective operators at the GUT scale. After the RGE running from the GUT scale to EW scale, we can dress those operators to obtain the $d=6$ effective operators. Evolving further to the proton decay scale 1 GeV, the lifetime for the different decay channels can be calculated using the Chiral Lagrangian technique. Details for the dressing, the RGE evolution~\cite{ellis:dim-5RGE,nihei:twoloop,hisano:dim-5RGE}, the hadronic matrix elements etc in the calculation of dimension-five operators induced proton decay can be found in~\cite{Goto:dim-5 RRRR, nath:proton decay,ellis:proton decay}.

The SUSY contributions to muon $g-2$ are proportional to the value of $\tan\beta$, so large $\tan\beta$ is always preferred to give large $\Delta a_\mu$. On the other hand, as noted in \cite{Goto:dim-5 RRRR,GUT:large tanbeta}, proton decay is stringently constrained by RRRR operators for large $\tan\beta$. 
The dominant decay mode  for $p\to K^+ \bar{\nu}_\tau$ is given by higgsino dressing diagrams. We should also note that the gluino is much heavier than the electroweakinoes in the gluino-SUGRA scenarios, which may enhance gluino dressing contributions. However, the gluino dressing contributions cancel each other and give subleading contributions to $p\to K^+ \bar{\nu}$ decay because the up and down scalar quarks have almost degenerate masses for the first two generations~\cite{gluino:dressing cancelation} in our scenarios. 
Our numerical calculations indeed confirm that the higgsino dressing contributions are dominant in $p\to K^+ \bar{\nu}$ decay modes. So we only take into account the higgsino dressing contributions in our following numerical results.

 The effective color-triplet Higgs mass $M^{eff}_{H_C}$, which is the key parameter for dimension-five proton decay calculation, is intimately connected to D-T splitting mechanism and GUT symmetry breaking sector. Unlike the minimal SUSY SU(5) GUT, within which the threshold correction receives a negligible contribution from the GUT breaking sector and $M^{eff}_{H_C}$ is required to take a very low value, $M^{eff}_{H_C}$ can be very large if we adopt proper D-T splitting mechanism and require significant corrections from the GUT breaking sector. So the value of $M^{eff}_{H_C}$ can be seen as a free parameter, which can possibly take the form $M_{GUT}^2/M_X$ by some mechanism with $m_X\ll M_{GUT}$ and take a value much larger than $M_{GUT}$ (even can be much higher than the Planck scale). We fix $M^{eff}_{H_C}$ to lie at the order of the Planck scale $M^{eff}_{H_C}=1.0\tm 10^{19}{\rm GeV}$ in our numerical discussions.

\eit

\subsection{Numerical Results}
To discuss the implications of recent muon $g-2$ measurments on $\tl{g}$SUGRA, we use the packages SuSpect2.52~\cite{SuSpect252} and MicrOMEGAs5.2.7.a~\cite{MicrOMEGAs527a} to scan the whole parameter space. The parameter space of $\tilde{g}$SUGRA is given by the following set of free parameters $$m_0,M_1,M_2,M_3,A_0,\tan \beta, sign(\mu),$$
at the GUT scale. Here $m_0$ is the universal soft SUSY breaking scalar mass, $M_1, M_2, M_ 3$ are the gaugino masses, $A_0$ is the universal trilinear coupling, and $\tan \beta$ is the ratio of $H_u^0,H_d^0$ VEVs.
In addition to the constraints already encoded in the packages, we also impose the following constraints:
 \bit
 \item[(i)] The lightest CP-even Higgs boson should act as the SM-like Higgs boson with a mass of $125\pm 2$ GeV~\cite{ATLAS:higgs,CMS:higgs}.
  \item[(ii)] Direct searches for low mass and high mass resonances at LEP, Tevatron and LHC by using the package HiggsBounds-5.1.1beta~\cite{higgsbounds511} and HiggsSignal~\cite{higgssignal}.
  \item[(iii)] Constraints on gluino and squark masses from the LHC~\cite{LHCmass}:
  \beqa
  m_{\tl{g}}\gtrsim 2.2 ~{\rm TeV},~~~~m_{\tl{q}}\gtrsim 1.4 ~{\rm TeV},
  \eeqa
and the lower mass bounds of charginos and sleptons from the LEP~\cite{LEPmass}:
  \beqa
  m_{\tl{\tau}}\gtrsim 93.2 ~{\rm GeV},~~~~  m_{\tl{\chi}^\pm} \gtrsim 103.5 ~{\rm GeV}.
  \eeqa
  Bounds on typical soft SUSY breaking parameters from vaccum stability are also imposed~\cite{Vaccumstability}.
  \item[(iv)] Constraints from $B \to X_s \gamma$, $B_s \to \mu^+ \mu^-$and $B^+ \to \tau^+ \nu_\tau$ etc \cite{BaBar-Bph, LHCb-BsMuMu, Btaunu}
      \begin{eqnarray}
        3.15\times10^{-4} <&Br(B_s\to X_s \gamma)&< 3.71\times10^{-4} \\
         1.7\times10^{-9} <&Br(B_s\to\mu^+\mu^-)&< 4.5\times10^{-9}  \\
        0.78\times10^{-4} <&Br(B^+\to\tau^+\nu_\tau)&< 1.44\times10^{-4}~.
      \end{eqnarray}

  \item[(v)] For dark matter relic density, we impose only the upper bound $0<\Omega h^2<0.1188$ from Planck~\cite{Planck} to permit additional dark matter components.

  \item[(vi)] All relevant EW SUSY searches are taken into account, via CheckMATE2~\cite{CheckMATE}, MadGraph5\_aMC@NLO~\cite{MG5} and PYTHIA8.2~\cite{pythia}. We discard the parameter points whose the $R$ values obtained from the CheckMATE2 are larger than 1, i.e., excluded at $95\%$ confidence level. The main constraints considered come from the searches for electroweak productions of charginos and neutralinos~\cite{CMSchi}, as well as the lepton final states from slepton pair production at the LHC~\cite{ATLASsl}.

\eit

 We need to specify those inputs at the GUT scale for our numerical results, especially the gaugino mass inputs. To obtain a heavy gluino, we require $M_3\gtrsim 3\max(M_1,M_2)$ at the GUT scale, which will lead to at least $M_3\gtrsim 9\max(M_1,M_2)$ at the EW scale. There are still several possibilities for the choices of $M_1$ and $M_2$. It is well known that $M_i/g_i^2$ is a RGE constant at one-loop level. For $M_1/M_2\lesssim 1.5$ at $M_{GUT}$, the lightest gaugino at EW scale will be the bino; while for $M_1/M_2\gtrsim 1.5$ at $M_{GUT}$, the lightest gaugino at EW scale will be the wino.
 The electroweakino mass hierarchy will partially determine the properties of the dark matter particle.

 The parameters $m_0,A_0,\tan\beta$ are chosen to lie in the following ranges:
  \begin{eqnarray}
          m_0 \in [50,2000]~\gev, ~~~~A_0 \in [-5000,5000]~\gev, ~~~~\tan \beta \in [2,60].
 \end{eqnarray}

We discuss several examples of the $M_1,M_2$ gaugino mass inputs, which will predict different types of dark matter:
\begin{itemize}
\item [(1)]  Scenario with
 \begin{eqnarray}
M_1=M_2,~~~~~ M_1 \in [50,1000]~\gev,
 \end{eqnarray}
which is the most economical $\tl{g}$SUGRA extension to mSUGRA.
\begin{figure}[htb]
\begin{center}
\includegraphics[width=2.9in]{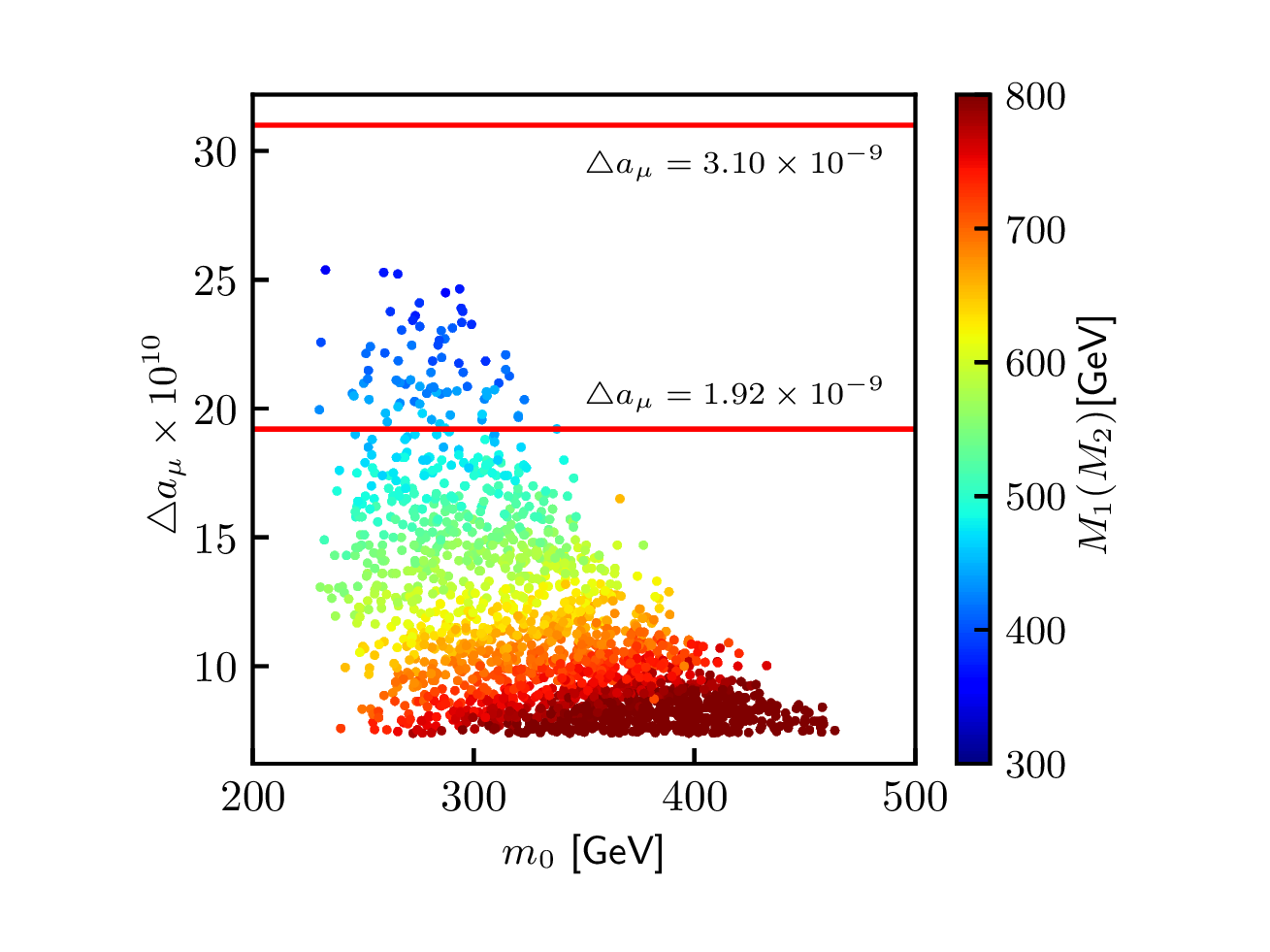}
\includegraphics[width=2.9in]{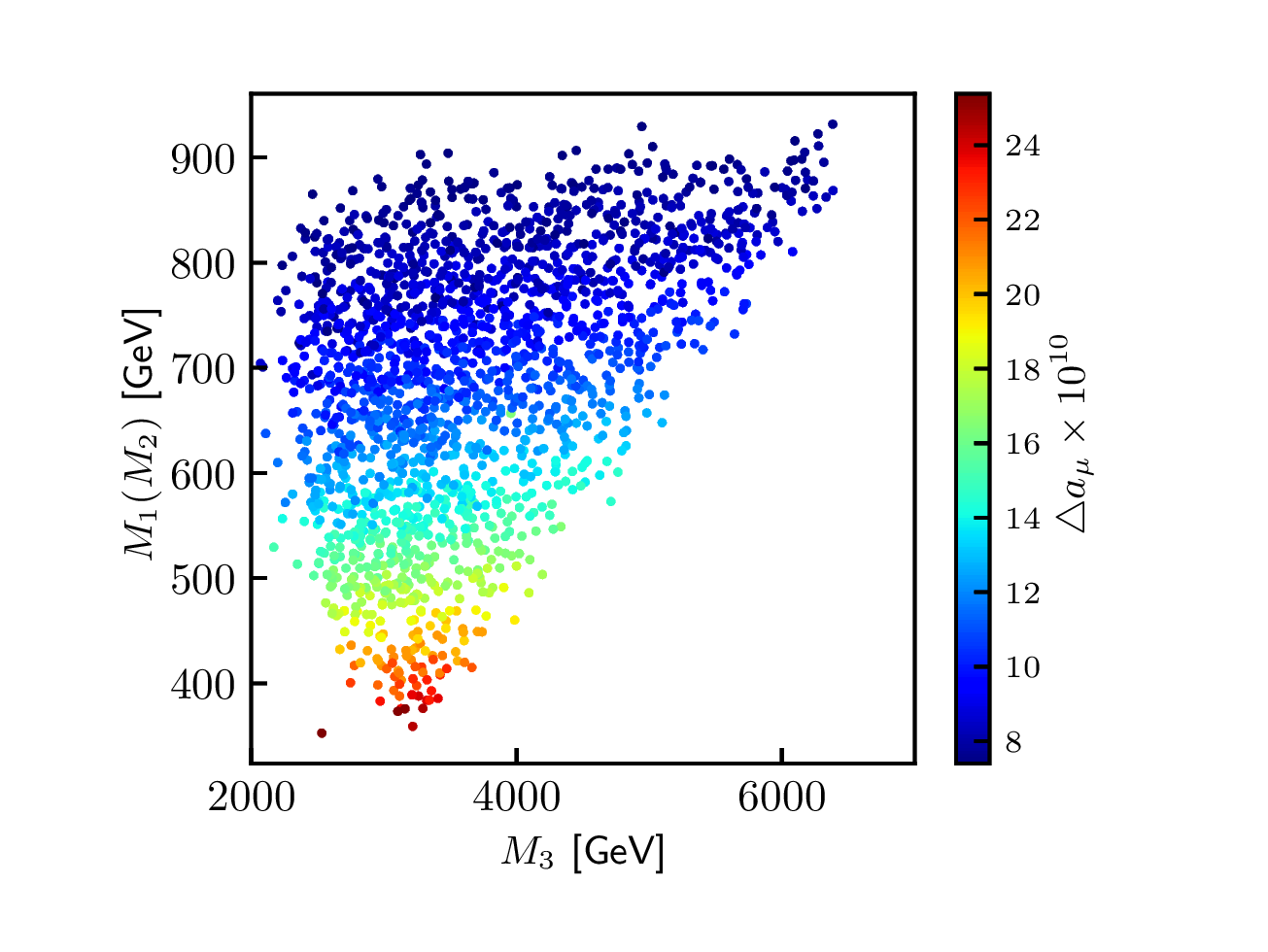}\\
\includegraphics[width=2.9in]{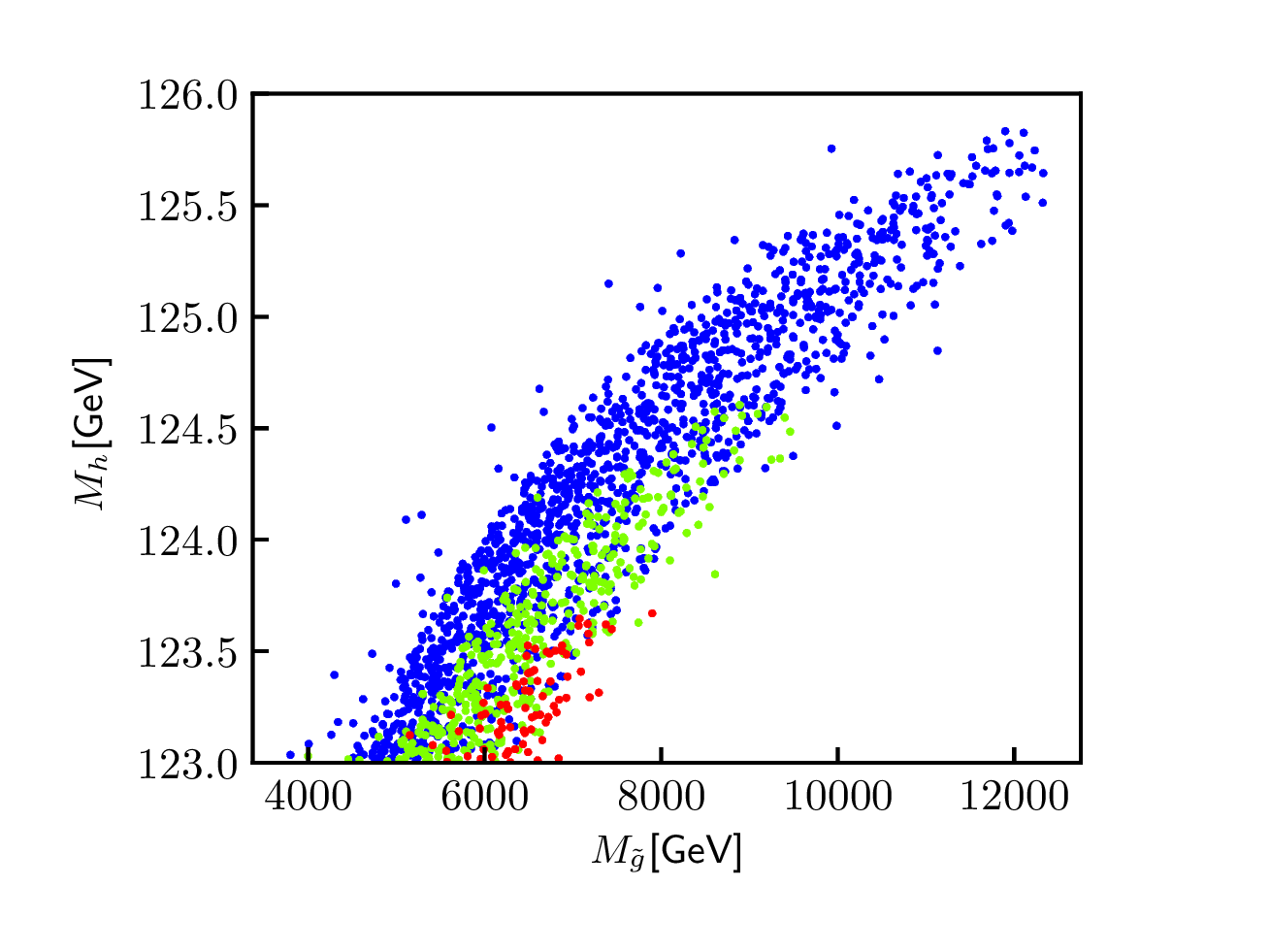}
\includegraphics[width=2.9in]{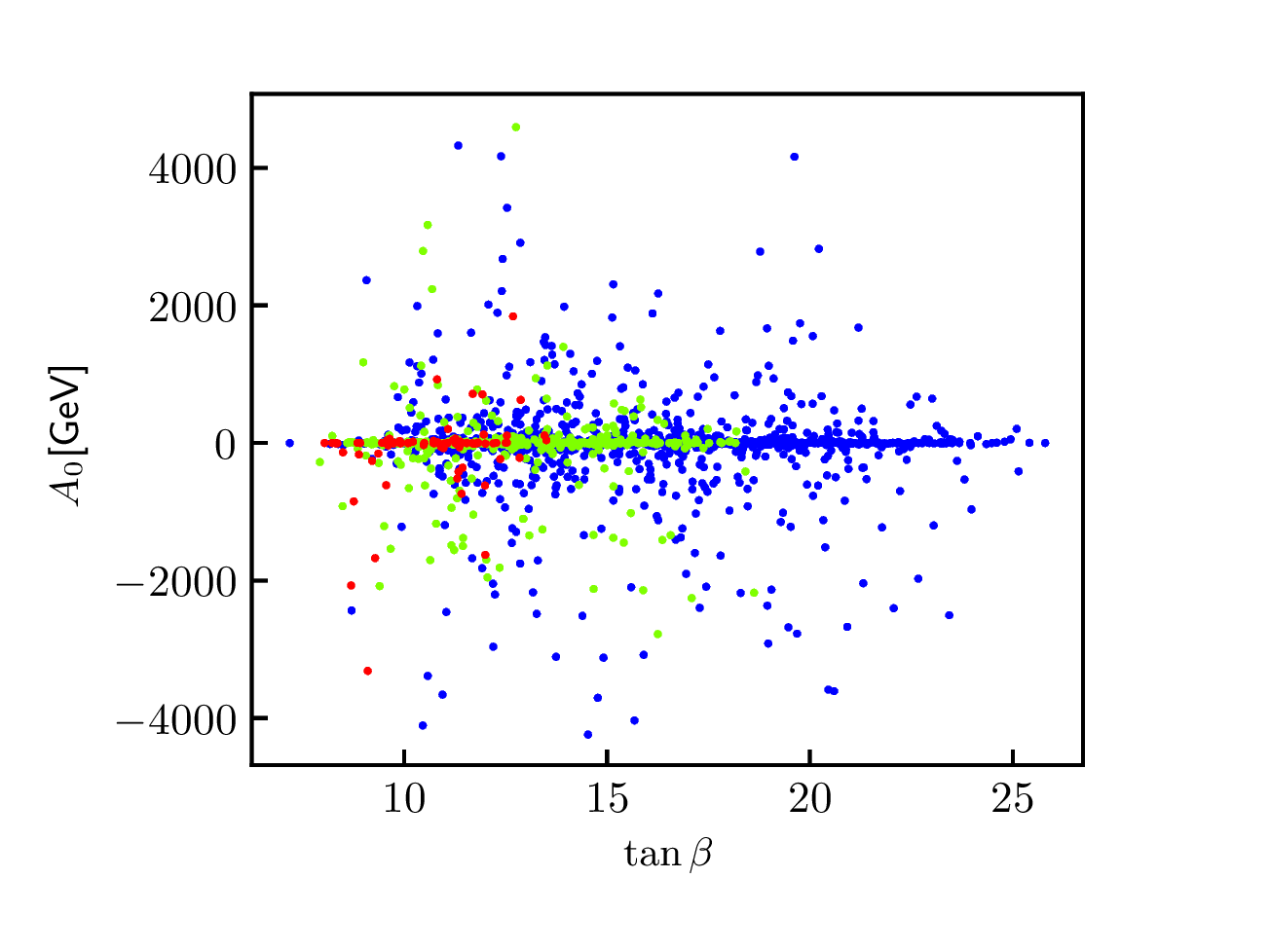}\\
\includegraphics[width=2.9in]{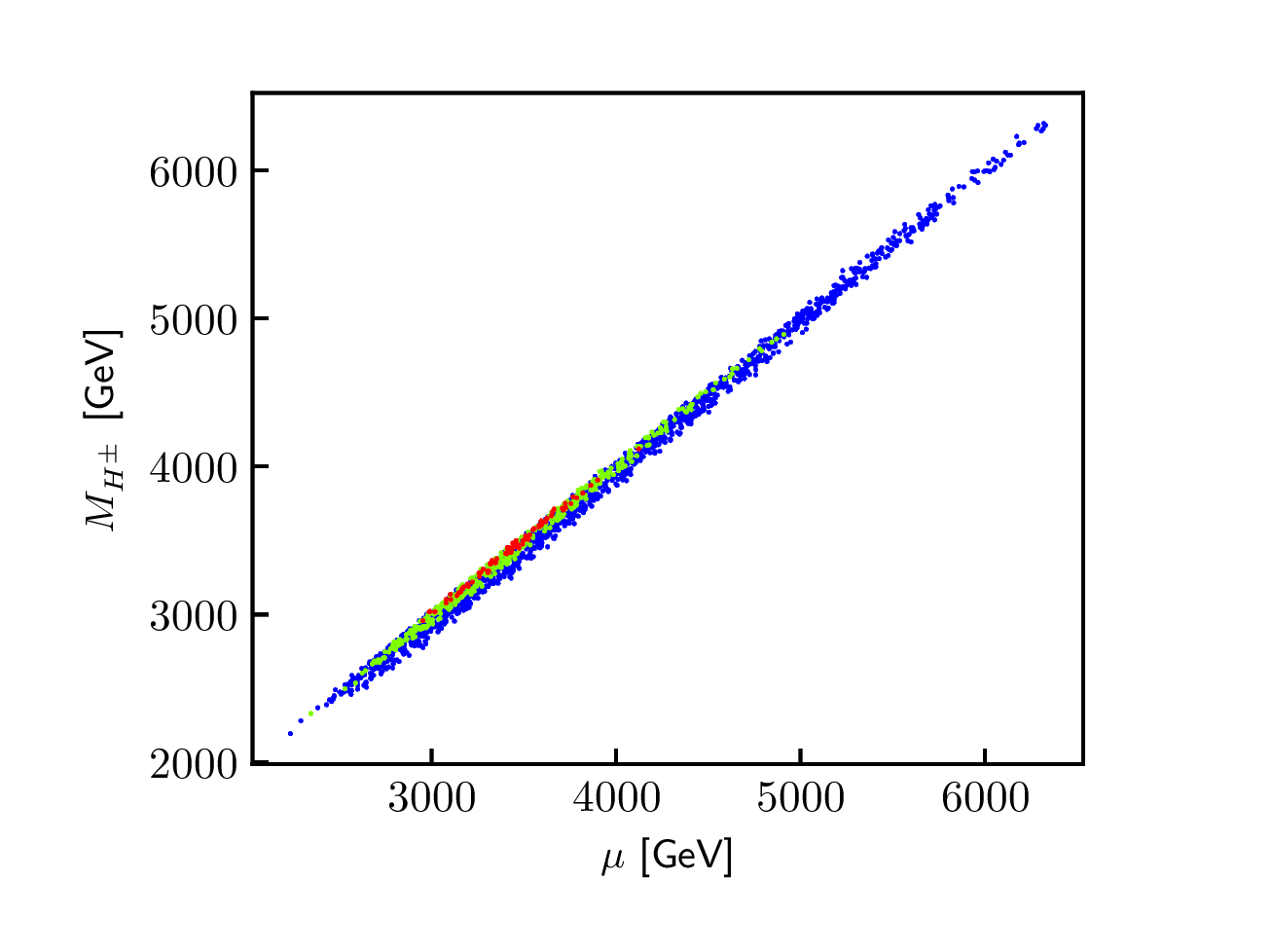}
\includegraphics[width=2.9in]{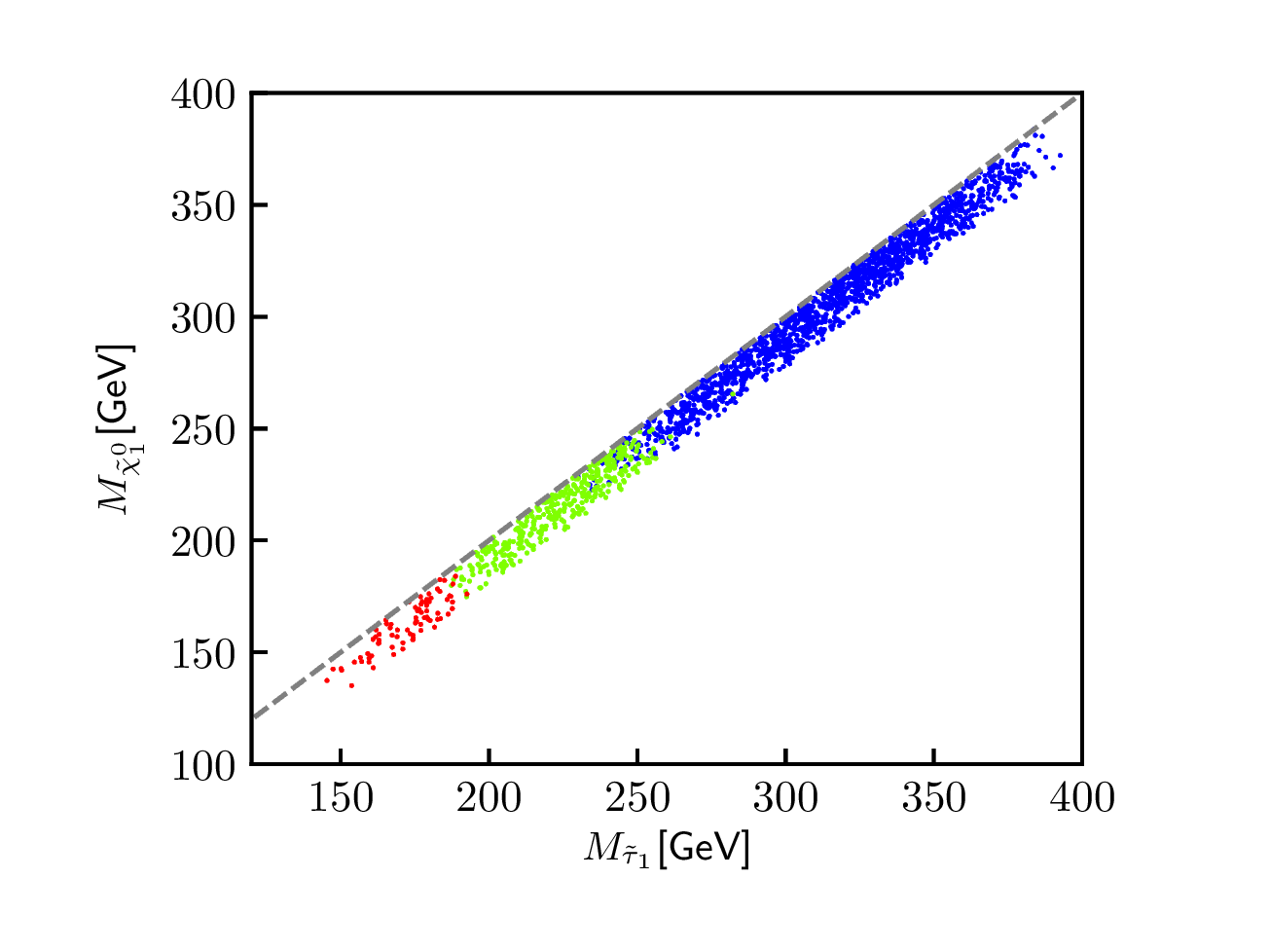}\\
\vspace{-.5cm}\end{center}
\caption{Survived points in scenario (1) that can satisfy the constraints (i-vi) and give SUSY contributions to $\Delta a_\mu$ up to the $1\sigma$ range of $\Delta a_\mu^{combine}$. In the middle and lower panels, the red, green and blue points denote the survived samples that can give $\Delta a_\mu^{SUSY}$ up to the $1\sigma$, $2\sigma$ and $3\sigma$ levels, respectively.
}
\label{fig1}
\end{figure}
In Fig.\ref{fig1}, we project the predictions of the surviving samples in various planes. It is obvious from this figure that the muon $g-2$ anomaly can be explained elegantly in this economical $\tl{g}$SUGRA scenario. From the upper-left panel of Fig.\ref{fig1}, we can see that the SUSY contributions $\Delta a_\mu^{SUSY}$ in this simplest $\tl{g}$SUGRA scenario can reach the $1\sigma$ range, which is $\Delta a_\mu^{SUSY} \in [1.92,3.10]\tm 10^{-9}$. To give $\Delta a_\mu^{SUSY}$ in the $1\sigma$ range, the parameter $M_1(M_2)$ is constrained to lie in the $[352,470]$ GeV range while the parameter $m_0$ is constrained to lie in the $[232,337]$ GeV range. In \cite{gSUGRA:nath}, to efficiently split the colored sparticles and the colorless sparticles, the hierarchy between $M_3$ and $M_1(M_2)$ is chosen to be $M_3/M_1\approx 10$. From the upper-right panel of Fig.\ref{fig1}, we can see that explaining the muon $g-2$ anomaly in the $1\sigma$ range still permits a wide range of the gaugino mass ratio $6\lesssim M_3/M_1\lesssim 9$ at the GUT scale, with the central value located near 7.5. A even larger mass hierarchy between electroweakinos and gluino at the EW scale, which is given by $M_1:M_2:M_3\approx 1:2:45$ at the previous central value, will lead to a very heavy gluino for a light bino of order ${\cal O}$(100) GeV. From the middle left panel, it can be seen that the allowed gluino mass is constrained to lie between 5.5 TeV to 7.8 TeV (4.5 TeV to 9.4 TeV) in the $1\sigma$  ($2\sigma$) range of $\Delta a_\mu^{combine}$, which cannot be discovered in the near future LHC experiment (however, may potentially be discovered in the future 100 TeV hadron collider experiments). The squarks, including the stops, can be pushed to be heavy through RGE evolution and easily be consistent with the current LHC exclusion bounds. So, the 125 GeV Higgs can be accommodated easily with heavy stops in this scenario with a small $A_t$, as shown in the middle panels of Fig.\ref{fig1}. The values of $\mu$ versus the charged Higgs masses for the survived points are given in the lower left panel of Fig.\ref{fig1}. It can be seen from the figure that the values of $\mu$, which characterizes the higgsino masses and the left-right handed smuon mixing for chirality flip, should be heavier than 2 TeV.

  In mSUGRA model, it is almost impossible to reconcile the 125 GeV Higgs or the $BR(B \ra X_s\ga)$ constraints with a large $\Delta a_\mu^{SUSY}$ (see Fig.\ref{fig0}). For example, the SUSY contributions to $BR(B \ra X_s\ga)$, which are dominated by loops involving $\tl{t}\tl{W}$ and $t H^\pm$, should be suppressed to agree with the experimental results, which requires heavy stops, winos and $H^\pm$. However, large $m_0,m_{1/2}$ tend to give a too small $\Delta a_\mu^{SUSY}$, although the 125 GeV Higgs can be achieved. In the $\tl{g}$SUGRA scenario, the squark masses will be split from the slepton masses by RGE effects from a heavy gluino. The winos need not be heavy because of the relaxed gaugino mass ratio. Besides, the bounds on $H^\pm$ from the (suppressed) SUSY contributions to $BR(B \ra X_s\ga)$ will only mildly constrain the value of $m_0$. Therefore, large SUSY contributions to $\Delta a_\mu$ will not be spoiled by $BR(B \ra X_s\ga)$ in our scenario.

The dark matter particle in this scenario is almost pure bino. It is well known that the bino dark matter will typically overclose the universe unless co-annihilations or resonant annihilations are efficient. In this scenario with light sleptons and heavy squarks, the co-annihilation of the bino LSP with stau is fairly efficient and it can reduce the dark matter relic abundance to correct values. From the lower right panel, except the already excluded region (by the LHC), it is clear that the allowed masses for stau and bino indeed satisfy the stau co-annihilation requirement $m_{\tl{\tau_1}}\lesssim 1.2 m_{\tl{\chi}_1^0}$.
As the co-annihilation implies that the mass of NLSP must be close to the LSP with a small mass gap to ensure efficient annihilation of the LSP, the small mass gap implies that the
final states in the decay of the NLSP will be very soft, making them difficult to detect. Such a co-annihilation region can survive current LHC exclusion bounds on sleptons~\cite{ATLASsl} and can potentially be discovered in the future LHC experiments. Discussions on the discovery potential for such compressed spectrums at the LHC for $\tl{g}$SUGRA can be found in~\cite{1704.04669}.

Electroweakino productions and slepton pair producitons at the LHC will in general set stringent constraints on the allowed parameter space. Although $\tl{\tau}_1$ is almost degenerate with bino-like $\tl{\chi}_1^0$, it is possible that the mild mass splittings between the first two generations of sleptons and $\tl{\chi}_1^0$ will lead to energetic lepton final states. Besides, chargino mass can be pushed to 1.4 TeV by LHC searches, assuming $\tl{\chi}_1^{\pm}$ and $\tl{\chi}_2^0$ ($100\%$) decay into the fist-two-generation sleptons.
Careful analyses on the status of several benchmark points have been given in our previous work~\cite{2104.03262}. In this work, with detailed simulations, we find that all the points (the red points) shown in the figures can pass the latest LHC constraints if $\Delta a_\mu$ is to be explained within the $3\sigma$ ($1\sigma$) range.

\begin{figure}[htb]
\begin{center}
\includegraphics[width=2.9in]{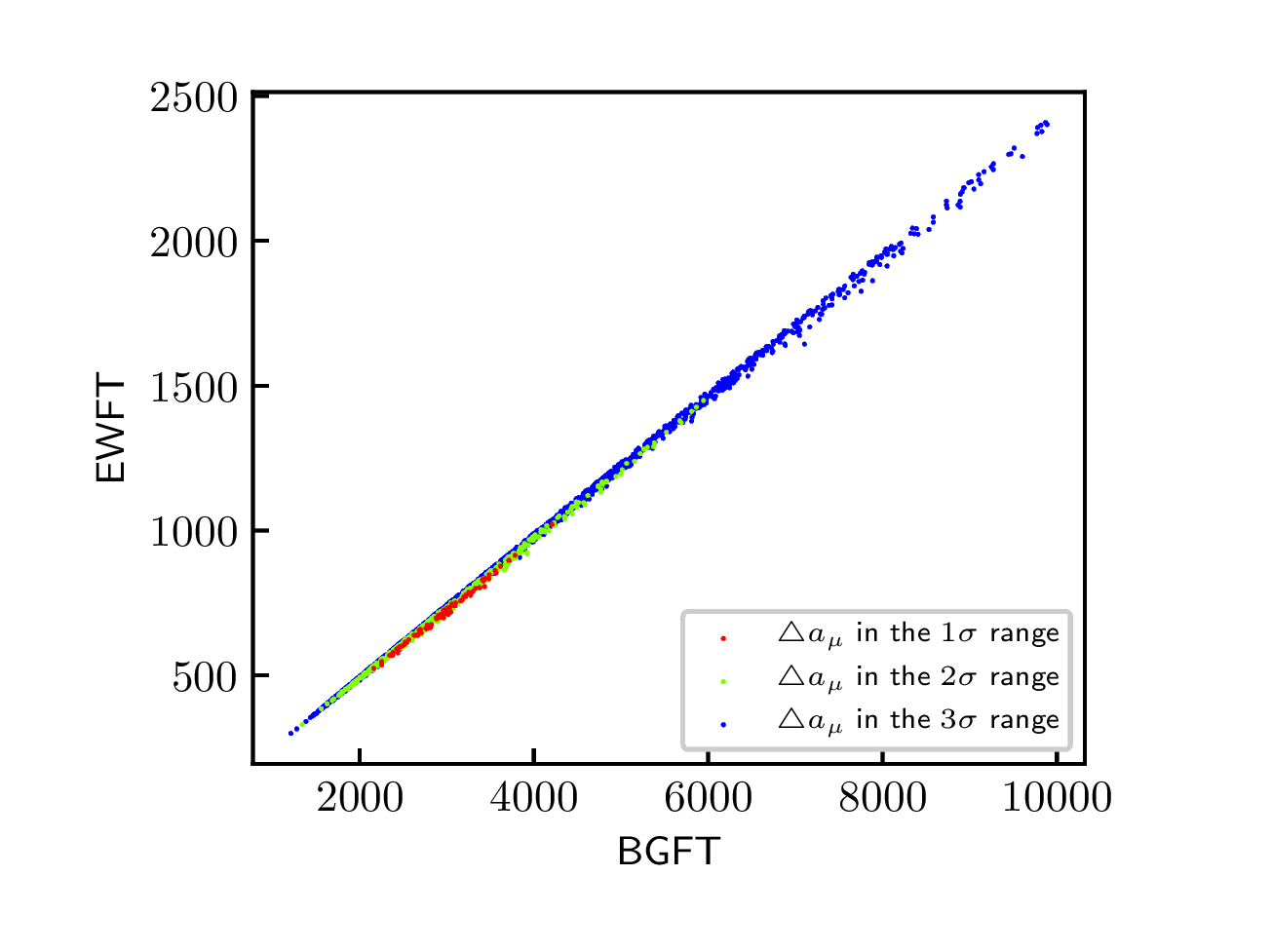}
\includegraphics[width=2.9in]{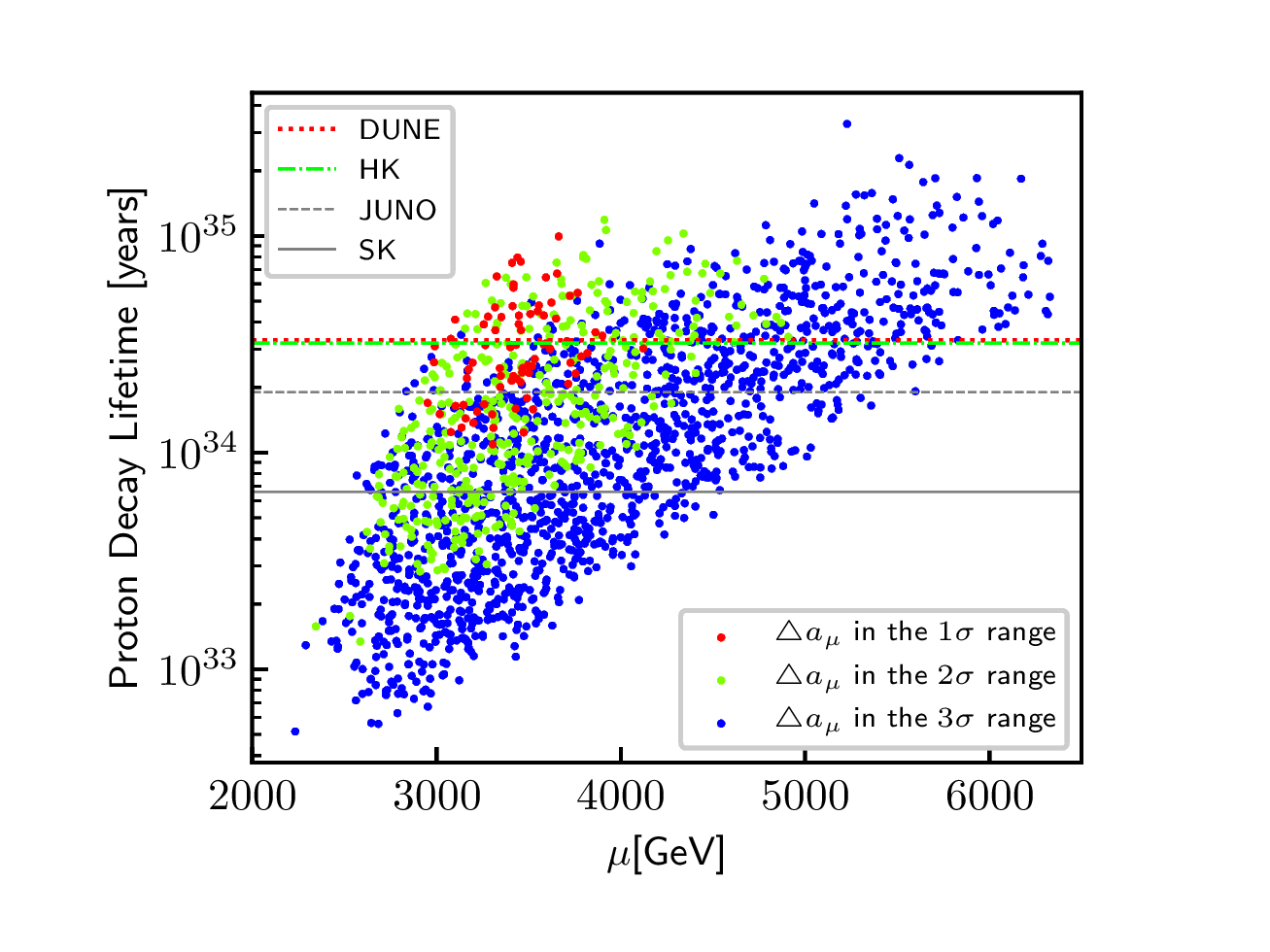}\\
\includegraphics[width=2.9in]{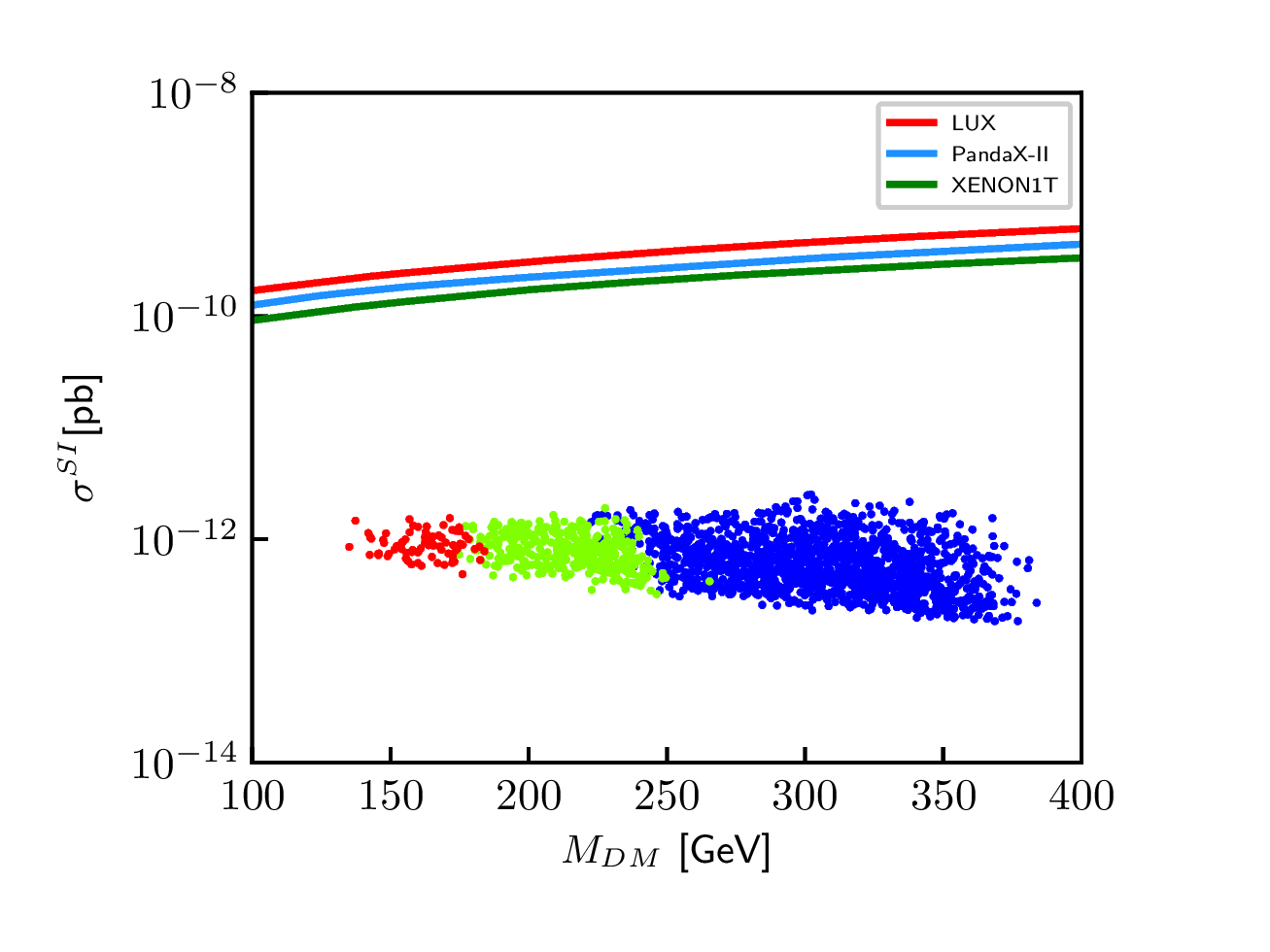}
\includegraphics[width=2.9in]{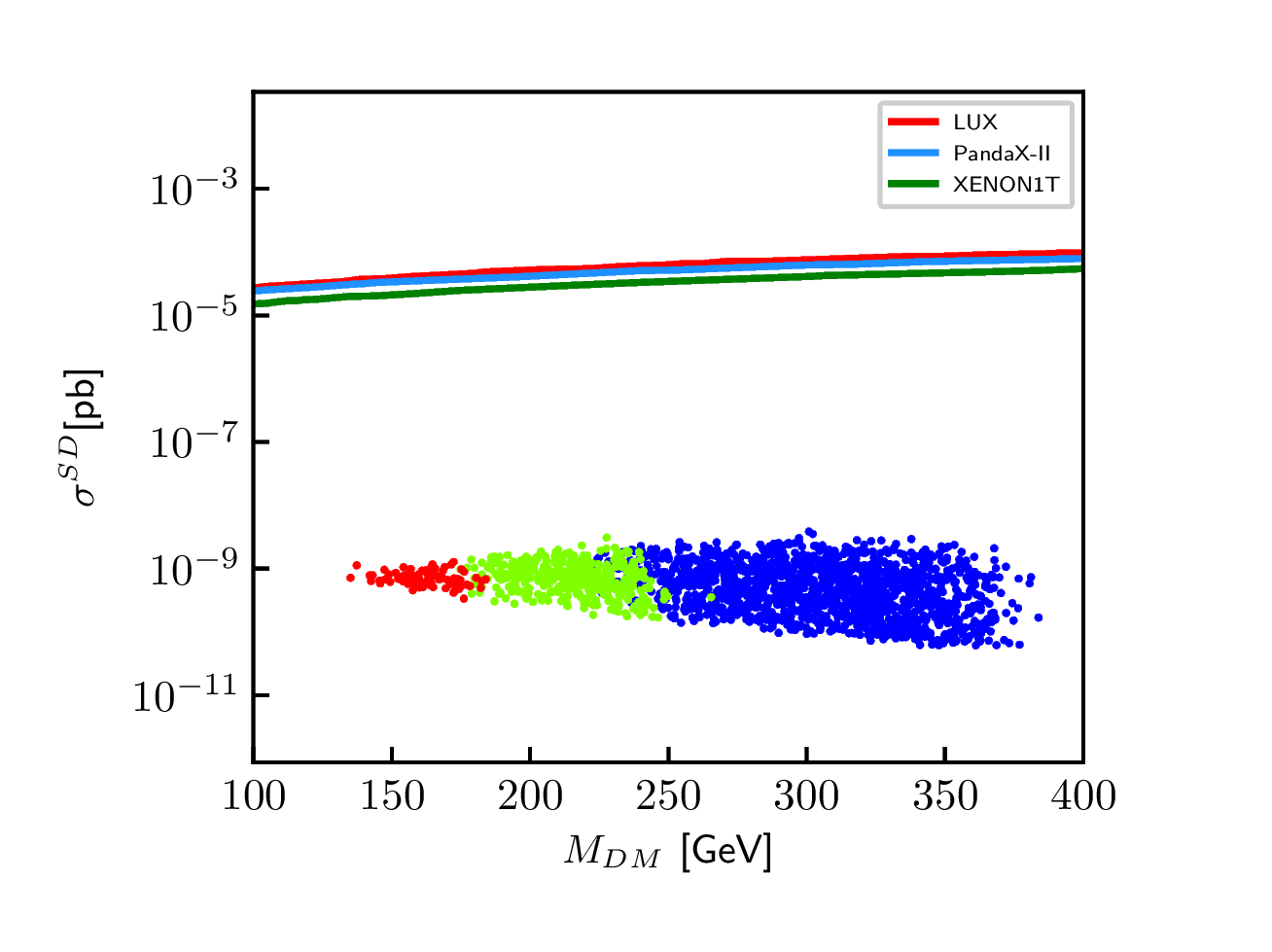}\\
\end{center}
\vspace{-.5cm}
\caption{In the upper left panel, the FTs of the survived points for scenario (1) with the corresponding range of SUSY contributions to $\Delta a_\mu$ are shown. The proton decay lifetimes via dimension-five operators for the survived points  are shown in the upper right panel. Possible proton decay bounds by existing super-Kamiokande, the forthcoming JUNO, DUNE and Hyper-Kamiokande experiments are also shown. The SI~(left) and SD~(right) direct detection cross sections with bino dark matter are shown in the lower panels. The red, green and blue points denote the survived samples that can give $\Delta a_\mu^{SUSY}$ above the $1\sigma$, $2\sigma$ and $3\sigma$ bounds, respectively. The exclusion limits for DM direct detection from LUX~\cite{lux}, XENON1T~\cite{xenon1t} and PandaX~\cite{pandax} are also shown.}
\label{fig2}
\end{figure}

We show the FT of the survived points in the upper left panel of Fig.\ref{fig2}. Both the BGFT and EWFT measures are given with the corresponding range of SUSY contributions to $\Delta a_\mu$. As the BGFT in general will over estimate the FT involved, the region with large $\Delta_{BG}$ is still not very fine-tuned by the criterion of $\Delta_{EW}$. It can be seen in the panel that the FTs satisfy $520 \leq\Delta_{EW}\leq 1020 $ ($2160\leq \Delta_{BG}\leq 4200$) in the $1\sigma$ range of $\Delta a_\mu^{combine}$. It is also obvious that much larger FTs are needed in most of the $3\sigma$ range of $\Delta a_\mu^{combine}$. From our previous discussions, we know that it is the consequence of larger gluino masses.

The lifetimes of proton decay via dimension-five operators are shown in the upper-right panel of Fig.\ref{fig2}, in which the lifetime of $p\to K^+ \bar{\nu}$ is plotted in the $\mu$ vs $\tau_p$ plane. As noted previously, we adopt $M^{eff}_{H_C}=1.0\tm 10^{19}{\rm GeV}$ in our numerical calculations. We can see that a small portion of survived parameter spaces in the $2\sigma$/$3\sigma$ range of $\Delta a_\mu^{combine}$ had already been ruled out by the results of super-Kamiokande~\cite{superK}. We should note that possible choice of larger $M^{eff}_{H_C}$ can still revive such parameter regions. A large portion of the allow region can be tested by the future proton decay experiments, such as DUNE, Hyper-K and JUNO. In most of the $1\sigma$ range of $\Delta a_\mu^{combine}$, the proton decay lifetime can reach $1.0\tm 10^{34}$ year. A large portion of the $1\sigma$ (and $2\sigma$) range of $\Delta a_\mu^{combine}$ will still not be covered by the upcoming proton decay experiments.

Spin-independent~(SI) interactions between the bino dark matter and the nucleons are primarily mediated by t-channel scalar Higgs bosons or by s-channel squarks~\cite{SI}, while the spin-dependent~(SD) interactions between bino dark matter and the nucleons are primarily mediated by t-channel $Z$ gauge boson or by s-channel squarks~\cite{SD}. In the $\tl{g}$SUGRA scenario, the squarks are pushed to be heavy by RGE. Besides, the higgsino component of the dark matter is tiny, leading to suppressed Higgs-neutralino-neutralino and  $Z$-neutralino-neutralino couplings. Therefore, such bino dark matter can easily survived the current direct detection experiments. We show the exclusion limits from LUX~\cite{lux}, XENON1T~\cite{xenon1t} and PandaX~\cite{pandax} for our scenarios. It can be seen that all the survived samples can pass the current SI and SD direct detection limits. We should note that, for dark matter-nucleon scattering cross section, we do not re-scale the original values by $\Omega/\Omega_0$ with $\Omega_0 h^2=0.1188$ to impose the direct detection constraints. If the cross sections are re-scaled, the direct detection limits can be further weakened.

\item[(2)]   Scenario with
\begin{eqnarray}
M_1:M_2=5:1, ~~~~~M_1 \in [50,2000]~\gev.
\end{eqnarray}
This gaugino mass ratio can be realized by choosing $\Phi_2$ as {\bf 200} representation Higgs and $\Phi_1$ as a proper combination of {\bf 24} and an additional singlet in eq.(\ref{case3}), with the choice of $\Phi_1$ generating the same wavefunction normalization factors $Z_1$ and $Z_2$.

    From the upper-left panel of Fig.\ref{fig3}, we can see that the SUSY contributions  $\Delta a_\mu^{SUSY}$  can reach the value $1.46\tm 10^{-9}$, which lies in the $2\sigma$ range of $\Delta a_\mu^{combine}$. Therefore, the muon $g-2$ anomaly can be explained at $2\sigma$ level in this scenario. The parameter $M_1$, which satisfies $M_1=5M_2$, is constrained to lie in the $[1040,1080]$ GeV range while the parameter $m_0$ is constrained to lie in the $[280,310]$ GeV range to give $\Delta a_\mu^{SUSY}$ in the $2\sigma$ range. As anticipated from the discussion below eq.(\ref{moroi3}), a smaller value of $m_0$ is preferred to obtain large SUSY contributions $\Delta a_\mu^{SUSY}$.

    From the upper-right panel of Fig.\ref{fig3}, we can see that the muon $g-2$ anomaly permits a wide range of gaugino mass ratio $3\lesssim M_3/M_1\lesssim 3.2$ at the GUT scale, with the central value located near $3.1$. Consequently, the mass hierarchy between electroweakinos and gluino at the EW scale will be given by $M_1:M_2:M_3\approx 5:2:93$ at the previous central value.
    From the lower left panel, it is clear that the allowed gluino mass is constrained to lie between 6.4 TeV to 6.7 TeV in the $2\sigma$ range of $\Delta a_\mu^{combine}$, which cannot be discovered in the near future LHC experiments. A large $M_3$ will push the colored sparticles heavy, including the stops. So, the 125 GeV Higgs can easily be accommodated in this scenario without the need of a very large $A_t$.

\begin{figure}[htb]
\begin{center}
\includegraphics[width=2.9in]{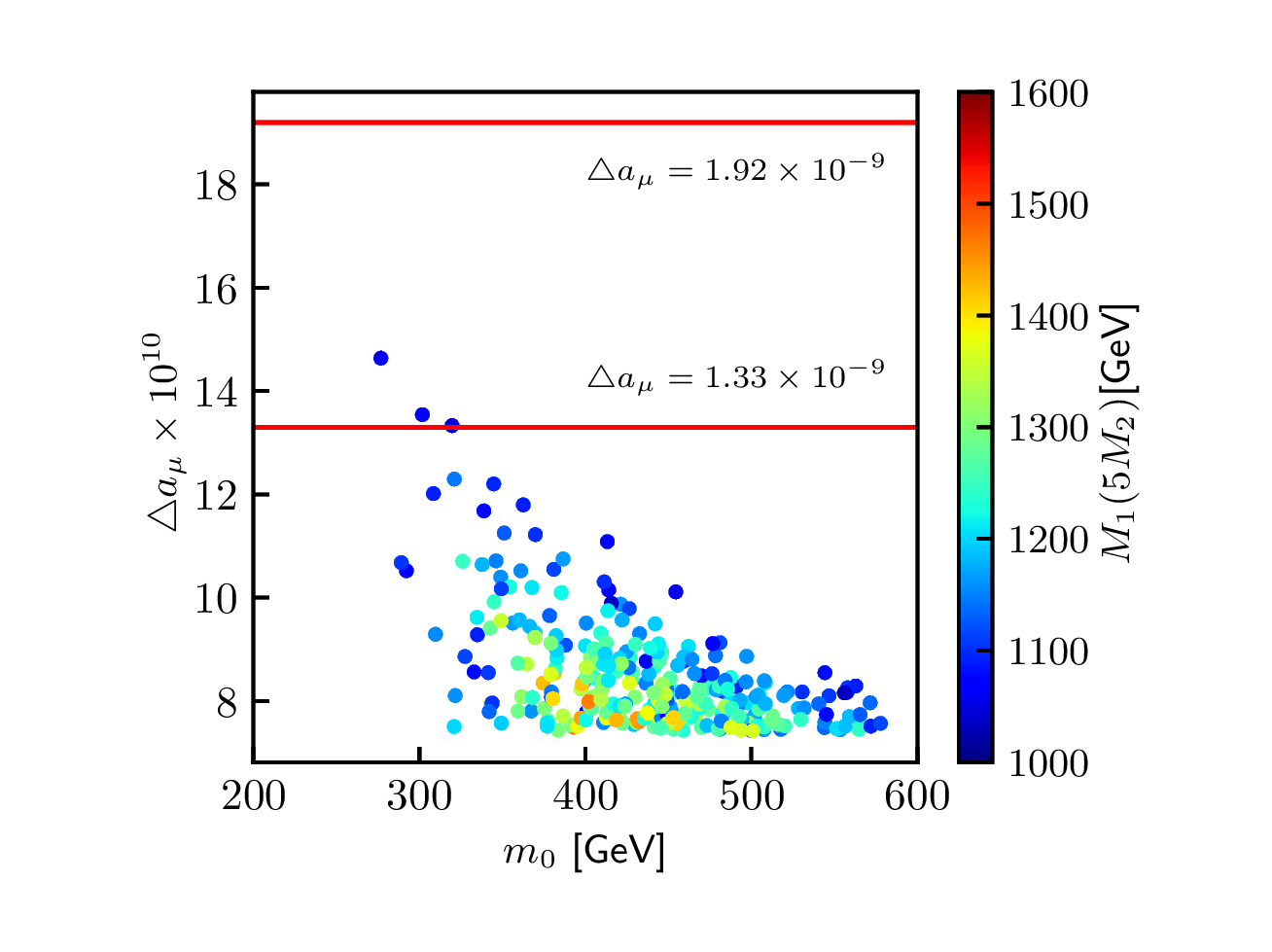}
\includegraphics[width=2.9in]{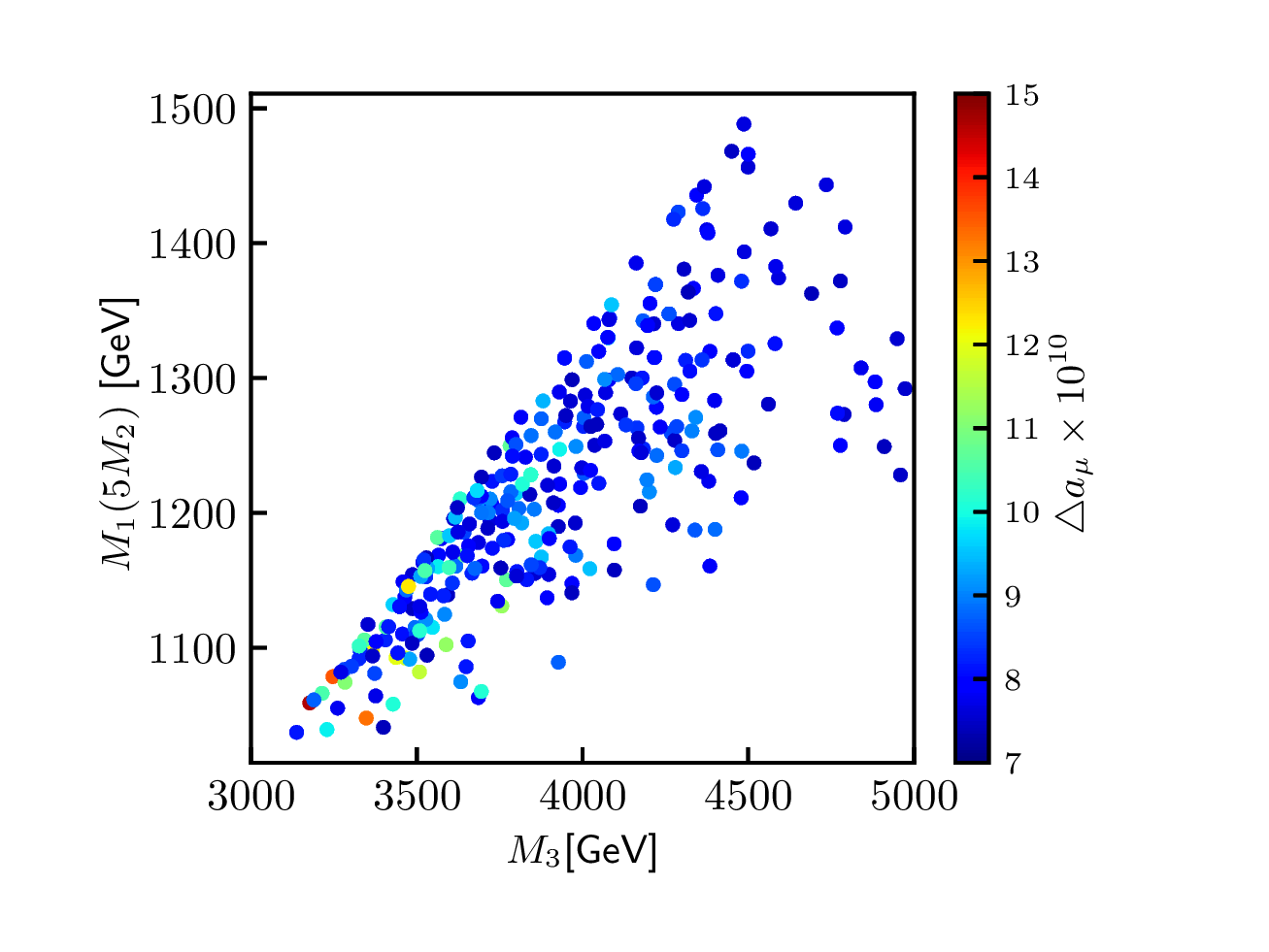}\\
\includegraphics[width=2.9in]{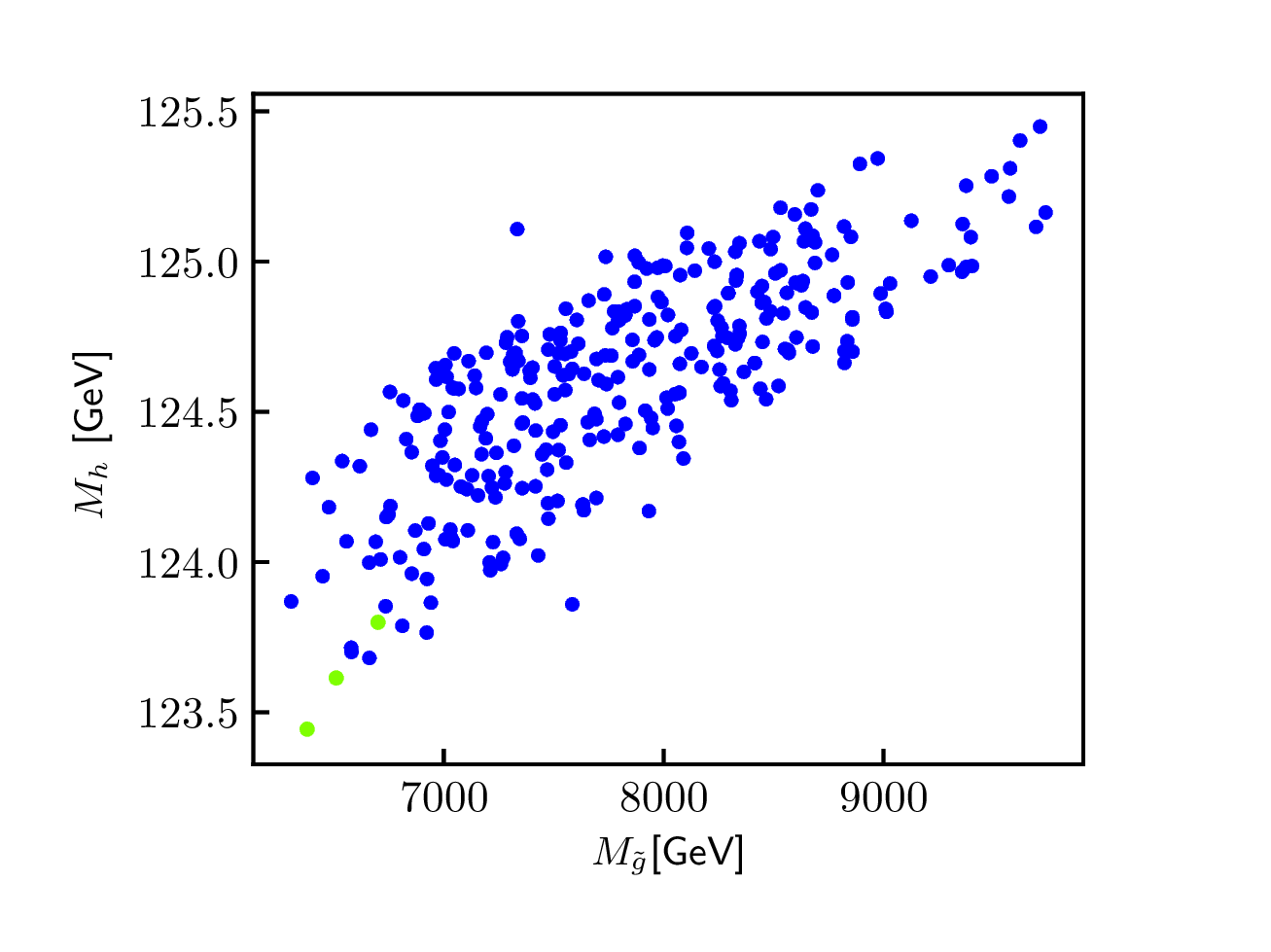}
\includegraphics[width=2.9in]{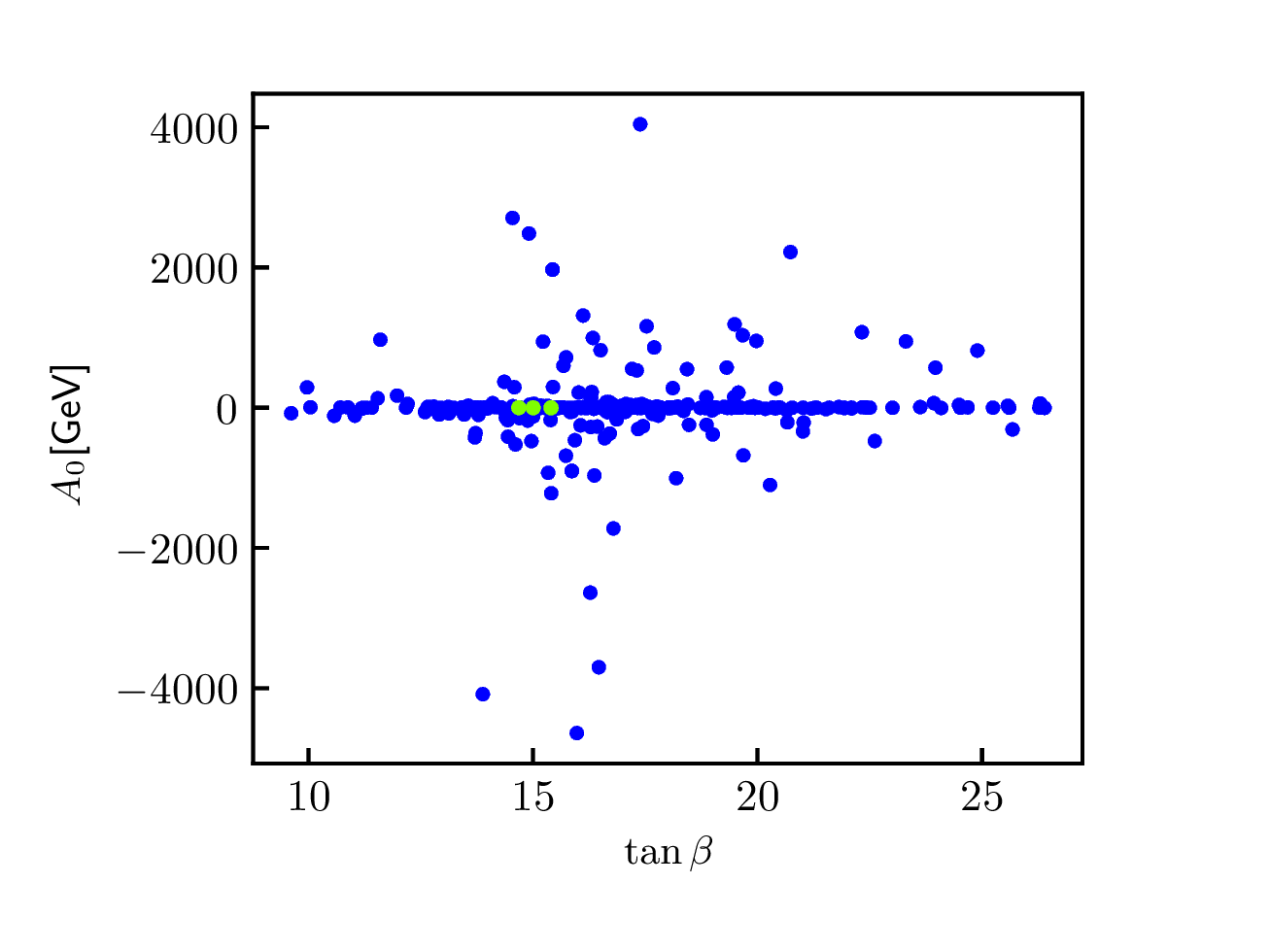}\\
\includegraphics[width=2.9in]{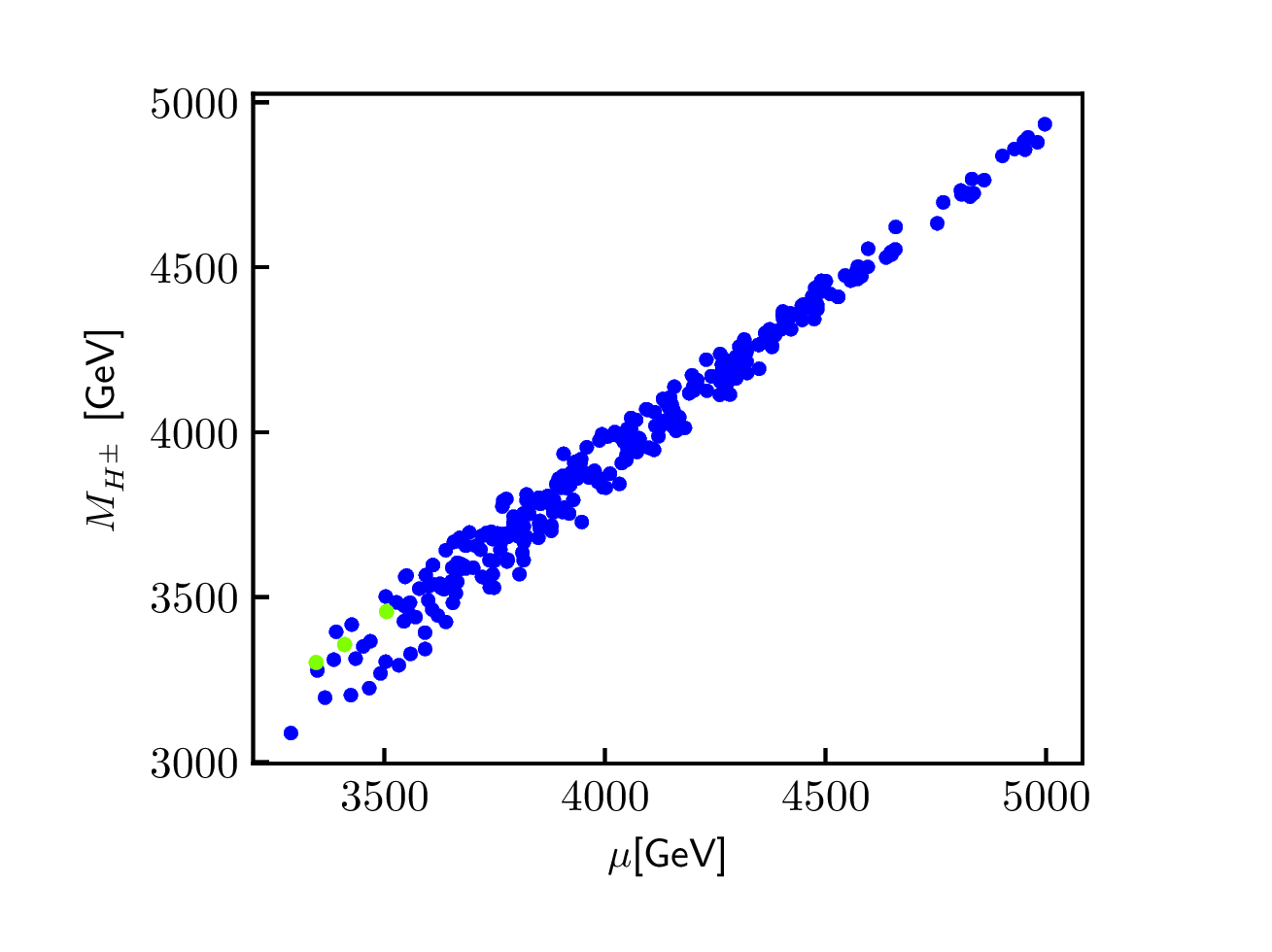}
\includegraphics[width=2.9in]{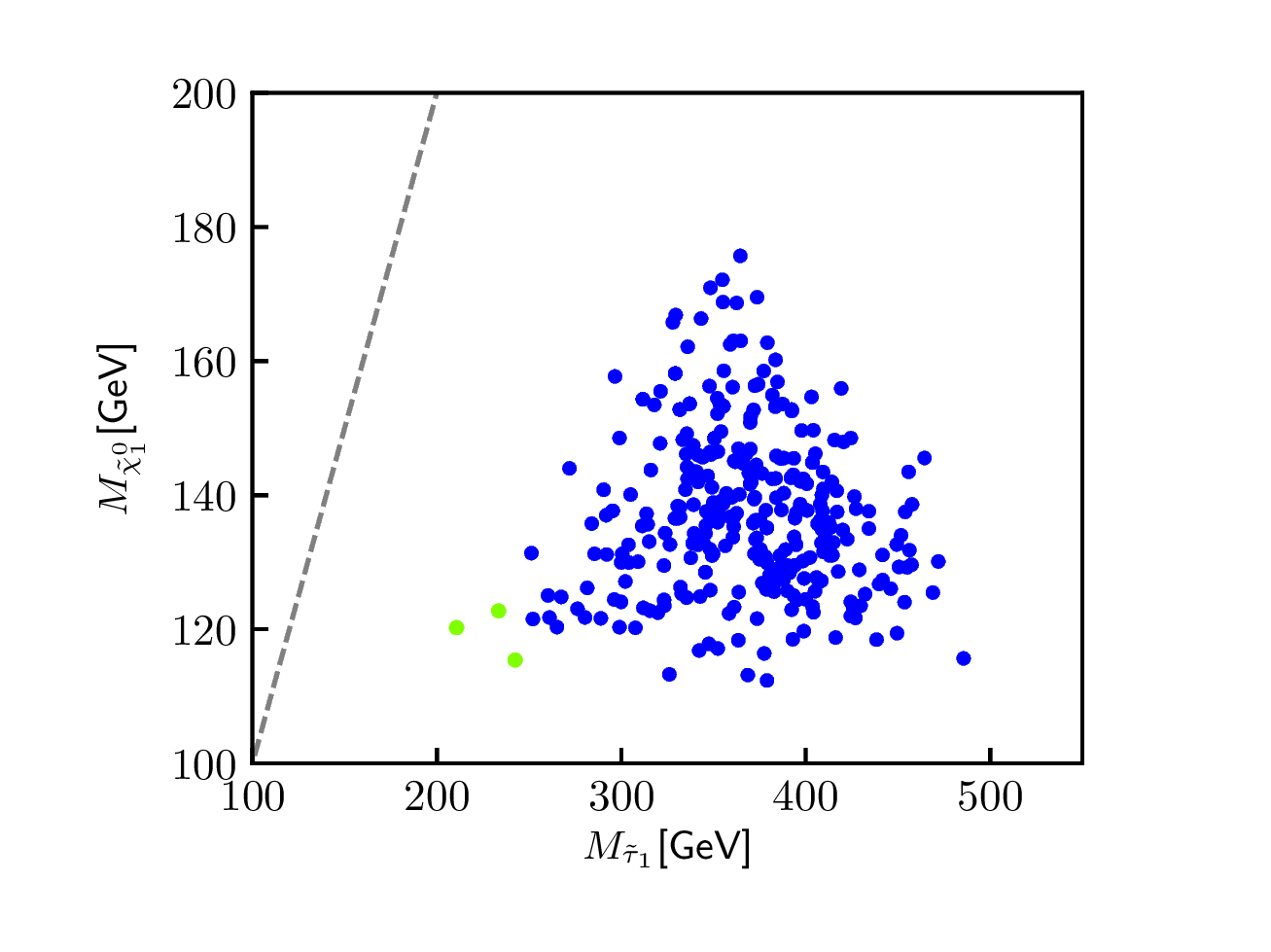}\\
\vspace{-.5cm}\end{center}
\caption{ Survived points in scenario (2) that can satisfy the constraints (i-vi) and give SUSY contributions to $\Delta a_\mu$ up to the $2\sigma$ range of $\Delta a_\mu^{combine}$.
In the middle and lower panels, the green (blue) points denote the survived samples that can give $\Delta a_\mu^{SUSY}$ in the $2\sigma$ ($3\sigma$) range.}
\label{fig3}
\end{figure}
The dark matter in this scenario is almost pure (neutral) wino. The wino dark matter will annihilate very efficiently so that it always provide under-abundance of cosmic dark matter if wino is lighter than approximately 3 TeV.  As the explanation of muon $g-2$ requires light electroweakinos of order ${\cal O}(100)$ GeV, the wino dark matter always provides insufficient dark matter relic abundance. So, additional dark matter species are necessary. The allowed masses for stau versus wino-like neutralino are shown in the lower right panel of Fig.\ref{fig3}. Much of the allowed region will be excluded by the LHC exclusion bounds on sleptons~\cite{ATLASsl}.
\begin{figure}[htb]
\begin{center}
\includegraphics[width=2.9in]{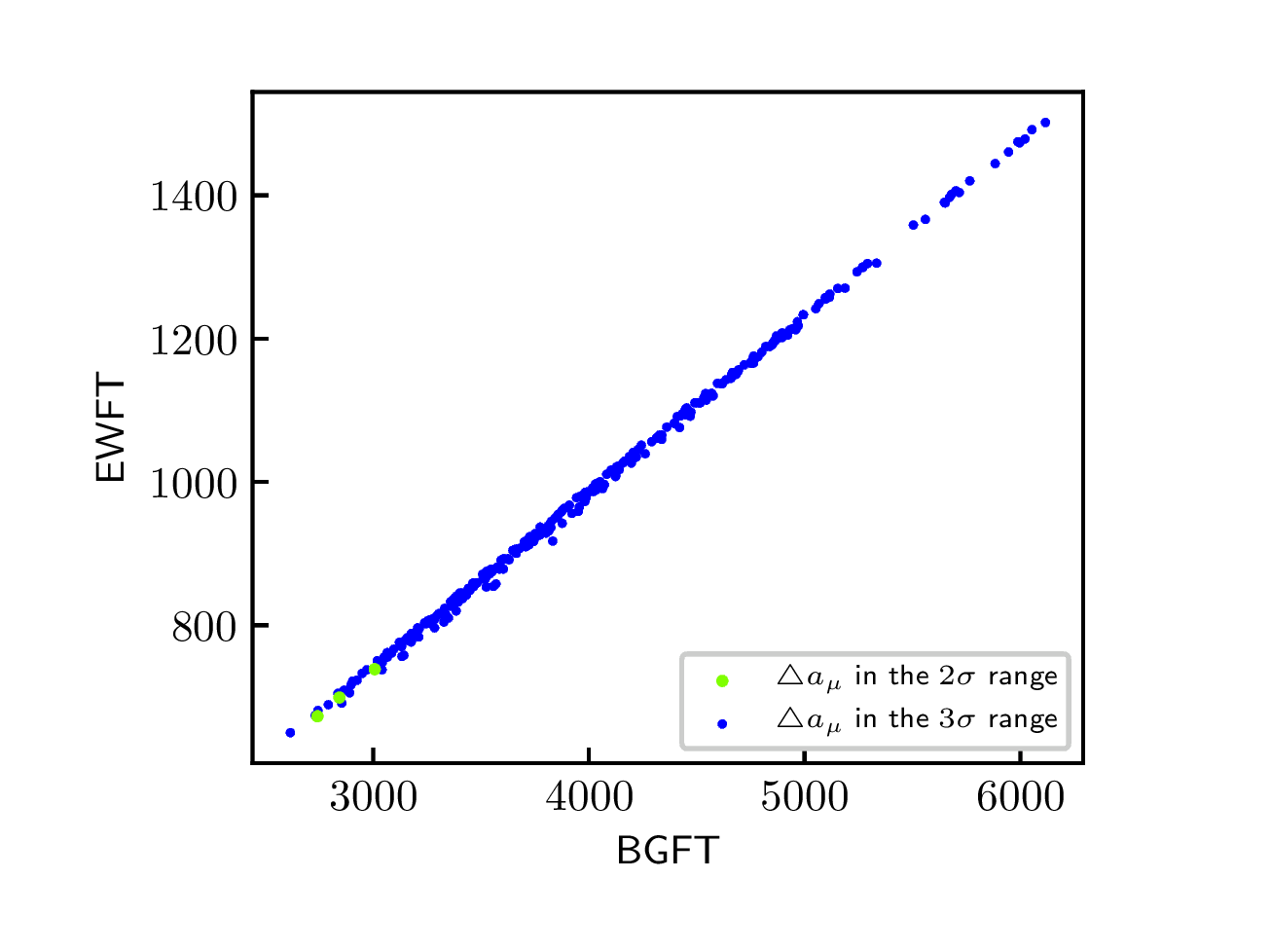}
\includegraphics[width=2.9in]{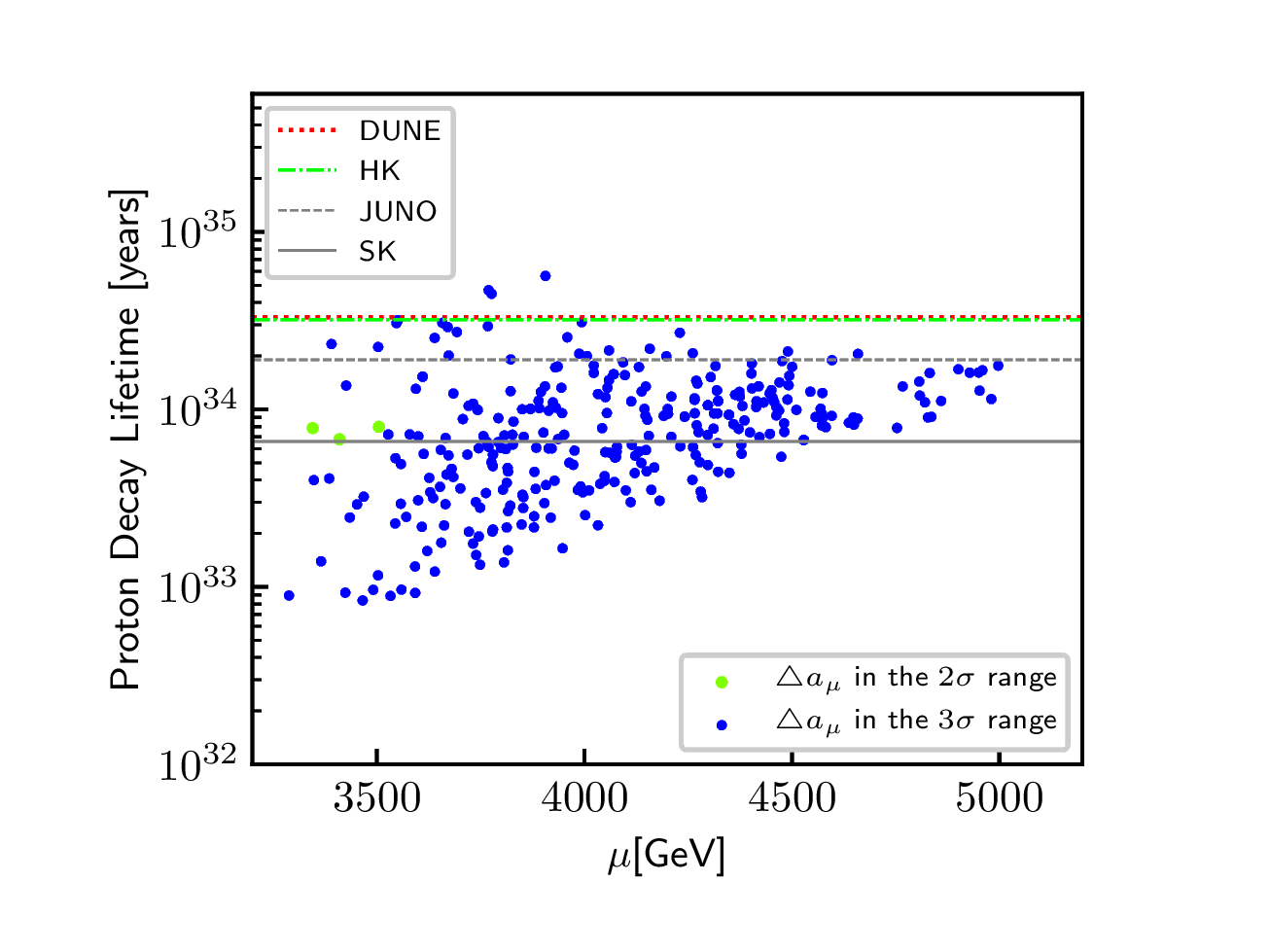}\\
\includegraphics[width=2.9in]{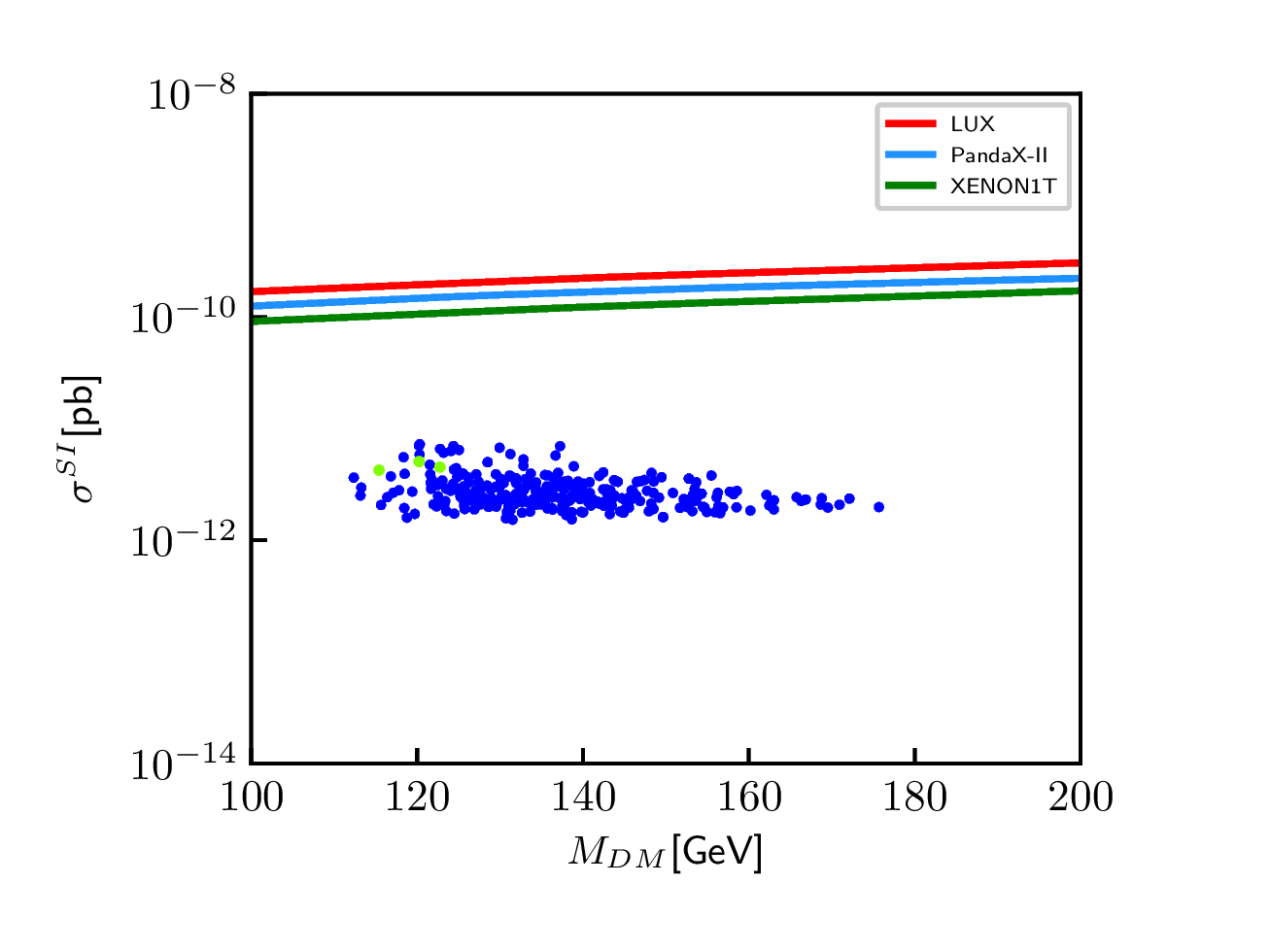}
\includegraphics[width=2.9in]{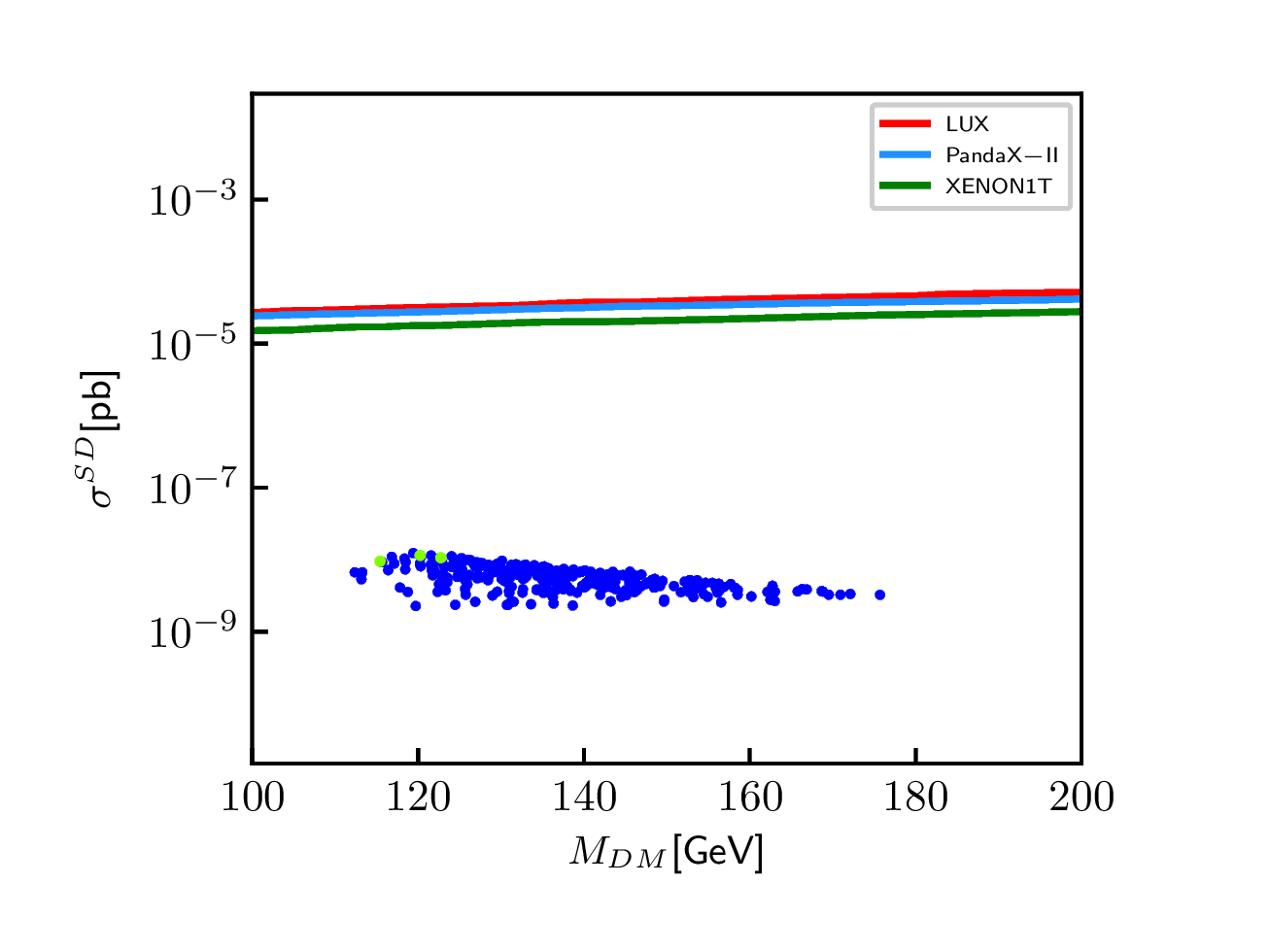}\\
\end{center}
\vspace{-.5cm}
\caption{Same as Fig.3, but for scenario (2) with the wino dark matter. }
\label{fig4}
\end{figure}

From Fig.\ref{fig4}, it can be seen that the EWFTs can be as low as 670 in the $2\sigma$ range of $\Delta a_\mu^{combine}$, which indicates that some survived points are fairy natural. It is also obvious that much larger FTs (up to $\Delta_{EW}\sim 1500$) are needed in most of the $3\sigma$ range of $\Delta a_\mu^{combine}$.

This scenario can also be constrained by the lifetime of proton decay via dimension-five operators. All the $2\sigma$ range of $\Delta a_\mu^{combine}$ can marginally survive the existing bounds by super-Kamiokande.  Most of the survived points in the $3\sigma$ range of $\Delta a_\mu^{combine}$, which predict $\tau_p\lesssim 4.0\tm 10^{34}$ year, can be tested by future proton decay experiments, such as DUNE, Hyper-Kamiokande and JUNO.

The interactions between wino dark matter and the nucleons are mediated by the t-channel scalar Higgs bosons and the s-channel squarks. Both the squark-mediated and Higgs-mediated neutralino-proton scattering amplitudes will in general be enhanced by more than one order of magnitude due to the wino nature of the dark matter particle. The reason for the enhancement is traced back to the structure of the neutralino-quark-squark and neutralino-neutralino-Higgs couplings, where the wino fraction is determined by the $SU(2)_L$ coupling, while the bino fraction by the (smaller) $U(1)_Y$ coupling. However, in this $\tl{g}$SUGRA scenario, the higgsino component of the dark matter is tiny and squarks are very heavy, leading to suppressed Higgs-neutralino-neutralino and neutralino-quark-squark couplings.
The SI cross section of wino-nucleon scattering from weak gauge boson loop is typically of order $10^{-10}\sim 10^{-12}$ pb if the tree-level contributions are suppressed. With an under-abundance of DM, direct detection of the wino dark matter is difficult. We find that all the survived samples can pass the current SI and SD direct detection limits,
even without re-scaling the original values by $\Omega/\Omega_0$.

\item[(3)]  Scenario with
\begin{eqnarray}
M_1:M_2=3:2, ~~~~~M_1 \in [50,2000]~\gev.
\end{eqnarray}
From the upper-left panel of Fig.\ref{fig5}, we can see that this scenario can marginally solve the muon $g-2$ anomaly in the 1$\sigma$ range, without conflicting with the constraints (i-vi). The parameter $M_1$ is bounded to lie within $[346,450]$ GeV and the parameter $m_0$ is bounded to lie within $[300,350]$ GeV if the muon $g-2$ anomaly is solved at the $1\sigma$ level. From the upper-right panel of Fig.\ref{fig5}, we can find the gaugino mass ratio should lie in the $6.9\lesssim M_3/M_1\lesssim 7.5$ range at the GUT scale with the central value at $7.2$. The gaugino mass ratios, which are pushed larger through RGE evolution, are given approximately by $M_1:M_2:M_3\approx 3:4:130$ at the weak scale. From the lower left panel, it is clear that the muon $g-2$ anomaly can be explained within 1$\sigma$ range with the gluino mass region of  $4.8~{\rm TeV} \lesssim M_{\tilde{g}}\lesssim 6.5~{\rm TeV}$, which cannot be discovered in the near future LHC experiments. Again, due to the heavy stops from a large $M_3$, the 125 GeV Higgs can be accommodated easily.

\begin{figure}[htb]
\begin{center}
\includegraphics[width=2.9in]{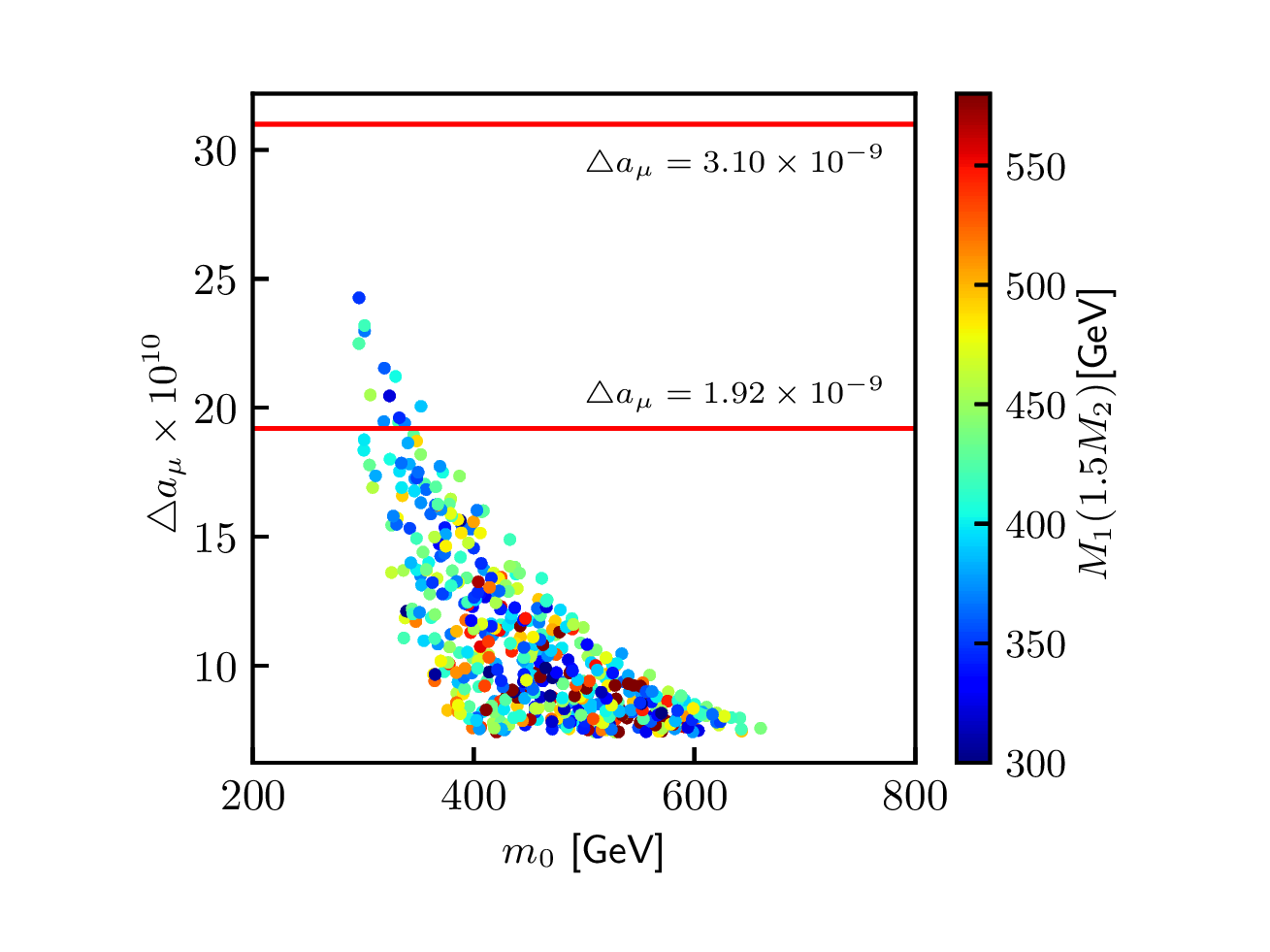}
\includegraphics[width=2.9in]{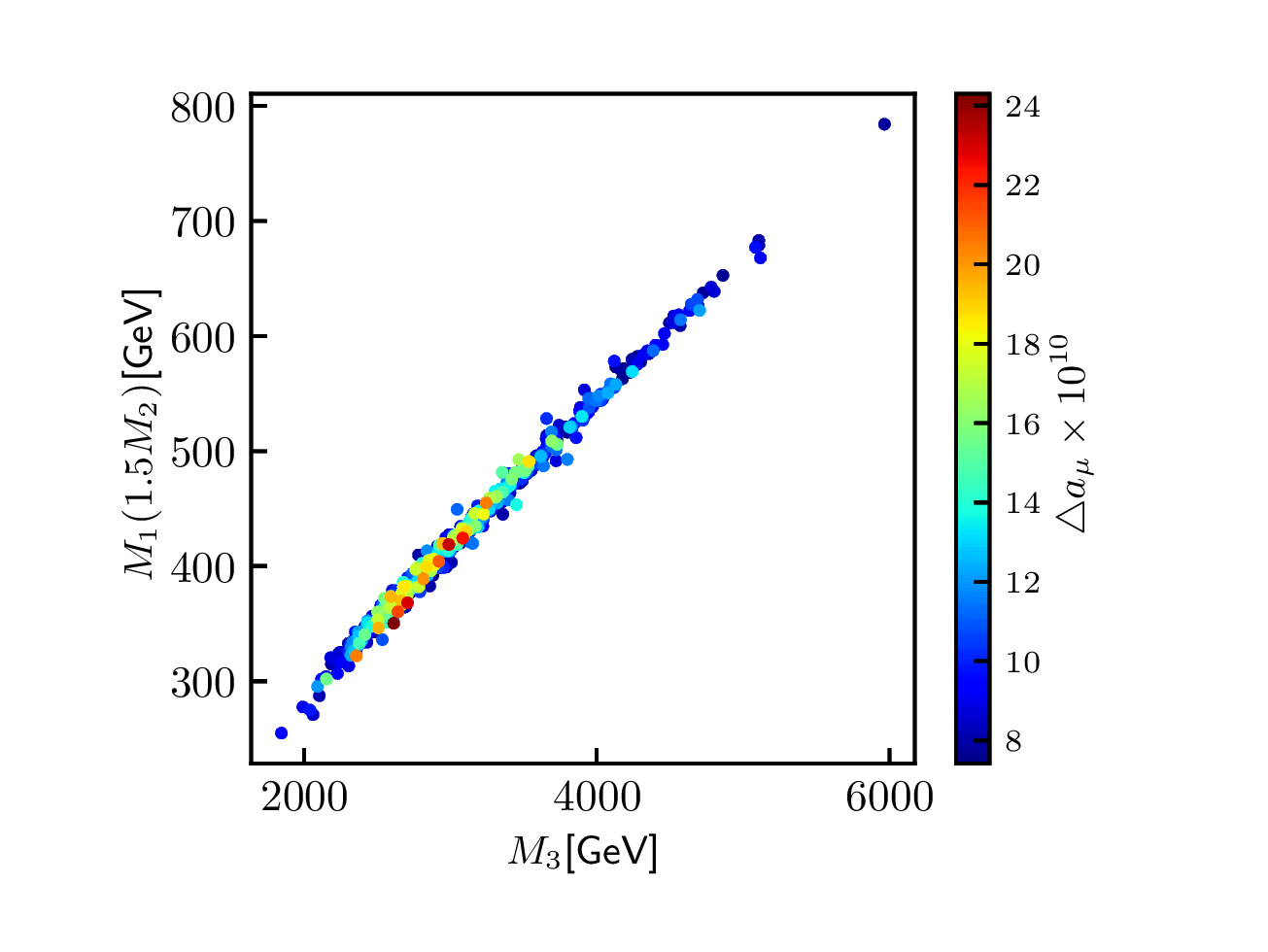}\\
\includegraphics[width=2.9in]{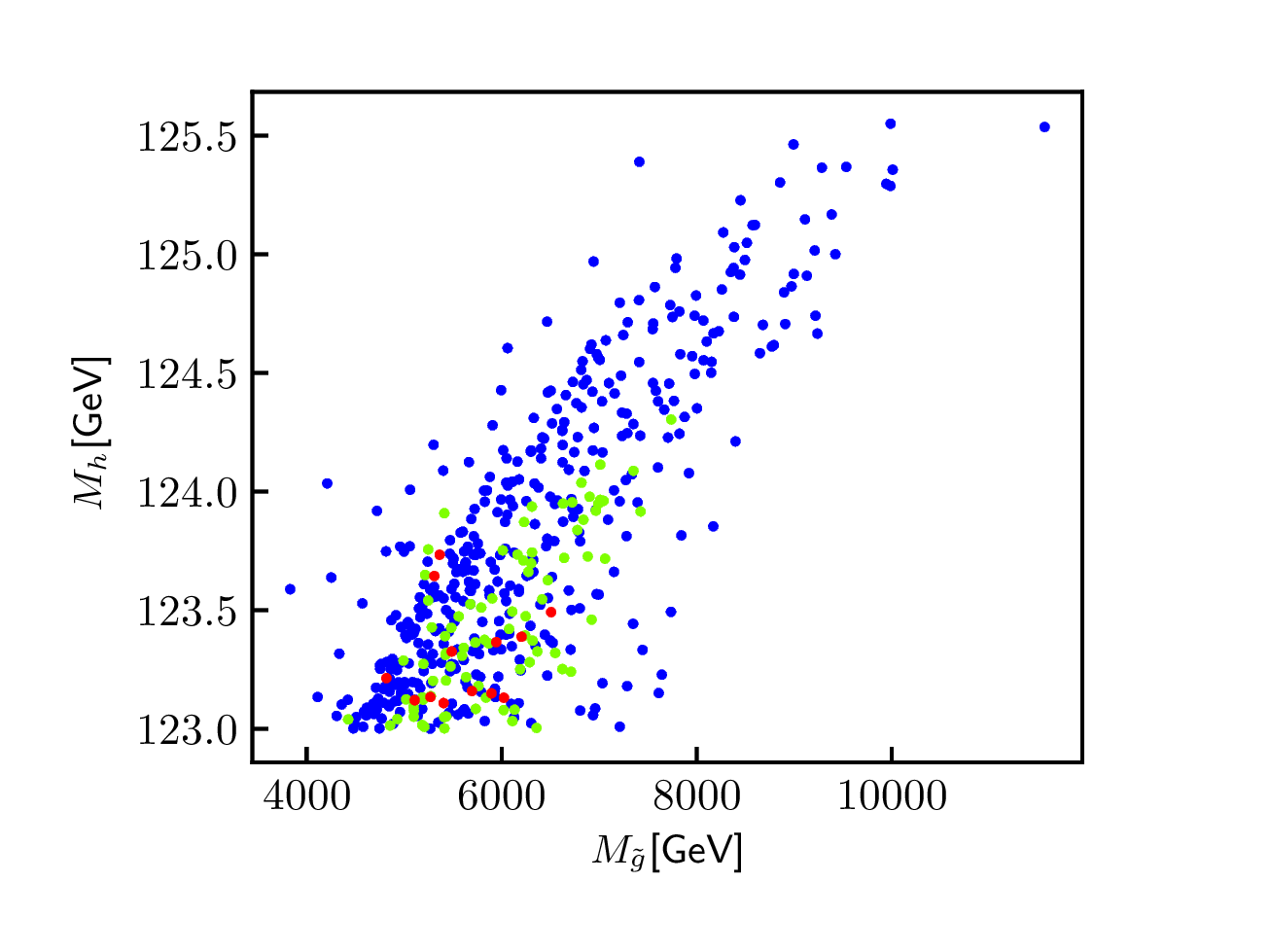}
\includegraphics[width=2.9in]{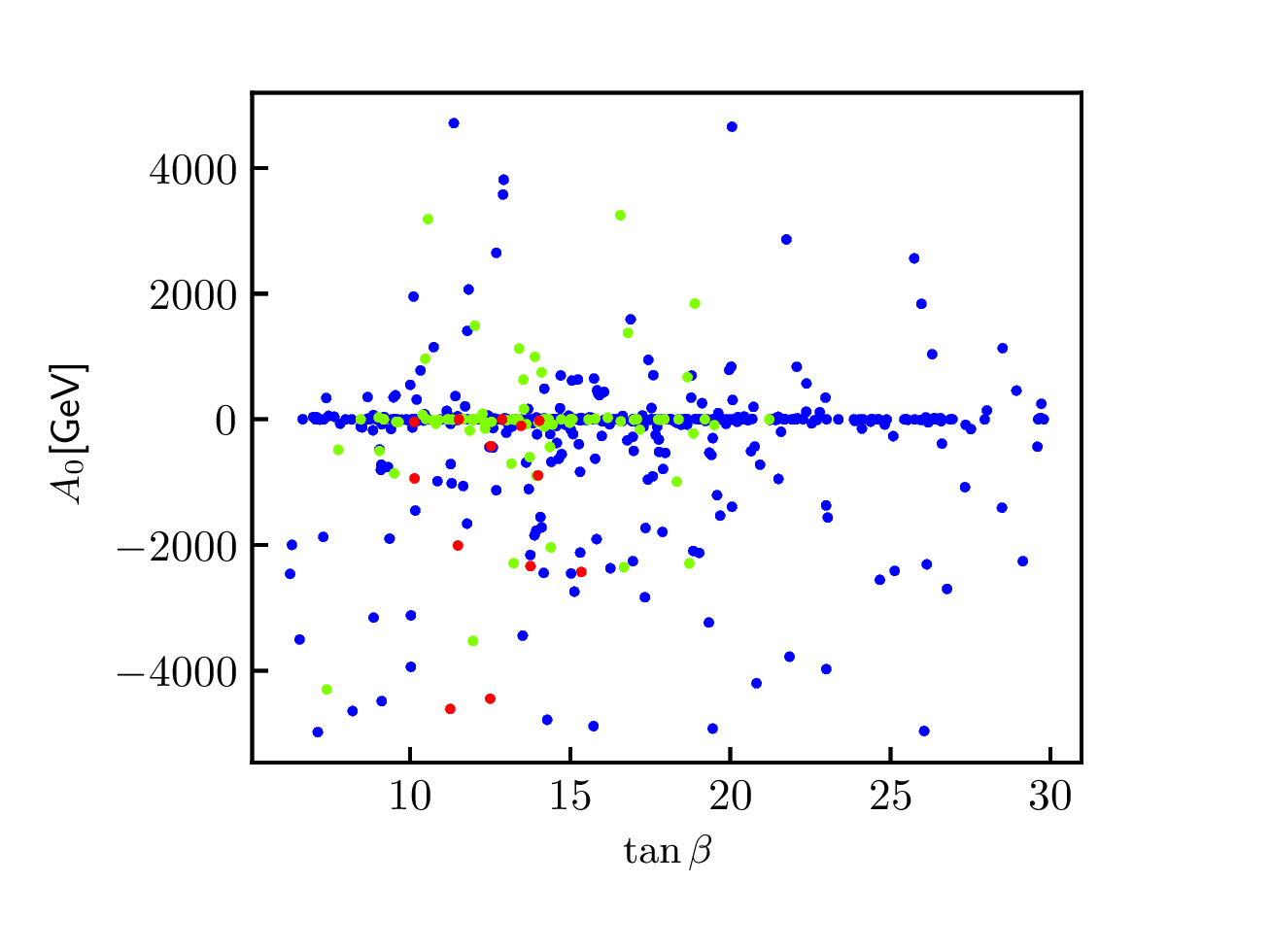}\\
\includegraphics[width=2.9in]{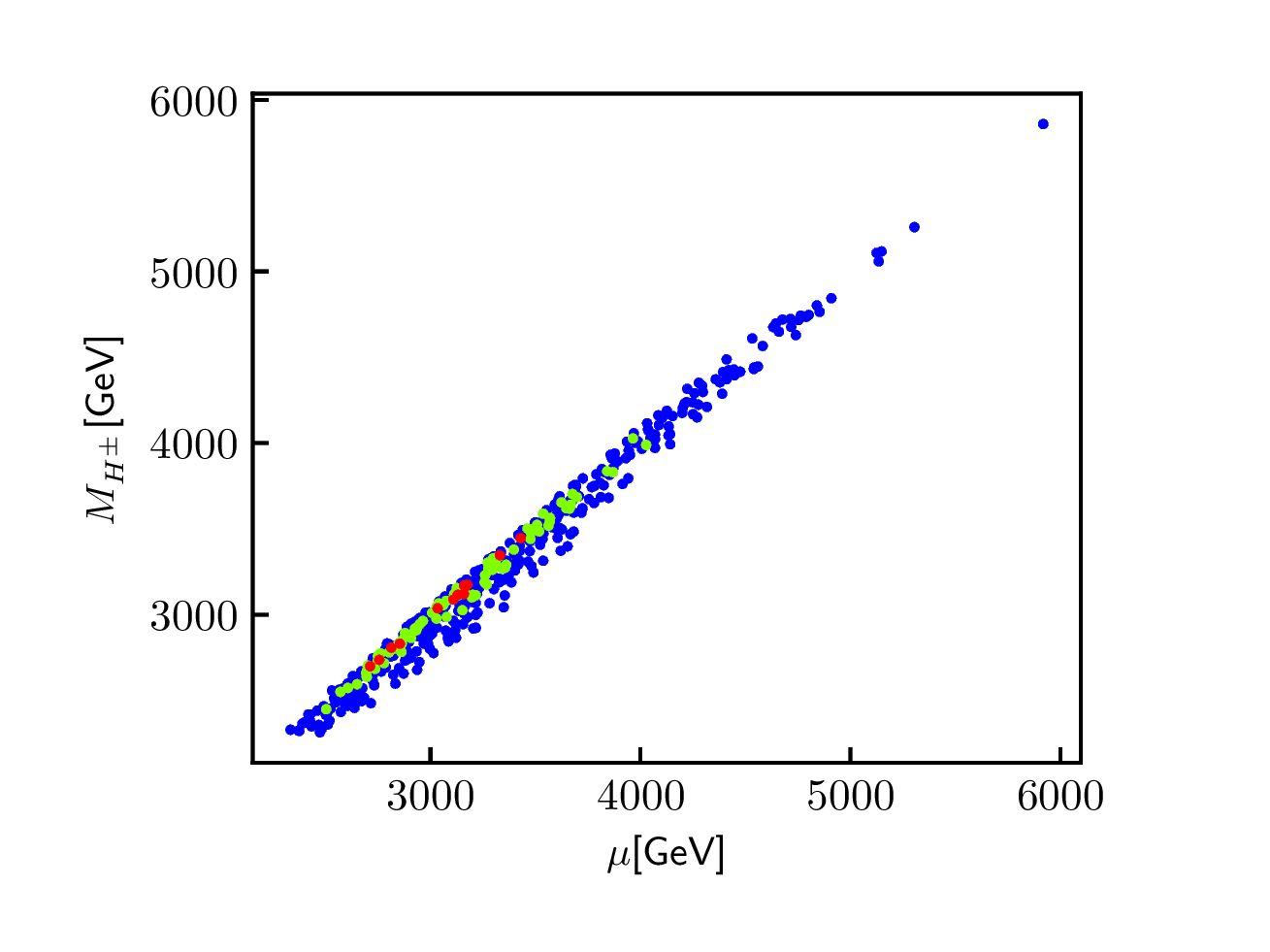}
\includegraphics[width=2.9in]{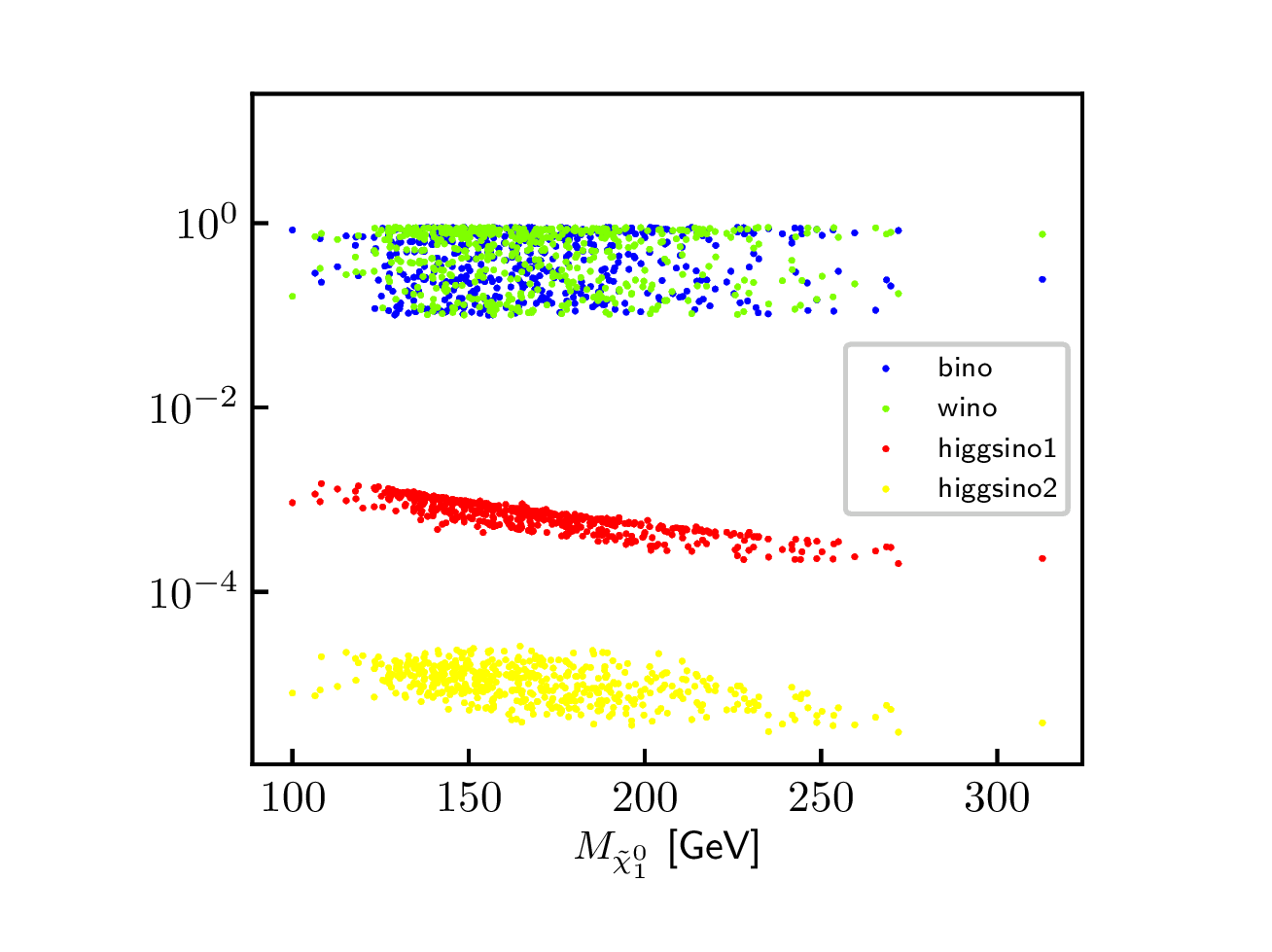}\\
\vspace{-.5cm}\end{center}
\caption{ Survived points in scenario (3) that can satisfy the constraints from (i-vi) and give SUSY contributions to $\Delta a_\mu$ up to the $1\sigma$ range of $\Delta a_\mu^{combine}$. In the middle and lower panels, the red, green and blue points denote the survived samples that can give $\Delta a_\mu^{SUSY}$ above the $1\sigma$, $2\sigma$ and $3\sigma$ bounds, respectively. }
\label{fig5}
\end{figure}
In this scenario, we find that the neutralino dark matter is a bino-wino mixture (see lower-right panel of Fig.\ref{fig5}). Because of the small mass difference between the bino and wino, a correct relic abundance of dark matter can be satisfied. The $\tl{\chi}_1^0-\tl{\chi}^\pm-W^\mp$ coupling becomes large when $\tl{\chi}_1^0$ becomes more and more wino-like, resulting in enhanced $\tl{\chi}_1^0\tl{\chi}_1^0 \ra W^+W^-$ annihilations. Co-annihilations with the lightest chargino and with the next-to-lightest neutralino help to further suppress the thermal relic abundance of the dark matter.

From Fig.\ref{fig6}, the EWFTs can be seen to lie between $440 \lesssim  \Delta_{EW}\lesssim 700$ in the $1\sigma$ range of $\Delta a_\mu^{combine}$. The EWFT can be as low as 370 for certain points in the $2\sigma$ range of $\Delta a_\mu^{combine}$, which indicate that it is still fairly natural to interpret $\Delta a_\mu^{combine}$ in this scenario. Again, it is also obvious that much larger FTs (up to $\Delta_{EW}\sim 2100$) are needed in most of the $3\sigma$ range of $\Delta a_\mu^{combine}$.

With the choice of $M^{eff}_{H_C}=1.0\tm 10^{19}{\rm GeV}$, current bounds for the lifetime of proton decay $p\to K^+ \bar{\nu}$ (via dimension-five operators) by super-Kamiokande can already rule out some of the survived points.  Most of the survived points in the $1\sigma$ range of $\Delta a_\mu^{combine}$, which predict $\tau_p\lesssim 3.1\tm 10^{34}$ year, can be tested by future DUNE, Hyper-Kamiokande proton decay experiments.

Similar to the previous two scenarios, dark matter can easily survive the current direct detection experiments as the higgsino component of the dark matter is tiny and squarks are heavy. We find that all the survived samples can pass the current SI and SD direct detection limits. Again, we do not rescale the original values by $\Omega/\Omega_0$.
\begin{figure}[htb]
\begin{center}
\includegraphics[width=2.9in]{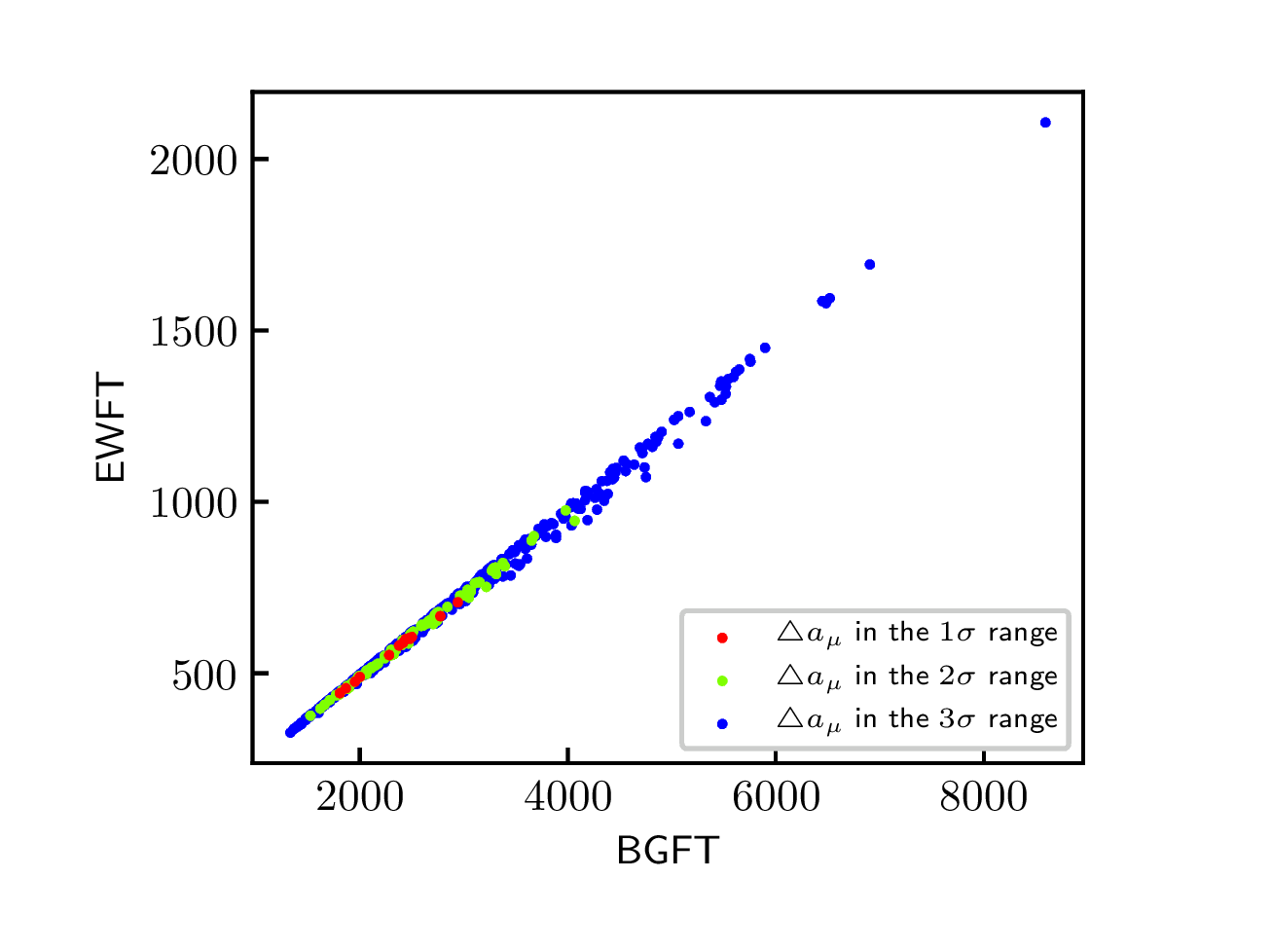}
\includegraphics[width=2.9in]{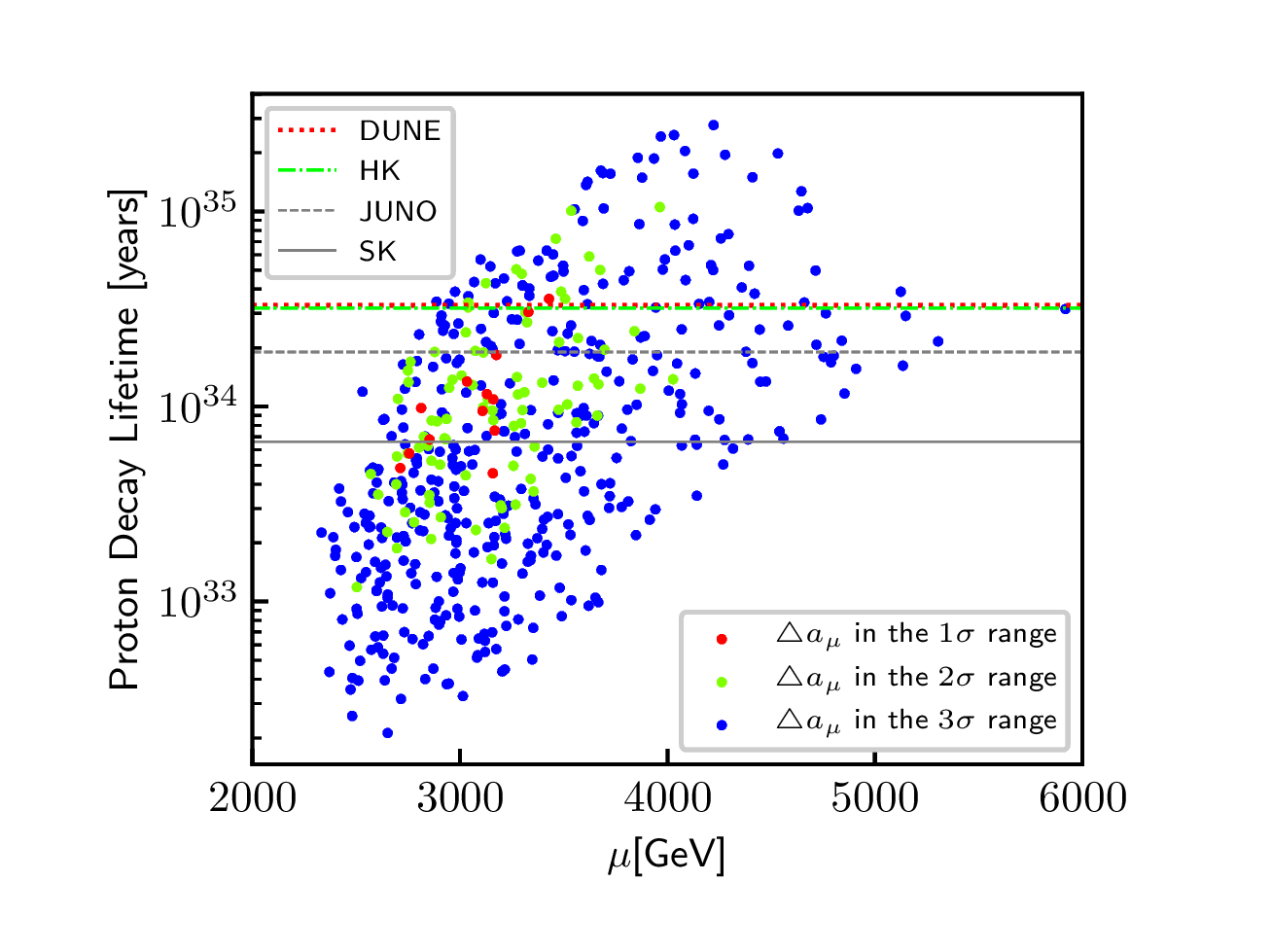}\\
\includegraphics[width=2.9in]{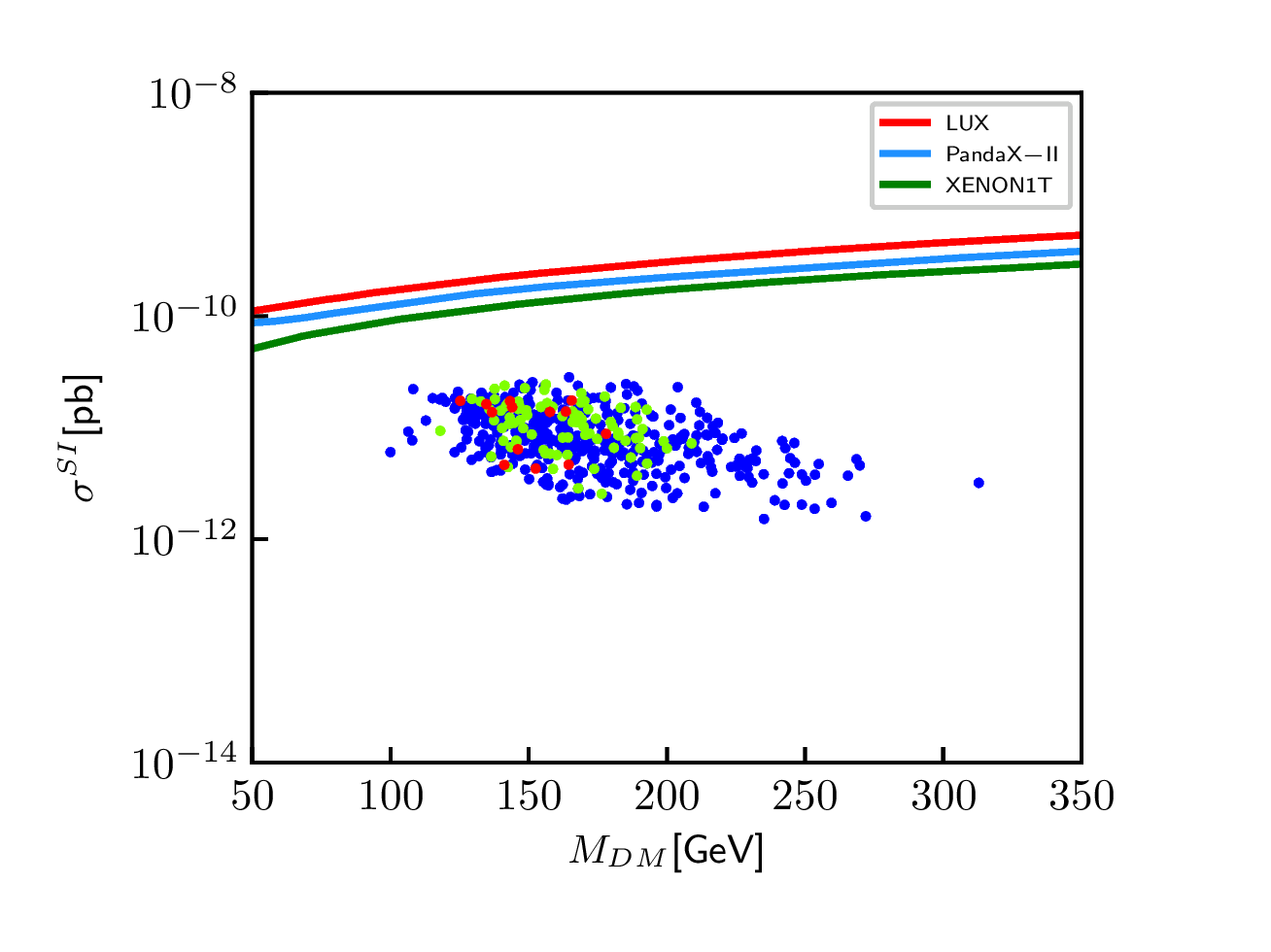}
\includegraphics[width=2.9in]{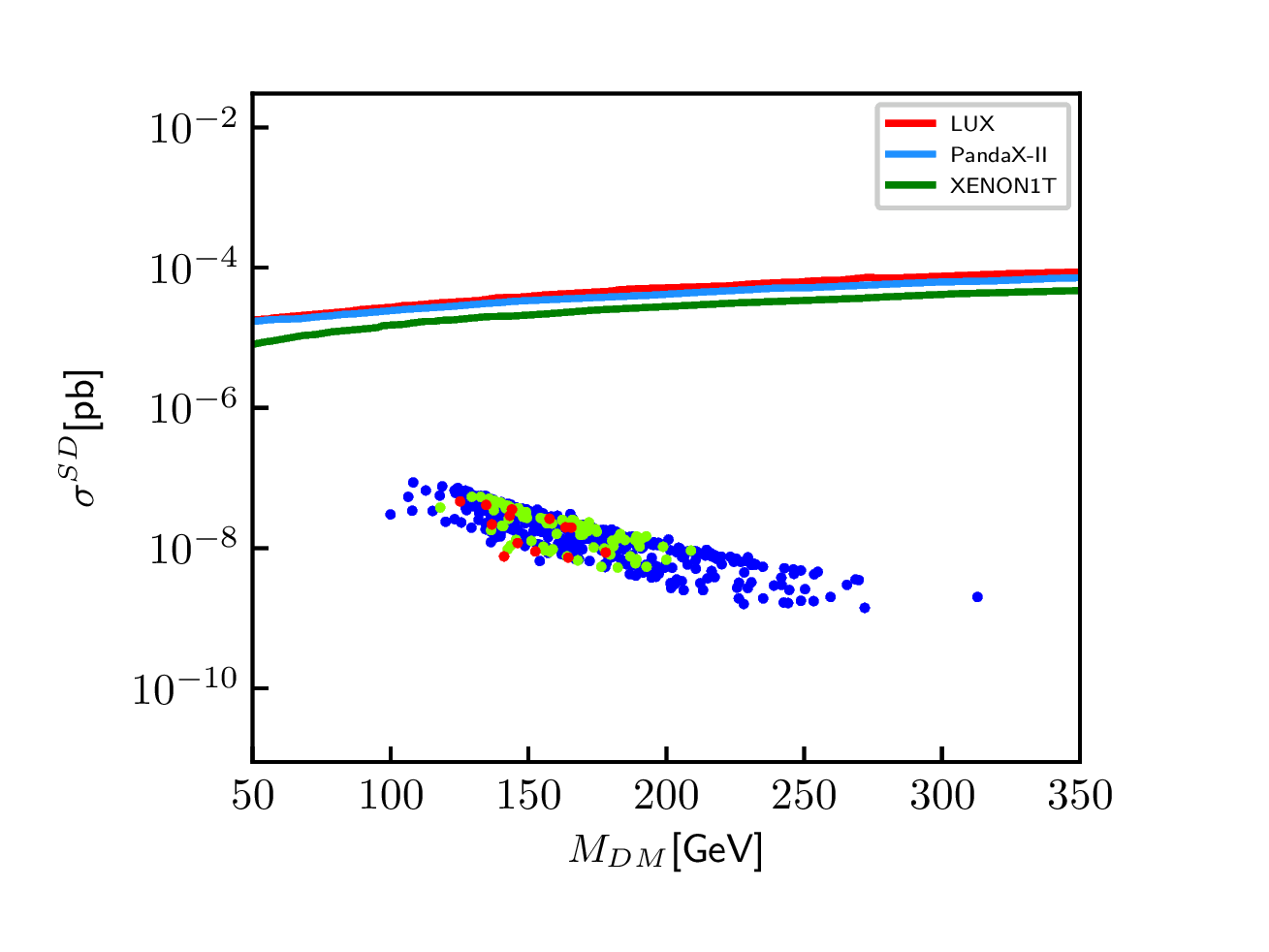}\\
\end{center}
\vspace{-.5cm}
\caption{Same as Fig.2, but for scenario (3) with bino-wino mixture dark matter. }
\label{fig6}
\end{figure}
\end{itemize}
Before we finish our numerical discussions, we would like to compare our results (based on SU(5) GUT) with some partial unification SUSY models, which can naturally allow non-universal soft SUSY breaking masses at the GUT scale $M_{GUT}$, for example, the partial unification $SU(4)_c\tm SU(2)_L\tm SU(2)_R$ Pati-Salam model. It is therefore interesting to see if new features can appear in such non-universal gaugino realization SUSY Pati-Salam model, taking into account also the constraints from LHC, cold dark matter searches and muon $g-2$ constraints etc. The authors in~\cite{Gomez:2018zzw} found that, for Pati-Salam 422 model, the particular relations between the gaugino masses in 422 result in relatively light gluinos with gluino-coannihilations to reduce the DM relic density, which is a very particular feature that does not appear in other GUT schemes. Besides, those points, which can account for the muon $g-2$ anomaly, prefer a very special ratio of GUT value $M_2/M_3\simeq 0.4$. This pattern is also different to our numerical results (for SU(5) GUT), in which the $1\sigma/2\sigma$ survived points do not show any preferences for the ratio of GUT value $M_2/M_3$.

\section{Conclusion}
\label{conclusion}
Gluino-SUGRA ($\tl{g}$SUGRA), which is an economical extension of the predictive mSUGRA, adopts much heavier gluino mass parameter than other gauginos mass parameters and universal scalar mass parameter at the unification scale. It can elegantly reconcile the experimental results on the Higgs boson mass, the muon $g-2$,
the null results in search for supersymmetry at the LHC and the results from B-physics.
In this work, we propose several new ways to generate large gaugino hierarchy (i.e. $M_3\gg M_1,M_2$) for $\tl{g}$SUGRA model building and then discuss in detail the implications of the new muon $g-2$ results with the updated LHC constraints on such $\tl{g}$SUGRA scenarios. We obtain the following observations: (i) For the most interesting $M_1=M_2$ case at the GUT scale with a viable bino-like dark matter, the $\tl{g}$SUGRA can explain the muon $g-2$ anomaly at $1\sigma$ level and be consistent with the updated LHC constraints for $6\leq M_3/M_1 \leq 9$ at the GUT scale; (ii) For $M_1:M_2=5:1$ at the GUT scale with wino-like dark matter, the $\tl{g}$SUGRA model can explain the muon $g-2$ anomaly at $2\sigma$ level and be consistent with the updated LHC constraints for $3\leq M_3/M_1 \leq 3.2$ at the GUT scale; (iii) For $M_1:M_2=3:2$ at the GUT scale with mixed bino-wino dark matter, the $\tl{g}$SUGRA model can explain the muon $g-2$ anomaly at $1\sigma$ level and be consistent with the updated LHC constraints for $6.9\leq M_3/M_1 \leq 7.5$ at the GUT scale. Although the choice of heavy gluino will always increase the FT involved, some of the $1\sigma/2\sigma$ survived points of $\Delta a_\mu^{combine}$ can still allow low EWFT of order several hundreds and be fairly natural. Constraints from (dimension-five operator induced) proton decay are also discussed.

We see that although the elegant and predictive mSUGRA can not explain the muon $g-2$ anomaly, its most economical extension $\tl{g}$SUGRA with $M_1=M_2$ can consistently explain the muon $g-2$ anomaly up to $1\sigma$ range. Other $\tl{g}$SUGRA scenarios can also consistently explain the muon $g-2$ anomaly. So, the recent FNAL muon $g-2$ experimental results still prefer such interesting SUGRA-type scenarios.

\addcontentsline{toc}{section}{Acknowledgments}
\acknowledgments
We are very grateful to the referee for helpful suggestions. This work was supported by the National Natural Science Foundation of China
(NNSFC) under grant Nos. 12075213,  11821505 and 12075300,
by the Key Research Project of Henan Education Department for colleges and universities under grant number 21A140025,
by Peng-Huan-Wu Theoretical Physics Innovation Center (12047503),
by the CAS Center for Excellence in Particle Physics (CCEPP),
by the CAS Key Research Program of Frontier Sciences,
and by a Key R\&D Program of Ministry of Science and Technology of China
under number 2017YFA0402204.


\begin{thebibliography}{99}
\vspace{-1mm}
  \bibitem{FNAL:gmuon}
  B. Abi et al. [Muon g-2], Phys. Rev. Lett. 126,  141801 (2021), 
 arXiv:2104.03281.
  \bibitem{BNL:gmuon}
  G. W. Bennett et al. [Muon g-2 collaboration], 
  Phys. Rev. D 73, 072003  (2006), hep-ex/0602035;\\
 P. Zyla et al. [Particle Data Group collaboration], 
 PTEP 2020,  083C01 (2020).
 \bibitem{SMmuong-2}
  T. Aoyama et al.  Phys. Rept. 887 (2020) 1-166, arXiv: 2006.04822 [hep-ph];\\
  A. Czarnecki, W. J. Marciano, A. Vainshtein,  Phys. Rev. D67 (2003) 073006, arXiv:hep-ph/0212229 [hep-ph];\\
  M. Davier, A. Hoecker, B. Malaescu, Z. Zhang, Eur. Phys. J. C80, 241(2020), arXiv:1908.00921 [hep-ph];\\
  M. Davier, A. Hoecker, B. Malaescu, Z. Zhang, Eur. Phys. J. C77, 827(2017), arXiv:1706.09436 [hep-ph];\\
  M. Davier, A. Hoecker, B. Malaescu, Z. Zhang, Eur. Phys. J. C71, 1515(2011), arXiv:1010.4180 [hep-ph];\\
  T. Aoyama, M. Hayakawa, T. Kinoshita, M. Nio, Phys. Rev. Lett. 109 (2012) 111808, arXiv:1205.5370 [hep-ph];\\
  T. Blum, N. Christ, M. Hayakawa, T. Izubuchi, L. Jin, C. Jung, C. Lehner, Phys. Rev. Lett. 124 no. 13, (2020) 132002, arXiv:1911.08123 [hep-lat];\\
  A. Kurz, T. Liu, P. Marquard, M. Steinhauser, Phys. Lett. B734 (2014)144, arXiv:1403.6400 [hep-ph].
\bibitem{Athron:muong-2}
P.~Athron, C.~Bal\'azs, D.~H.~Jacob, W.~Kotlarski, D.~St\"ockinger and H.~St\"ockinger-Kim,
arXiv:2104.03691.
 \bibitem{Run2} A. Canepa, Rev. Phys. 4, 100033 (2019), doi:10.1016/j.revip.2019.100033
 \bibitem{CMSSM:gluino} M. Aaboud et al. [ATLAS Collaboration], Phys. Rev. D 97, 112001 (2018),
                        arXiv:1712.02332;\\
                         T. A. Vami et al. [ATLAS and CMS Collaborations], PoS LHCP 2019, 168 (2019),
                         arXiv:1909.11753.
  \bibitem{CMSSM:stop} ATLAS collaboration [ATLAS Collaboration], ATLAS-CONF-2019-017;\\
  A. M. Sirunyan et al. [CMS Collaboration], CMS-SUS-19-009, JHEP 05, 032(2020). arXiv:1912.08887.
  \bibitem{SUGRA}
                 A.~H.~Chamseddine, R.~L.~Arnowitt and P.~Nath,
                 Phys.\ Rev.\ Lett.\ {\bf 49}, 970 (1982);\\
                 H.~P.~Nilles,
                 Phys.\ Lett.\ B {\bf 115}, 193 (1982);\\
                 L.~E.~Ibanez,
                 Phys.\ Lett.\ B {\bf 118}, 73 (1982);\\
                 R.~Barbieri, S.~Ferrara and C.~A.~Savoy,
                 Phys.\ Lett.\ B {\bf 119}, 343 (1982);\\
                 H.~P.~Nilles, M.~Srednicki and D.~Wyler,
                  Phys.\ Lett.\ B {\bf 120}, 346 (1983);\\
                 J.~R.~Ellis, D.~V.~Nanopoulos and K.~Tamvakis,
                 Phys.\ Lett.\ B {\bf 121}, 123 (1983);\\
                 J.~R.~Ellis, J.~S.~Hagelin, D.~V.~Nanopoulos and K.~Tamvakis,
                 Phys.\ Lett.\ B {\bf 125}, 275 (1983);\\
                 N. Ohta,
                 Prog.\ Theor.\ Phys.\ 70, 542 (1983);\\
                 L.~J.~Hall, J.~D.~Lykken and S.~Weinberg,
                 Phys.\ Rev.\ D {\bf 27}, 2359 (1983);\\
                 F.~Wang, K.~Wang, J.~M.~Yang and J.~Zhu,
                JHEP \textbf{12}, 041 (2018),
                arXiv:1808.10851; \\
                F.~Wang, W.~Wang and J.~M.~Yang,
                JHEP \textbf{03}, 050 (2015),
                arXiv:1501.02906;\\
                 K. Wang, F. Wang, J. Zhu, Q. Jie, Chinese Physics C 42, 103109 (2018).
  \bibitem{GMSB}
                M.~Dine, W.~Fischler and M.~Srednicki,
                Nucl.\ Phys.\ B {\bf 189}, 575 (1981);\\
                S.~Dimopoulos and S.~Raby,
                Nucl.\ Phys.\ B {\bf 192}, 353 (1981);\\
                M.~Dine and W.~Fischler, Phys.\ Lett.\ B {\bf 110}, 227 (1982);\\
                M. Dine and A. E. Nelson, Phys. Rev. {\bf D48}, 1277 (1993);\\
                M. Dine, A. E. Nelson and Y. Shirman, Phys. Rev. {\bf D51}, 1362 (1995);\\
                M. Dine, A. E. Nelson, Y. Nir and Y. Shirman, Phys. Rev. {\bf D53}, 2658 (1996);\\
                G. F. Giudice and R. Rattazzi, Phys. Rept. {\bf 322}, 419 (1999);\\
                J.~Dai, T.~Liu and J.~M.~Yang, [arXiv:2104.12656 [hep-ph]];
  \bibitem{AMSB}
                L.~Randall and R.~Sundrum,
                Nucl.\ Phys.\ B {\bf 557}, 79 (1999),
                hep-th/9810155;
                G.~F.~Giudice, M.~A.~Luty, H.~Murayama and R.~Rattazzi,
                JHEP {\bf 9812}, 027 (1998),
                 hep-ph/9810442;
                F. Wang, Phys. Lett. B 751, 402 (2015);
                F. Wang, W. Wang, J.M. Yang, Y. Zhang, JHEP 07, 138 (2015);
                X. Ning, F. Wang, JHEP 08, 089 (2017);
                D. Xiaokang, F. Wang, Eur. Phys. J. C 78, 431 (2018);
                F. Wang, J.M. Yang, Y. Zhang, JHEP 04, 177 (2016);
                F. Wang, W. Wang, J.M. Yang, Phys. Rev. D 96, 075025 (2017);
                Zhuang Li, Fei Wang, Eur. Phys. J. C (2020) 80:798.
   \bibitem{muong-2inSUSY}
    A. Crivellin, M. Hoferichter, arXiv:2104.03202 [hep-ph];\\
    M. Endo, K. Hamaguchi, S. Iwamoto, T. Kitahara, arXiv:2104.03217 [hep-ph];\\
    Y. Gu, N. Liu, L. Su, D. Wang, arXiv:2104.03239 [hep-ph];\\
    M. Van Beekveld, W. Beenakker, M. Schutten, J. De Wit, arXiv:2104.03245 [hep-ph];\\
    W. Yin, arXiv:2104.03259 [hep-ph];\\
    M. Abdughani, Y.-Z. Fan, L. Feng, Y.-L. Sming Tsai, L. Wu, Q. Yuan, arXiv:2104.03274 [hep-ph];\\
    M. Ibe, S. Kobayashi, Y. Nakayama, S. Shirai, arXiv:2104.03289 [hep-ph];\\
    P. Cox, C. Han, and T. T. Yanagida, arXiv:2104.03290 [hep-ph];\\
    C. Han, arXiv:2104.03292 [hep-ph];\\
    S. Baum, M. Carena, N. R. Shah, C. E. M. Wagner, arXiv:2104.03302 [hep-ph];\\
    H.B. Zhang, C.X. Liu, J.L. Yang, T.-F. Feng, arXiv:2104.03489 [hep-ph];\\
    W. Ahmed, I. Khan, J. Li, T. Li, S. Raza, W. Zhang, arXiv:2104.03491 [hep-ph];\\
    J.L. Yang, H.B. Zhang, C.X. Liu, X.X. Dong, T.F. Feng, arXiv:2104.03542 [hep-ph];\\
    A. Aboubrahim, M. Klasen, P. Nath, arXiv:2104.03839 [hep-ph];\\
    M. Chakraborti, L. Roszkowski, S. Trojanowski, arXiv:2104.04458 [hep-ph];\\
    H. Baer, V. Barger, H. Serce, arXiv:2104.07597 [hep-ph];\\
    W. Altmannshofer, S. A. Gadam, S. Gori, N. Hamer, arXiv:2104.08293 [hep-ph];\\
    A. Aboubrahim, P. Nath, and R. M. Syed, arXiv:2104.10114 [hep-ph];\\
    M. Chakraborti, S. Heinemeyer, I. Saha, arXiv:2105.06408 [hep-ph];\\
    Z.N. Zhang, H.B. Zhang, J.L. Yang, S.-M. Zhao,T.F. Feng, arXiv:2105.09799 [hep-ph];\\
    K.S. Jeong, J. Kawamaura, C.B. Park, arXiv:2106.04238 [hep-ph].
   \bibitem{2104.03262}
    F.~Wang, L.~Wu, Y.~Xiao, J.~M.~Yang and Y.~Zhang,
    Nucl. Phys. B \textbf{970}, 115486 (2021)
    [arXiv:2104.03262 [hep-ph]].
   \bibitem{GUT-SUSY}
    M.~Chakraborti, L.~Roszkowski and S.~Trojanowski,
    JHEP \textbf{05} (2021), 252,
    arXiv:2104.04458;
    A.~Aboubrahim, P.~Nath and R.~M.~Syed,
    JHEP \textbf{06} (2021), 002,
    arXiv:2104.10114.
   \bibitem{Abdughani:2019wai}
    See, e.g., M.~Abdughani, K.~I.~Hikasa, L.~Wu, J.~M.~Yang and J.~Zhao,
    JHEP \textbf{11} (2019), 095
    [arXiv:1909.07792 [hep-ph]];
    P.~Cox, C.~Han and T.~T.~Yanagida,
    Phys. Rev. D \textbf{98} (2018) no.5, 055015
    [arXiv:1805.02802 [hep-ph]].
   \bibitem{MSSM:mg2}
    M.~D.~Zheng and H.~H.~Zhang,
    [arXiv:2105.06954 [hep-ph]];
    H.~Baer, V.~Barger and H.~Serce,
    [arXiv:2104.07597 [hep-ph]];
    W.~Altmannshofer, S.~A.~Gadam, S.~Gori and N.~Hamer,
    [arXiv:2104.08293 [hep-ph]];
    J.~Cao, J.~Lian, Y.~Pan, D.~Zhang and P.~Zhu,
    [arXiv:2104.03284 [hep-ph]].
   \bibitem{gSUGRA:nath}
                       S.~Akula and P.~Nath,
                       Phys. Rev. D \textbf{87}, 115022  (2013),
                       arXiv:1304.5526.
   \bibitem{NUGM} B. Ananthanarayan, P. N. Pandita, Int. J. Mod. Phys. A22, 3229-3259 (2007);
    S. Bhattacharya, A. Datta and B. Mukhopadhyaya, JHEP 0710, 080 (2007);
    S. P. Martin, Phys. Rev. D79, 095019 (2009); J. Chakrabortty and A. Raychaudhuri, Phys. Lett. B 673, 57 (2009); Stephen P. Martin, Phys. Rev. D 89, 035011 (2014).
   \bibitem{Gogoladze:2014cha}
    I.~Gogoladze, F.~Nasir, Q.~Shafi and C.~S.~Un,
    Phys. Rev. D \textbf{90}, no.3, 035008 (2014)
    doi:10.1103/PhysRevD.90.035008
    [arXiv:1403.2337 [hep-ph]].
   \bibitem{murayama:pierce} H. Murayama and A. Pierce, Phys. Rev. D 65 (2002) 055009 [hep-ph/0108104].
   \bibitem{Babu:2020ncc}
    K.~S.~Babu, I.~Gogoladze and C.~S.~Un,
    [arXiv:2012.14411 [hep-ph]].
   \bibitem{Bajc:2002pg}
    B.~Bajc, P.~Fileviez Perez and G.~Senjanovic,
    [arXiv:hep-ph/0210374 [hep-ph]].
   \bibitem{Ibe:2013oha}
    M.~Ibe, T.~T.~Yanagida and N.~Yokozaki,
    JHEP \textbf{08}, 067 (2013)
    doi:10.1007/JHEP08(2013)067
    [arXiv:1303.6995 [hep-ph]].
   \bibitem{Babu:2014lwa}
    K.~S.~Babu, I.~Gogoladze, Q.~Shafi and C.~S.~\"Un,
    Phys. Rev. D \textbf{90}, no.11, 116002 (2014)
    doi:10.1103/PhysRevD.90.116002
    [arXiv:1406.6965 [hep-ph]].
   \bibitem{super-GUT:mSUGRA}
    J.~Ellis, A.~Mustafayev and K.~A.~Olive,
    Eur. Phys. J. C \textbf{69}, 201-217 (2010)
    doi:10.1140/epjc/s10052-010-1373-8
    [arXiv:1003.3677 [hep-ph]].
   \bibitem{sub-GUT:mSUGRA}
    J.~R.~Ellis, K.~A.~Olive and P.~Sandick,
    Phys. Lett. B \textbf{642}, 389-399 (2006)
    doi:10.1016/j.physletb.2006.09.043
    [arXiv:hep-ph/0607002 [hep-ph]].
   \bibitem{gSUGRA:WWY} F.~Wang, W.~Wang and J.~M.~Yang,
                      JHEP \textbf{06}, 079 (2015),
                     arXiv:1504.00505.
  \bibitem{gSUGRA:WWYZ}
   F.~Wang, K.~Wang, J.~M.~Yang and J.~Zhu,
    JHEP \textbf{12} (2018), 041
    arXiv:1808.10851.
\bibitem{hagiwara}  K. Hagiwara and Y. Yamada, Phys. Rev. Lett 70,709(1993).

\bibitem{Chkareuli:1998wi}
J.~L.~Chkareuli and I.~G.~Gogoladze,
Phys. Rev. D \textbf{58}, 055011 (1998)
doi:10.1103/PhysRevD.58.055011
[arXiv:hep-ph/9803335 [hep-ph]].
   \bibitem{Bajc:2002bv}
    B.~Bajc, P.~Fileviez Perez and G.~Senjanovic,
    Phys. Rev. D \textbf{66}, 075005 (2002)
    doi:10.1103/PhysRevD.66.075005
    [arXiv:hep-ph/0204311 [hep-ph]].
\bibitem{Zheng:2012pt}
J.~h.~Zheng and D.~X.~Zhang,
JHEP \textbf{02}, 046 (2012)
doi:10.1007/JHEP02(2012)046
[arXiv:1202.5072 [hep-ph]].
\bibitem{fix-point}  D.F. Litim, F. Sannino, JHEP12(2014)178 [arXiv:1406.2337]; \\
D.F. Litim, M. Mojaza, F. Sannino, JHEP01(2016)081 [arXiv:1501.03061].
  \bibitem{gmuon}
T. Moroi, 
Phys. Rev. D 53, 6565 (1996), 
hep-ph/9512396;\\
D. Stockinger,
 J. Phys. G 34, R45 (2007), hep-ph/0609168.
\bibitem{radiative:natural} H. Baer, V. Barger, P. Huang, A. Mustafayev, X. Tata, Phys. Rev.
Lett. 109, 161802 (2012);\\
 H. Baer, V. Barger, P. Huang, D. Mickelson, A. Mustafayev, X.
Tata, Phys. Rev. D 87, 115028 (2013).
\bibitem{Baer:2013gva}
H.~Baer, V.~Barger and D.~Mickelson,
Phys. Rev. D \textbf{88}, no.9, 095013 (2013)
doi:10.1103/PhysRevD.88.095013
[arXiv:1309.2984 [hep-ph]].
\bibitem{JUNO} F. An et al., [JUNO Collaboration], J. Phys. G 43(3), , 030401
(2016). arXiv:1507.05613 [physics.ins-det].
\bibitem{DUNE} R. Acciarri et al., [DUNE Collaboration], arXiv:1512.06148
[physics.ins-det]; B. Abi et al., [DUNE Collaboration], arXiv:1807.10334
[physics.ins-det].
\bibitem{hyperK} K. Abe et al., [Hyper-Kamiokande Collaboration],
arXiv:1805.04163 [physics.ins-det].
\bibitem{ellis:proton decay} John Ellis, L. Evans, Natsumi Nagata, Keith A. Olive, Liliana Velasco-Sevilla, Eur. Phys. J. C (2020) 80 :332.
\bibitem{superK} K. Abe et al., [Super-Kamiokande Collaboration], Phys. Rev. D
90(7), 072005 (2014) [arXiv:1408.1195]; V. Takhistov (for the Super-Kamiokande Collaboration),  [Super-Kamiokande Collaboration], [arXiv:1605.03235].
\bibitem{ellis:dim-5RGE} John ELLIS, D.V. NANOPOULOS, Serge RUDAZ, Nucl. Phys. B202 (1982) 43-62.
\bibitem{nihei:twoloop} Takeshi Nihei, Jiro Arafune, Prog.Theor.Phys. 93 (1995) 665-669.
\bibitem{hisano:dim-5RGE}J. Hisano, [arXiv:hep-ph/0004266 [hep-ph]];  Junji Hisano, Daiki Kobayashi, Takumi Kuwahara, Natsumi Nagata, JHEP 1307 (2013) 038.
\bibitem{Goto:dim-5 RRRR}  T. Goto, T. Nihei, Phys. Rev. D59, 115009 (1999).
\bibitem{nath:proton decay} P. Nath, P. Fileviez Perez, Physics Reports 441 (2007) 191-317.
\bibitem{GUT:large tanbeta} V. Lucas and S. Raby, Phys. Rev. D55, 6986 (1997);
 K.S. Babu and M.J. Strassler, hep-ph/9808447.
\bibitem{gluino:dressing cancelation} Pran Nath, A. H. Chamseddine, R. Arnowitt,
Phys. Rev. D 32, 2348 (1985); J. McDonald, C.E. Vayonakis, Phys. Lett. B 163, 148(1985).
  \bibitem{SuSpect252}
                      A.Djouadi, J.-L. Kneur and G. Moultaka. Comput. Phys. Commun. 176, 426 (2007), 
                      hep-ph/0211331.
  \bibitem{MicrOMEGAs527a} G. Belanger, F. Boudjema, A. Pukhov and A. Semenov, arXiv:1005.4133;
                          Comput. Phys. Commun. 180, 5 (2009),
                           arXiv:0803.2360;
                           Comput. Phys. Commun. 176, 367 (2007), hep-ph/0607059.
\bibitem{ATLAS:higgs} G. Aad et al. [ATLAS Collaboration], Phys. Lett. B 710, 49 (2012).
  \bibitem{CMS:higgs} S. Chatrachyan et al. [CMS Collaboration], Phys. Lett.B 710, 26 (2012).
  \bibitem{higgsbounds511} P. Bechtle, O. Brein, S. Heinemeyer, O. Stal, T. Stefaniak, G. Weiglein and K. E. Williams,
                           Eur. Phys. J. C 74, 2693 (2014), arXiv:1311.0055.
  \bibitem{higgssignal} P. Bechtle, S. Heinemeyer, O. Stal, T. Stefaniak and G. Weiglein, Eur. Phys. J. C 74, 2711 (2014),
  arXiv:1305.1933; JHEP 1411, 039 (2014), arXiv:1403.1582.
  \bibitem{LHCmass}
ATLAS Collboration, ATLAS-CONF-2017-020; CMS Collboration, CMS-SUS-16-051, CMSSUS-16-049;ATLAS Collboration, ATLAS-CONF-2017-021.
  \bibitem{LEPmass}
S. Schael et al. [ALEPH and DELPHI and L3 and OPAL and
SLD and LEP Electroweak Working Group and SLD Electroweak Group and SLD Heavy Flavour Group Collaborations],
Phys. Rept. 427, 257 (2006).
  \bibitem{Vaccumstability}
   T. Kitahara, T. Yoshinaga, JHEP 1305, 035(2013), arXiv:1303.0461 [hep-ph];\\
   M. Endo, K. Hamaguchi, T. Kitahara, T. Yoshinaga, JHEP 1311, 013(2013), arXiv:1309.3065 [hep-ph].
  \bibitem{BaBar-Bph}
                     J.~P.~Lees  et al. [BaBar Collaboration],
                       Phys.\ Rev.\ Lett.\  {\bf 109}, 191801 (2012), arXiv:1207.2690;
                       Phys.\ Rev.\ Lett.\  {\bf 109}, 101802 (2012), arXiv:1205.5442.
  \bibitem{LHCb-BsMuMu}
                       R. Aaij  et al. [LHCb Collaboration],
                       Phys.\ Rev.\ Lett.\  {\bf 110}, 021801  (2013), arXiv:1211.2674.
  \bibitem{Btaunu}
                   A.G. Akeroyd, S. Recksiegel,
                   J. Phys. G 29, 2311 (2003), hep-ph/0306037.
  \bibitem{Planck}
                   P. A. R. Ade, et. al. [Planck Collaboration],
A$\&$A, 594, A13 (2016), arXiv:1502.01589.
  \bibitem{CheckMATE}
  M. Drees, H. Dreiner, D. Schmeier, J. Tattersall and J. S. Kim, Comput. Phys. Commun. 187, 227 (2015), 
  arXiv:1312.2591;\\
  J. S. Kim, D. Schmeier, J. Tattersall and K. Rolbiecki, Comput. Phys. Commun. 196, 535 (2015), 
  arXiv:1503.01123;\\
  D. Dercks, N. Desai, J. S. Kim, K. Rolbiecki, J. Tattersall and T. Weber, Comput. Phys. Commun. 221, 383 (2017), 
  arXiv:1611.09856.
\bibitem{MG5}
J.~Alwall, R.~Frederix, S.~Frixione, V.~Hirschi, F.~Maltoni, O.~Mattelaer, H.~S.~Shao, T.~Stelzer, P.~Torrielli and M.~Zaro,
JHEP \textbf{07}, 079 (2014),
arXiv:1405.0301;\\
R.~Frederix, S.~Frixione, V.~Hirschi, D.~Pagani, H.~S.~Shao and M.~Zaro,
JHEP \textbf{07}, 185 (2018),
arXiv:1804.10017.
\bibitem{pythia}
T.~Sj\"ostrand, S.~Ask, J.~R.~Christiansen, R.~Corke, N.~Desai, P.~Ilten, S.~Mrenna, S.~Prestel, C.~O.~Rasmussen and P.~Z.~Skands,
Comput. Phys. Commun. \textbf{191}, 159  (2015), 
arXiv:1410.3012.
  \bibitem{CMSchi}
A. M. Sirunyan et al. [CMS collaboration],  
JHEP 03, 166 (2018), arXiv:1709.05406.
  \bibitem{ATLASsl}
G. Aad et al. [ATLAS collaboration],  
Eur. Phys. J. C 80, 123 (2020), arXiv:1908.08215.

  \bibitem{1704.04669} A. Aboubrahim, P. Nath, A. B. Spisak, Phys. Rev. D 95, 115030 (2017), arXiv:1704.04669.
  \bibitem{SI}
M. Badziak, M. Olechowski and P. Szczerbiak,
arXiv 1601.00768;\\
A. Pierce, N. R. Shah and K. Freese,
arXiv:1309.7351.
  \bibitem{SD}
M. Badziak, M. Olechowski and P. Szczerbiak,
JHEP 03 (2016) 179, arXiv:1512.02472;\\
M. Badziak, M. Olechowski and P. Szczerbiak, 
JHEP 07, 050 (2017), arXiv:1705.00227.
  \bibitem{lux} D.S. Akerib et al. Phys. Rev. Lett. 118, 021303 (2017). arXiv:1608.07648.
  \bibitem{xenon1t} E. Aprile et al. [XENON Collaboration],
  Phys. Rev. Lett. 121, 111302 (2018),
  arXiv:1805.12562;
  Phys. Rev. Lett. 122, 141301 (2019),
                    arXiv:1902.03234.
  \bibitem{pandax} C. Fu et al., Phys. Rev. Lett. 118, 071301 (2017), arXiv:1611.06553.
\bibitem{Gomez:2018zzw}
M.~E.~Gomez, S.~Lola, R.~Ruiz De Austri and Q.~Shafi,
JHEP \textbf{10}, 062 (2018)
doi:10.1007/JHEP10(2018)062
[arXiv:1806.06220 [hep-ph]].

\end{thebibliography}
\end{document}